\documentclass[aps,prl,nofootinbib,twocolumn,superscriptaddress,preprintnumbers]{revtex4-1}
\usepackage[nodisplayskipstretch]{setspace}
\usepackage{cancel}
\usepackage{xkeyval}

\usepackage{tikz,fp}
\usetikzlibrary{external,fixedpointarithmetic,decorations.pathmorphing,positioning,calc,fadings,decorations.markings,decorations.pathreplacing,patterns,arrows}

\tikzset{
  label distance=-3pt,
  >=stealth,
  inner sep=1.5pt,
  boson/.style={
    decoration={snake, segment length=2mm, amplitude=0.5mm},
    decorate,
  },
  graviton/.style={
  	line width=0.1pt,
    decoration={snake, segment length=0.7mm, amplitude=0.2mm},
    decorate,
  },
  quark/.style={
    postaction={decorate},
    decoration={markings,mark=at position .5 with {\draw[->] (-0.01,0) -- (0.07,0);}}
  },
  aquark/.style={
    postaction={decorate},
    decoration={markings,mark=at position .5 with {\draw[->] (0.01,0) -- (-0.07,0);}}
  },
  gluonOld/.style={
    line width=0.7pt,
    decorate,
    decoration={coil,aspect=-0.5,amplitude=-2pt,segment length=2.2pt}
  },
  gluon/.style={
    line width=0.5pt,
    decorate,
    decoration={coil,aspect=-0.75,amplitude=-1.5pt,segment length=2pt}
  },
  agluon/.style={
    line width=0.5pt,
    decorate,
    decoration={coil,aspect=0.75,amplitude=1.5pt,segment length=2pt}
  },
  scalar/.style={
    dashed
  },
  dscalar/.style={
    dashed,
    postaction={decorate},
    decoration={markings,mark=at position 0.65 with {\draw[->] (-0.01,0) -- (0.07,0);}}
  },
  adscalar/.style={
    dashed,
    postaction={decorate},
    decoration={markings,mark=at position 0.65 with {\draw[->] (-0.01,0) -- (0.07,0);}}
  },
  line/.style={
  },
  doubleline/.style={
    double
  },
  middleArrow/.style={
    draw=white,
    decoration={markings,mark=at position 0.65 with{\arrow[scale=1,black]{>}}},
    decorate
  },
  dot/.style={
  	postaction={decorate},
    decoration={markings,mark=between positions 0.5 and 0.5 step 1 with {\draw[fill] (0,0) circle [radius=1.5pt];}}  
  },
  dots/.style={
  	postaction={decorate},
    decoration={markings,mark=between positions 0.33 and 0.66 step 0.33 with {\draw[fill] (0,0) circle [radius=1.5pt];}}  
  },
  dotss/.style={
  	postaction={decorate},
    decoration={markings,mark=between positions 0.25 and 0.75 step 0.25 with {\draw[fill] (0,0) circle [radius=1.5pt];}}  
  },
  blob/.style={
  	pattern=north west lines,
    draw=black
  },
  blobBlack/.style={
    draw=black,
    fill=black
  },
  blobWhite/.style={
    draw=black,
    fill=white
  }
}

\makeatletter
\def\gSetStdExtTypes#1{\presetkeys{gTypes}{eA=#1,eB=#1,eC=#1,eD=#1,eE=#1}{}}
\def\gSetStdIntTypes#1{\presetkeys{gTypes}{iA=#1,iB=#1,iC=#1,iD=#1,iE=#1,iF=#1,iG=#1,iH=#1,iI=#1,iJ=#1}{}}
\def\gSetStdTypes#1{\gSetStdExtTypes{#1}\gSetStdIntTypes{#1}}

\define@cmdkeys{general}{scale}
\define@cmdkeys{general}{rotate}
\define@cmdkeys{general}{font}
\define@cmdkeys{general}{yshift}
\presetkeys{general}{scale=1, rotate=0, font=\scriptsize, yshift=0pt}{}

\define@cmdkeys{gTypes}{eA,eB,eC,eD,eE,iA,iB,iC,iD,iE,iF,iG,iH,iI,iJ}
\gSetStdTypes{line}
\define@key{pre}{all}{\gSetStdTypes{#1}}
\define@key{pre}{allE}{\gSetStdExtTypes{#1}}
\define@key{pre}{allI}{\gSetStdIntTypes{#1}}

\define@cmdkeys{gLabels}{eLA,eLB,eLC,eLD,eLE,iLA,iLB,iLC,iLD,iLE,iLF,iLG,iLH,iLI,iLJ}
\presetkeys{gLabels}{eLA={},eLB={},eLC={},eLD={},eLE={},iLA={},iLB={},iLC={},iLD={},iLE={},iLF={},iLG={},iLH={},iLI={},iLJ={}}{}

\define@boolkeys{dots}{lDots,rDots}
\presetkeys{dots}{lDots=false, rDots=false}{}

\define@cmdkeys{blobs}{blobA,blobB,blobC}
\presetkeys{blobs}{blobA=blob,blobB=blob,blobC=blob}{}

\def\gTreeTri{\@ifnextchar[\@gTreeTri{\@gTreeTri[]}}
\def\@gTreeTri[#1]#2{%
  \setkeys*{pre}{#1}%
  \setrmkeys{general,gTypes,gLabels}%
  \begin{tikzpicture}
    [line width=1pt,
    baseline={([yshift=-0.5ex+\cmdKV@general@yshift]current bounding box.center)},
    scale=\cmdKV@general@scale,
    rotate=\cmdKV@general@rotate,
    font=\cmdKV@general@font]
    \draw[\cmdKV@gTypes@eA] (0.7,0.3) -- (0.4,0.3) node[label=(-180+\cmdKV@general@rotate):\cmdKV@gLabels@eLA] {};
    \draw[\cmdKV@gTypes@eB] (0.7,0.3) -- (0.9,0.6) node[label=(\cmdKV@general@rotate):\cmdKV@gLabels@eLB] {};
    \draw[\cmdKV@gTypes@eC] (0.7,0.3) -- (0.9,0) node[label=(\cmdKV@general@rotate):\cmdKV@gLabels@eLC] {};
    #2
  \end{tikzpicture}
  \gSetStdTypes{line}
}

\def\gTreeS{\@ifnextchar[\@gTreeS{\@gTreeS[]}}
\def\@gTreeS[#1]#2{%
  \setkeys*{pre}{#1}%
  \setrmkeys{general,gTypes,gLabels}%
  \begin{tikzpicture}
    [line width=1pt,
    baseline={([yshift=-0.5ex+\cmdKV@general@yshift]current bounding box.center)},
    scale=\cmdKV@general@scale,
    rotate=\cmdKV@general@rotate,
    font=\cmdKV@general@font]
    \draw[\cmdKV@gTypes@eA] (0.2,0.3) -- (0,0) node[label=(-180+\cmdKV@general@rotate):\cmdKV@gLabels@eLA] {};
    \draw[\cmdKV@gTypes@eB] (0.2,0.3) -- (0,0.6) node[label=(180+\cmdKV@general@rotate):\cmdKV@gLabels@eLB] {};
    \draw[\cmdKV@gTypes@eC] (0.7,0.3) -- (0.9,0.6) node[label=(\cmdKV@general@rotate):\cmdKV@gLabels@eLC] {};
    \draw[\cmdKV@gTypes@eD] (0.7,0.3) -- (0.9,0) node[label=(\cmdKV@general@rotate):\cmdKV@gLabels@eLD] {};    
    \draw[label distance=0,\cmdKV@gTypes@iA] (0.2,0.3) -- node[label=(90+\cmdKV@general@rotate):\cmdKV@gLabels@iLA] {} (0.7,0.3);
    #2
  \end{tikzpicture}
  \gSetStdTypes{line}
}

\def\gTreeT{\@ifnextchar[\@gTreeT{\@gTreeT[]}}
\def\@gTreeT[#1]#2{%
  \setkeys*{pre}{#1}%
  \setrmkeys{general,gTypes,gLabels}%
  \begin{tikzpicture}
    [line width=1pt,
    baseline={([yshift=-0.5ex+\cmdKV@general@yshift]current bounding box.center)},
    scale=\cmdKV@general@scale,
    rotate=\cmdKV@general@rotate,
    font=\cmdKV@general@font]
    \draw[\cmdKV@gTypes@eA] (0.45,0.1) -- (0,0) node[label=(-180+\cmdKV@general@rotate):\cmdKV@gLabels@eLA] {};
    \draw[\cmdKV@gTypes@eB] (0.45,0.5) -- (0,0.6) node[label=(180+\cmdKV@general@rotate):\cmdKV@gLabels@eLB] {};
    \draw[\cmdKV@gTypes@eC] (0.45,0.5) -- (0.9,0.6) node[label=(\cmdKV@general@rotate):\cmdKV@gLabels@eLC] {};
    \draw[\cmdKV@gTypes@eD] (0.45,0.1) -- (0.9,0) node[label=(\cmdKV@general@rotate):\cmdKV@gLabels@eLD] {};
    \draw[label distance=0,\cmdKV@gTypes@iA] (0.45,0.5) -- node[label=(\cmdKV@general@rotate):\cmdKV@gLabels@iLA] {} (0.45,0.1);
    #2
  \end{tikzpicture}
  \gSetStdTypes{line}
}

\def\gTreeU{\@ifnextchar[\@gTreeU{\@gTreeU[]}}
\def\@gTreeU[#1]#2{%
  \setkeys*{pre}{#1}%
  \setrmkeys{general,gTypes,gLabels}%
  \begin{tikzpicture}
    [line width=1pt,
    baseline={([yshift=-0.5ex+\cmdKV@general@yshift]current bounding box.center)},
    scale=\cmdKV@general@scale,
    rotate=\cmdKV@general@rotate,
    font=\cmdKV@general@font]
    \draw[\cmdKV@gTypes@eA] (0.45,0.5) -- (0,0) node[label=(-180+\cmdKV@general@rotate):\cmdKV@gLabels@eLA] {};
    \foreach \x in {5,...,14} \fill[opacity=0.12,white] (0.27,0.3) circle (\x/100);
    \draw[\cmdKV@gTypes@eB] (0.45,0.1) -- (0,0.6) node[label=(180+\cmdKV@general@rotate):\cmdKV@gLabels@eLB] {};
    \draw[\cmdKV@gTypes@eC] (0.45,0.5) -- (0.9,0.6) node[label=(\cmdKV@general@rotate):\cmdKV@gLabels@eLC] {};
    \draw[\cmdKV@gTypes@eD] (0.45,0.1) -- (0.9,0) node[label=(\cmdKV@general@rotate):\cmdKV@gLabels@eLD] {};
    \draw[label distance=0,\cmdKV@gTypes@iA] (0.45,0.5) -- node[label=(\cmdKV@general@rotate):\cmdKV@gLabels@iLA] {} (0.45,0.1);
    #2
  \end{tikzpicture}
  \gSetStdTypes{line}
}

\def\gTreeFourPt{\@ifnextchar[\@gTreeFourPt{\@gTreeFourPt[]}}
\def\@gTreeFourPt[#1]#2{%
  \setkeys*{pre}{#1}%
  \setrmkeys{general,gTypes,gLabels}%
  \begin{tikzpicture}
    [line width=1pt,
    baseline={([yshift=-0.5ex+\cmdKV@general@yshift]current bounding box.center)},
    scale=\cmdKV@general@scale,
    rotate=\cmdKV@general@rotate,
    font=\cmdKV@general@font]
    \draw[\cmdKV@gTypes@eA] (0,0) -- (-0.4,-0.4) node[label=(-180+\cmdKV@general@rotate):\cmdKV@gLabels@eLA] {};
    \draw[\cmdKV@gTypes@eB] (0,0) -- (-0.4,0.4) node[label=(180+\cmdKV@general@rotate):\cmdKV@gLabels@eLB] {};
    \draw[\cmdKV@gTypes@eC] (0,0) -- (0.4,0.4) node[label=(\cmdKV@general@rotate):\cmdKV@gLabels@eLC] {};
    \draw[\cmdKV@gTypes@eD] (0,0) -- (0.4,-0.4) node[label=(\cmdKV@general@rotate):\cmdKV@gLabels@eLD] {};
    #2
  \end{tikzpicture}
  \gSetStdTypes{line}
}

\def\gTadA{\@ifnextchar[\@gTadA{\@gTadA[]}}
\def\@gTadA[#1]#2{%
  \setkeys*{pre}{#1}%
  \setrmkeys{general,gTypes,gLabels}%
  \begin{tikzpicture}
    [line width=1pt,
    baseline={([yshift=-0.5ex+\cmdKV@general@yshift]current bounding box.center)},
    scale=\cmdKV@general@scale,
    rotate=\cmdKV@general@rotate,
    font=\cmdKV@general@font]
    \draw[\cmdKV@gTypes@eA] (-0.7,0.3) -- (-0.9,0) node[label=(180+\cmdKV@general@rotate):\cmdKV@gLabels@eLA] {};
    \draw[\cmdKV@gTypes@eB] (-0.7,0.3) -- (-0.9,0.6) node[label=(180+\cmdKV@general@rotate):\cmdKV@gLabels@eLB] {};
    \draw[\cmdKV@gTypes@eC] (0.15,0.45) -- (0.5,0.6) node[label=(\cmdKV@general@rotate):\cmdKV@gLabels@eLC] {};
    \draw[\cmdKV@gTypes@eD] (-0.3,0.3) -- (-0.2,0) node[label=(\cmdKV@general@rotate):\cmdKV@gLabels@eLD] {};
    \draw[label distance=0,\cmdKV@gTypes@iA] (-0.7,0.3) -- node[label=(90+\cmdKV@general@rotate):\cmdKV@gLabels@iLA] {} (-0.3,0.3);
    \draw[label distance=0,\cmdKV@gTypes@iB] (-0.3,0.3) -- node[label=(-80+\cmdKV@general@rotate):\cmdKV@gLabels@iLB] {} (0.15,0.45);
    \draw[label distance=0,\cmdKV@gTypes@iC] (0.15,0.45) -- node[label=(45+\cmdKV@general@rotate):\cmdKV@gLabels@iLC] {} (-0.18,0.69);
    \draw[label distance=-0.5,\cmdKV@gTypes@iD] (-0.18,0.69) .. controls ($(-0.18,0.69) + (-0.12,-0.12)$) and ($(-0.35,0.87) + (-0.12,-0.12)$) .. (-0.35,0.87) node[label=(180+\cmdKV@general@rotate):\cmdKV@gLabels@iLD] {} .. controls ($(-0.35,0.87) + (0.12,0.12)$) and ($(-0.18,0.69) + (0.12,0.12)$) .. (-0.18,0.69);
    #2
  \end{tikzpicture}
  \gSetStdTypes{line}
}

\def\gTadB{\@ifnextchar[\@gTadB{\@gTadB[]}}
\def\@gTadB[#1]#2{%
  \setkeys*{pre}{#1}%
  \setrmkeys{general,gTypes,gLabels}%
  \begin{tikzpicture}
    [line width=1pt,
    baseline={([yshift=-0.5ex+\cmdKV@general@yshift]current bounding box.center)},
    scale=\cmdKV@general@scale,
    rotate=\cmdKV@general@rotate,
    font=\cmdKV@general@font]
    \draw[\cmdKV@gTypes@eA] (0,0.3) -- (-0.2,0) node[label=(-180+\cmdKV@general@rotate):\cmdKV@gLabels@eLA] {};
    \draw[\cmdKV@gTypes@eB] (0,0.3) -- (-0.2,0.6) node[label=(180+\cmdKV@general@rotate):\cmdKV@gLabels@eLB] {};
    \draw[\cmdKV@gTypes@eC] (0.7,0.3) -- (0.9,0.6) node[label=(\cmdKV@general@rotate):\cmdKV@gLabels@eLC] {};
    \draw[\cmdKV@gTypes@eD] (0.7,0.3) -- (0.9,0) node[label=(\cmdKV@general@rotate):\cmdKV@gLabels@eLD] {};
    \draw[label distance=0,\cmdKV@gTypes@iA] (0,0.3) -- node[label=(-90+\cmdKV@general@rotate):\cmdKV@gLabels@iLA] {} (0.35,0.3);
    \draw[label distance=0,\cmdKV@gTypes@iB] (0.35,0.3) -- node[label=(\cmdKV@general@rotate):\cmdKV@gLabels@iLB] {} (0.35,0.6);
    \draw[label distance=-0.5,\cmdKV@gTypes@iC] (0.35,0.6) .. controls (0.18,0.6) and (0.18,0.85) .. (0.35,0.85) .. controls (0.52,0.85) and (0.52,0.6) .. node[label=(\cmdKV@general@rotate):\cmdKV@gLabels@iLC] {} (0.35,0.6);
    \draw[label distance=0,\cmdKV@gTypes@iD] (0.35,0.3) -- node[label=(-90+\cmdKV@general@rotate):\cmdKV@gLabels@iLD] {} (0.7,0.3);
    #2
  \end{tikzpicture}
  \gSetStdTypes{line}
}
\def\gBubA{\@ifnextchar[\@gBubA{\@gBubA[]}}
\def\@gBubA[#1]#2{%
  \setkeys*{pre}{#1}%
  \setrmkeys{general,gTypes,gLabels}%
  \begin{tikzpicture}
    [line width=1pt,
    baseline={([yshift=-0.5ex+\cmdKV@general@yshift]current bounding box.center)},
    scale=\cmdKV@general@scale,
    rotate=\cmdKV@general@rotate,
    font=\cmdKV@general@font]
    \draw[\cmdKV@gTypes@eA] (0.5,0.7) -- (0.2,0.7) node[label=(180+\cmdKV@general@rotate):\cmdKV@gLabels@eLA] {};
    \draw[\cmdKV@gTypes@eB] (0.9,0.7) -- (1.2,0.7) node[label=(\cmdKV@general@rotate):\cmdKV@gLabels@eLB] {};
    \draw[label distance=0,\cmdKV@gTypes@iA] (0.5,0.7) .. controls (0.5,0.95) and (0.9,0.95) .. node[label=(90+\cmdKV@general@rotate):\cmdKV@gLabels@iLA] {} (0.9,0.7);
    \draw[label distance=0,\cmdKV@gTypes@iB] (0.9,0.7) .. controls (0.9,0.45) and (0.5,0.45) .. node[label=(-90+\cmdKV@general@rotate):\cmdKV@gLabels@iLB] {} (0.5,0.7);
    #2
  \end{tikzpicture}
  \gSetStdTypes{line}
}
\def\gBubB{\@ifnextchar[\@gBubB{\@gBubB[]}}
\def\@gBubB[#1]#2{%
  \setkeys*{pre}{#1}%
  \setrmkeys{general,gTypes,gLabels}%
  \begin{tikzpicture}
    [line width=1pt,
    baseline={([yshift=-0.5ex+\cmdKV@general@yshift]current bounding box.center)},
    scale=\cmdKV@general@scale,
    rotate=\cmdKV@general@rotate,
    font=\cmdKV@general@font]
    \draw[\cmdKV@gTypes@eA] (0.2,0.7) -- (0,0.4) node[label=(-180+\cmdKV@general@rotate):\cmdKV@gLabels@eLA] {};
    \draw[\cmdKV@gTypes@eB] (0.2,0.7) -- (0,1) node[label=(180+\cmdKV@general@rotate):\cmdKV@gLabels@eLB] {};
    \draw[\cmdKV@gTypes@eC] (1.2,0.7) -- (1.4,1) node[label=(\cmdKV@general@rotate):\cmdKV@gLabels@eLC] {};
    \draw[\cmdKV@gTypes@eD] (1.2,0.7) -- (1.4,0.4) node[label=(\cmdKV@general@rotate):\cmdKV@gLabels@eLD] {};
    \draw[label distance=0,\cmdKV@gTypes@iA] (0.2,0.7) -- node[label=(90+\cmdKV@general@rotate):\cmdKV@gLabels@iLA] {} (0.5,0.7);
    \draw[label distance=0,\cmdKV@gTypes@iB] (0.5,0.7) .. controls (0.5,0.95) and (0.9,0.95) .. node[label=(90+\cmdKV@general@rotate):\cmdKV@gLabels@iLB] {} (0.9,0.7);
    \draw[label distance=0,\cmdKV@gTypes@iC] (0.9,0.7) .. controls (0.9,0.45) and (0.5,0.45) .. node[label=(-90+\cmdKV@general@rotate):\cmdKV@gLabels@iLC] {} (0.5,0.7);
    \draw[label distance=0,\cmdKV@gTypes@iD] (0.9,0.7) -- node[label=(90+\cmdKV@general@rotate):\cmdKV@gLabels@iLD] {} (1.2,0.7);
    #2
  \end{tikzpicture}
  \gSetStdTypes{line}
}

\def\gBubC{\@ifnextchar[\@gBubC{\@gBubC[]}}
\def\@gBubC[#1]#2{%
  \setkeys*{pre}{#1}%
  \setrmkeys{general,gTypes,gLabels}%
  \begin{tikzpicture}
    [line width=1pt,
    baseline={([yshift=-0.5ex+\cmdKV@general@yshift]current bounding box.center)},
    scale=\cmdKV@general@scale,
    rotate=\cmdKV@general@rotate,
    font=\cmdKV@general@font]
    \draw[\cmdKV@gTypes@eA] (-0.7,0.3) -- (-0.9,0) node[label=(180+\cmdKV@general@rotate):\cmdKV@gLabels@eLA] {};
    \draw[\cmdKV@gTypes@eB] (-0.7,0.3) -- (-0.9,0.6) node[label=(180+\cmdKV@general@rotate):\cmdKV@gLabels@eLB] {};
    \draw[\cmdKV@gTypes@eC] (0.3,0.5) -- (0.5,0.6) node[label=(\cmdKV@general@rotate):\cmdKV@gLabels@eLC] {};
    \draw[\cmdKV@gTypes@eD] (-0.3,0.3) -- (-0.2,0) node[label=(\cmdKV@general@rotate):\cmdKV@gLabels@eLD] {};
    \draw[label distance=0,\cmdKV@gTypes@iA] (-0.7,0.3) -- node[label=(90+\cmdKV@general@rotate):\cmdKV@gLabels@iLA] {} (-0.3,0.3);
    \draw[label distance=0,\cmdKV@gTypes@iB] (-0.3,0.3) -- node[label=(90+\cmdKV@general@rotate):\cmdKV@gLabels@iLB] {} (-0,0.4);
    \draw[label distance=0,\cmdKV@gTypes@iC] (0,0.4) .. controls (-0.03,0.6) and (0.22,0.65) .. node[label=(90+\cmdKV@general@rotate):\cmdKV@gLabels@iLC] {} (0.3,0.5);
    \draw[label distance=0,\cmdKV@gTypes@iD] (0.3,0.5) .. controls (0.38,0.3) and (0.08,0.21) .. node[label=(-90+\cmdKV@general@rotate):\cmdKV@gLabels@iLD] {} (0,0.4);
    #2
  \end{tikzpicture}
  \gSetStdTypes{line}
}
\def\gTriA{\@ifnextchar[\@gTriA{\@gTriA[]}}
\def\@gTriA[#1]#2{%
  \setkeys*{pre}{#1}%
  \setrmkeys{general,gTypes,gLabels}%
  \begin{tikzpicture}
    [line width=1pt,
    baseline={([yshift=-0.5ex+\cmdKV@general@yshift]current bounding box.center)},
    scale=\cmdKV@general@scale,
    rotate=\cmdKV@general@rotate,
    font=\cmdKV@general@font]
    \draw[\cmdKV@gTypes@eC] (0:0.3) -- (0:0.55) node[label=(\cmdKV@general@rotate):\cmdKV@gLabels@eLC] {};
    \draw[\cmdKV@gTypes@eA] (-120:0.3) -- (-120:0.55) node[label=(-180+\cmdKV@general@rotate):\cmdKV@gLabels@eLA] {};
    \draw[\cmdKV@gTypes@eB] (120:0.3) -- (120:0.55) node[label=(180+\cmdKV@general@rotate):\cmdKV@gLabels@eLB] {};    
    \draw[label distance=0,\cmdKV@gTypes@iA] (-120:0.3) -- node[label=(180+\cmdKV@general@rotate):\cmdKV@gLabels@iLA] {} (120:0.3);
    \draw[label distance=0,\cmdKV@gTypes@iB] (120:0.3) -- node[label=(60+\cmdKV@general@rotate):\cmdKV@gLabels@iLB] {} (0:0.3);
    \draw[label distance=0,\cmdKV@gTypes@iC] (0:0.3) -- node[label=(-60+\cmdKV@general@rotate):\cmdKV@gLabels@iLC] {} (-120:0.3);
    #2
  \end{tikzpicture}
  \gSetStdTypes{line}
}
\def\gTriB{\@ifnextchar[\@gTriB{\@gTriB[]}}
\def\@gTriB[#1]#2{%
  \setkeys*{pre}{#1}%
  \setrmkeys{general,gTypes,gLabels}%
  \begin{tikzpicture}
    [line width=1pt,
    baseline={([yshift=-0.5ex+\cmdKV@general@yshift]current bounding box.center)},
    scale=\cmdKV@general@scale,
    rotate=\cmdKV@general@rotate,
    font=\cmdKV@general@font]
    \draw[\cmdKV@gTypes@eA] (0.3,0.45) -- (0,0.3) node[label=(-180+\cmdKV@general@rotate):\cmdKV@gLabels@eLA] {};
    \draw[\cmdKV@gTypes@eB] (0.3,0.95) -- (0,1.1) node[label=(180+\cmdKV@general@rotate):\cmdKV@gLabels@eLB] {};
    \draw[\cmdKV@gTypes@eC] (1,0.7) -- (1.3,1.1) node[label=(\cmdKV@general@rotate):\cmdKV@gLabels@eLC] {};
    \draw[\cmdKV@gTypes@eD] (1,0.7) -- (1.3,0.3) node[label=(\cmdKV@general@rotate):\cmdKV@gLabels@eLD] {};
    \draw[label distance=0,\cmdKV@gTypes@iA] (0.3,0.45) -- node[label=(180+\cmdKV@general@rotate):\cmdKV@gLabels@iLA] {} (0.3,0.95);
    \draw[label distance=0,\cmdKV@gTypes@iB] (0.3,0.95) -- node[label=(70+\cmdKV@general@rotate):\cmdKV@gLabels@iLB] {} (0.7,0.7);
    \draw[label distance=0,\cmdKV@gTypes@iC] (0.7,0.7) -- node[label=(-70+\cmdKV@general@rotate):\cmdKV@gLabels@iLC] {} (0.3,0.45);
    \draw[label distance=0,\cmdKV@gTypes@iD] (1,0.7) -- node[label=(90+\cmdKV@general@rotate):\cmdKV@gLabels@iLD] {} (0.7,0.7);
    #2
  \end{tikzpicture}
  \gSetStdTypes{line}
}
\def\gBox{\@ifnextchar[\@gBox{\@gBox[]}}
\def\@gBox[#1]#2{%
  \setkeys*{pre}{#1}%
  \setrmkeys{general,gTypes,gLabels}%
  \begin{tikzpicture}
    [line width=1pt,
    baseline={([yshift=-0.5ex+\cmdKV@general@yshift]current bounding box.center)},
    scale=\cmdKV@general@scale,
    rotate=\cmdKV@general@rotate,
    font=\cmdKV@general@font]
    \draw[\cmdKV@gTypes@eA] (0.25,0.25) -- (0,0) node[label=(-180+\cmdKV@general@rotate):\cmdKV@gLabels@eLA] {};
    \draw[\cmdKV@gTypes@eB] (0.25,0.75) -- (0,1) node[label=(180+\cmdKV@general@rotate):\cmdKV@gLabels@eLB] {};
    \draw[\cmdKV@gTypes@eC] (0.75,0.75) -- (1,1) node[label=(\cmdKV@general@rotate):\cmdKV@gLabels@eLC] {};
    \draw[\cmdKV@gTypes@eD] (0.75,0.25) -- (1,0) node[label=(\cmdKV@general@rotate):\cmdKV@gLabels@eLD] {};
    \draw[label distance=0,\cmdKV@gTypes@iA] (0.25,0.25) -- node[label=(180+\cmdKV@general@rotate):\cmdKV@gLabels@iLA] {} (0.25,0.75);
    \draw[label distance=0,\cmdKV@gTypes@iB] (0.25,0.75) -- node[label=(90+\cmdKV@general@rotate):\cmdKV@gLabels@iLB] {} (0.75,0.75);
    \draw[label distance=0,\cmdKV@gTypes@iC] (0.75,0.75) -- node[label=(\cmdKV@general@rotate):\cmdKV@gLabels@iLC] {} (0.75,0.25);
    \draw[label distance=0,\cmdKV@gTypes@iD] (0.75,0.25) -- node[label=(-90+\cmdKV@general@rotate):\cmdKV@gLabels@iLD] {} (0.25,0.25);
    #2
  \end{tikzpicture}
  \gSetStdTypes{line}
}
\def\gPenta{\@ifnextchar[\@gPenta{\@gPenta[]}}
\def\@gPenta[#1]#2{%
  \setkeys*{pre}{#1}%
  \setrmkeys{general,gTypes,gLabels}%
  \begin{tikzpicture}
    [line width=1pt,
    baseline={([yshift=-0.5ex+\cmdKV@general@yshift]current bounding box.center)},
    scale=\cmdKV@general@scale,
    rotate=\cmdKV@general@rotate,
    font=\cmdKV@general@font]
    \draw[\cmdKV@gTypes@eA] (-144:0.35) -- (-144:0.75) node[label=(-144+\cmdKV@general@rotate):\cmdKV@gLabels@eLA] {};
    \draw[\cmdKV@gTypes@eB] (144:0.35) -- (144:0.75) node[label=(144+\cmdKV@general@rotate):\cmdKV@gLabels@eLB] {};
    \draw[\cmdKV@gTypes@eC] (72:0.35) -- (72:0.75) node[label=(72+\cmdKV@general@rotate):\cmdKV@gLabels@eLC] {};
    \draw[\cmdKV@gTypes@eD] (0:0.35) -- (0:0.75) node[label=(\cmdKV@general@rotate):\cmdKV@gLabels@eLD] {};
    \draw[\cmdKV@gTypes@eE] (-72:0.35) -- (-72:0.75) node[label=(-72+\cmdKV@general@rotate):\cmdKV@gLabels@eLE] {};
    \draw[label distance=0,\cmdKV@gTypes@iA] (-144:0.35) -- node[label=(180+\cmdKV@general@rotate):\cmdKV@gLabels@iLA] {} (144:0.35);
    \draw[label distance=0,\cmdKV@gTypes@iB] (144:0.35) -- node[label=(108+\cmdKV@general@rotate):\cmdKV@gLabels@iLB] {} (72:0.35);
    \draw[label distance=0,\cmdKV@gTypes@iC] (72:0.35) -- node[label=(36+\cmdKV@general@rotate):\cmdKV@gLabels@iLC] {} (0:0.35);
    \draw[label distance=0,\cmdKV@gTypes@iD] (0:0.35) -- node[label=(-36+\cmdKV@general@rotate):\cmdKV@gLabels@iLD] {} (-72:0.35);
    \draw[label distance=0,\cmdKV@gTypes@iE] (-72:0.35) -- node[label=(-108+\cmdKV@general@rotate):\cmdKV@gLabels@iLE] {} (-144:0.35);
    #2
  \end{tikzpicture}
  \gSetStdTypes{line}
}

\def\gBoxBox{\@ifnextchar[\@gBoxBox{\@gBoxBox[]}}
\def\@gBoxBox[#1]#2{%
  \setkeys*{pre}{#1}%
  \setrmkeys{general,gTypes,gLabels}%
  \begin{tikzpicture}
    [line width=1pt,
    baseline={([yshift=-0.5ex+\cmdKV@general@yshift]current bounding box.center)},
    scale=\cmdKV@general@scale,
    rotate=\cmdKV@general@rotate,
    font=\cmdKV@general@font]
    \draw[\cmdKV@gTypes@eA] (0.3,0.3) -- (0,0) node[label=(-180+\cmdKV@general@rotate):\cmdKV@gLabels@eLA] {};
    \draw[\cmdKV@gTypes@eB] (0.3,0.9) -- (0,1.2) node[label=(180+\cmdKV@general@rotate):\cmdKV@gLabels@eLB] {};
    \draw[\cmdKV@gTypes@eC] (1.5,0.9) -- (1.8,1.2) node[label=(\cmdKV@general@rotate):\cmdKV@gLabels@eLC] {};
    \draw[\cmdKV@gTypes@eD] (1.5,0.3) -- (1.8,0) node[label=(\cmdKV@general@rotate):\cmdKV@gLabels@eLD] {};
    \draw[label distance=0,\cmdKV@gTypes@iA] (0.3,0.3) -- node[label=(180+\cmdKV@general@rotate):\cmdKV@gLabels@iLA] {} (0.3,0.9);
    \draw[label distance=0,\cmdKV@gTypes@iB] (0.3,0.9) -- node[label=(90+\cmdKV@general@rotate):\cmdKV@gLabels@iLB] {} (0.9,0.9);
    \draw[label distance=0,\cmdKV@gTypes@iC] (0.9,0.9) -- node[label=(90+\cmdKV@general@rotate):\cmdKV@gLabels@iLC] {} (1.5,0.9);
    \draw[label distance=0,\cmdKV@gTypes@iD] (1.5,0.9) -- node[label=(\cmdKV@general@rotate):\cmdKV@gLabels@iLD] {} (1.5,0.3);
    \draw[label distance=0,\cmdKV@gTypes@iE] (1.5,0.3) -- node[label=(-90+\cmdKV@general@rotate):\cmdKV@gLabels@iLE] {} (0.9,0.3);
    \draw[label distance=0,\cmdKV@gTypes@iF] (0.9,0.3) -- node[label=(-90+\cmdKV@general@rotate):\cmdKV@gLabels@iLF] {} (0.3,0.3);
    \draw[label distance=0,\cmdKV@gTypes@iG] (0.9,0.9) -- node[label=(\cmdKV@general@rotate):\cmdKV@gLabels@iLG] {} (0.9,0.3);
    #2
  \end{tikzpicture}
  \gSetStdTypes{line}
}
\def\gBoxBoxNP{\@ifnextchar[\@gBoxBoxNP{\@gBoxBoxNP[]}}
\def\@gBoxBoxNP[#1]#2{%
  \setkeys*{pre}{#1}%
  \setrmkeys{general,gTypes,gLabels}%
  \begin{tikzpicture}
    [line width=1pt,
    baseline={([yshift=-0.5ex+\cmdKV@general@yshift]current bounding box.center)},
    scale=\cmdKV@general@scale,
    rotate=\cmdKV@general@rotate,
    font=\cmdKV@general@font]
    \draw[\cmdKV@gTypes@eA] (-1.9,0.3) -- (-2.2,0) node[label=(180+\cmdKV@general@rotate):\cmdKV@gLabels@eLA] {};
    \draw[\cmdKV@gTypes@eB] (-1.9,1.1) -- (-2.2,1.4) node[label=(180+\cmdKV@general@rotate):\cmdKV@gLabels@eLB] {};
    \draw[\cmdKV@gTypes@eC] (-0.3,0.7) -- (0,0.7) node[label=(\cmdKV@general@rotate):\cmdKV@gLabels@eLC] {};
    \draw[\cmdKV@gTypes@eD] (-1.1,0.7) -- (-1.4,0.7) node[label=(180+\cmdKV@general@rotate):\cmdKV@gLabels@eLD] {};
    \draw[label distance=0,\cmdKV@gTypes@iA] (-1.9,0.3) -- node[label=(180+\cmdKV@general@rotate):\cmdKV@gLabels@iLA] {} (-1.9,1.1);
    \draw[label distance=0,\cmdKV@gTypes@iB] (-1.9,1.1) -- node[label=(90+\cmdKV@general@rotate):\cmdKV@gLabels@iLB] {} (-0.7,1.1);
    \draw[label distance=0,\cmdKV@gTypes@iC] (-0.7,1.1) -- node[label=(45+\cmdKV@general@rotate):\cmdKV@gLabels@iLC] {} (-0.3,0.7);
    \draw[label distance=0,\cmdKV@gTypes@iD] (-0.3,0.7) -- node[label=(-45+\cmdKV@general@rotate):\cmdKV@gLabels@iLD] {} (-0.7,0.3);
    \draw[label distance=0,\cmdKV@gTypes@iE] (-0.7,0.3) -- node[label=(-90+\cmdKV@general@rotate):\cmdKV@gLabels@iLE] {} (-1.9,0.3);
    \draw[label distance=0,\cmdKV@gTypes@iF] (-0.7,1.1) -- node[label=(180+\cmdKV@general@rotate):\cmdKV@gLabels@iLF] {} (-1.1,0.7);
    \draw[label distance=0,\cmdKV@gTypes@iG] (-1.1,0.7) -- node[label=(-180+\cmdKV@general@rotate):\cmdKV@gLabels@iLG] {} (-0.7,0.3);
    #2
  \end{tikzpicture}
  \gSetStdTypes{line}
}

\def\gBoxBoxNPB{\@ifnextchar[\@gBoxBoxNPB{\@gBoxBoxNPB[]}}
\def\@gBoxBoxNPB[#1]#2{%
  \setkeys*{pre}{#1}%
  \setrmkeys{general,gTypes,gLabels}%
  \begin{tikzpicture}
    [line width=1pt,
    baseline={([yshift=-0.5ex+\cmdKV@general@yshift]current bounding box.center)},
    scale=\cmdKV@general@scale,
    rotate=\cmdKV@general@rotate,
    font=\cmdKV@general@font]
    \draw[\cmdKV@gTypes@eA] (0.3,0.3) -- (0,0) node[label=(-180+\cmdKV@general@rotate):\cmdKV@gLabels@eLA] {};
    \draw[\cmdKV@gTypes@eB] (0.3,0.9) -- (0,1.2) node[label=(180+\cmdKV@general@rotate):\cmdKV@gLabels@eLB] {};
    \draw[\cmdKV@gTypes@eC] (1.5,0.9) -- (1.8,1.2) node[label=(\cmdKV@general@rotate):\cmdKV@gLabels@eLC] {};
    \draw[\cmdKV@gTypes@eD] (1.5,0.3) -- (1.8,0) node[label=(\cmdKV@general@rotate):\cmdKV@gLabels@eLD] {};
    \draw[label distance=0,\cmdKV@gTypes@iA] (0.3,0.3) -- node[label=(180+\cmdKV@general@rotate):\cmdKV@gLabels@iLA] {} (0.3,0.9);
    \draw[label distance=0,\cmdKV@gTypes@iB] (0.3,0.9) -- node[label=(90+\cmdKV@general@rotate):\cmdKV@gLabels@iLB] {} (0.9,0.9);
    \draw[label distance=0,\cmdKV@gTypes@iC] (0.9,0.9) -- node[label=(90+\cmdKV@general@rotate):\cmdKV@gLabels@iLC] {} (1.5,0.3);
    \foreach \x in {5,...,14} \fill[opacity=0.12,white] (1.2,0.6) circle (\x/100);
    \draw[label distance=0,\cmdKV@gTypes@iD] (1.5,0.3) -- node[label=(\cmdKV@general@rotate):\cmdKV@gLabels@iLD] {} (1.5,0.9);
    \draw[label distance=0,\cmdKV@gTypes@iE] (1.5,0.9) -- node[label=(-90+\cmdKV@general@rotate):\cmdKV@gLabels@iLE] {} (0.9,0.3);
    \draw[label distance=0,\cmdKV@gTypes@iF] (0.9,0.3) -- node[label=(-90+\cmdKV@general@rotate):\cmdKV@gLabels@iLF] {} (0.3,0.3);
    \draw[label distance=-1,\cmdKV@gTypes@iG] (0.9,0.9) -- node[label=(180+\cmdKV@general@rotate):\cmdKV@gLabels@iLG] {} (0.9,0.3);
    #2
  \end{tikzpicture}
  \gSetStdTypes{line}
}

\def\gBoxBoxNPC{\@ifnextchar[\@gBoxBoxNPC{\@gBoxBoxNPC[]}}
\def\@gBoxBoxNPC[#1]#2{%
  \setkeys*{pre}{#1}%
  \setrmkeys{general,gTypes,gLabels}%
  \begin{tikzpicture}
    [line width=1pt,
    baseline={([yshift=-0.5ex+\cmdKV@general@yshift]current bounding box.center)},
    scale=\cmdKV@general@scale,
    rotate=\cmdKV@general@rotate,
    font=\cmdKV@general@font]
    \draw[\cmdKV@gTypes@eA] (0.3,0.3) -- (0,0) node[label=(-180+\cmdKV@general@rotate):\cmdKV@gLabels@eLA] {};
    \draw[\cmdKV@gTypes@eB] (0.3,0.9) -- (0,1.2) node[label=(180+\cmdKV@general@rotate):\cmdKV@gLabels@eLB] {};
    \draw[\cmdKV@gTypes@eC] (1.5,0.9) -- (1.8,1.2) node[label=(\cmdKV@general@rotate):\cmdKV@gLabels@eLC] {};
    \draw[\cmdKV@gTypes@eD] (1.5,0.3) -- (1.8,0) node[label=(\cmdKV@general@rotate):\cmdKV@gLabels@eLD] {};
    \draw[label distance=0,\cmdKV@gTypes@iA] (0.3,0.3) -- node[label=(180+\cmdKV@general@rotate):\cmdKV@gLabels@iLA] {} (0.3,0.9);
    \draw[label distance=0,\cmdKV@gTypes@iB] (0.3,0.9) -- node[label=(90+\cmdKV@general@rotate):\cmdKV@gLabels@iLB] {} (0.9,0.9);
    \draw[label distance=0,\cmdKV@gTypes@iC] (0.9,0.9) -- node[label=(90+\cmdKV@general@rotate):\cmdKV@gLabels@iLC] {} (1.5,0.9);
    \draw[label distance=0,\cmdKV@gTypes@iD] (1.5,0.9) -- node[label=(\cmdKV@general@rotate):\cmdKV@gLabels@iLD] {} (0.9,0.3);
    \foreach \x in {5,...,14} \fill[opacity=0.12,white] (1.2,0.6) circle (\x/100);
    \draw[label distance=0,\cmdKV@gTypes@iE] (1.5,0.3) -- node[label=(-90+\cmdKV@general@rotate):\cmdKV@gLabels@iLE] {} (0.9,0.3);
    \draw[label distance=0,\cmdKV@gTypes@iF] (0.9,0.3) -- node[label=(-90+\cmdKV@general@rotate):\cmdKV@gLabels@iLF] {} (0.3,0.3);
    \draw[label distance=-1,\cmdKV@gTypes@iG] (0.9,0.9) -- node[label=(180+\cmdKV@general@rotate):\cmdKV@gLabels@iLG] {} (1.5,0.3);
    #2
  \end{tikzpicture}
  \gSetStdTypes{line}
}

\def\gBoxTriNPA{\@ifnextchar[\@gBoxTriNPA{\@gBoxTriNPA[]}}
\def\@gBoxTriNPA[#1]#2{%
  \setkeys*{pre}{#1}%
  \setrmkeys{general,gTypes,gLabels}%
  \begin{tikzpicture}
    [line width=1pt,
    baseline={([yshift=-0.5ex+\cmdKV@general@yshift]current bounding box.center)},
    scale=\cmdKV@general@scale,
    rotate=\cmdKV@general@rotate,
    font=\cmdKV@general@font]
    \draw[\cmdKV@gTypes@eA] (0.3,0.6) -- (0,0) node[label=(-180+\cmdKV@general@rotate):\cmdKV@gLabels@eLA] {};
    \draw[\cmdKV@gTypes@eB] (0.3,0.6) -- (0,1.2) node[label=(180+\cmdKV@general@rotate):\cmdKV@gLabels@eLB] {};
    \draw[\cmdKV@gTypes@eC] (1.5,0.9) -- (1.8,1.2) node[label=(\cmdKV@general@rotate):\cmdKV@gLabels@eLC] {};
    \draw[\cmdKV@gTypes@eD] (1.5,0.3) -- (1.8,0) node[label=(\cmdKV@general@rotate):\cmdKV@gLabels@eLD] {};
    \draw[label distance=0,\cmdKV@gTypes@iA] (0.3,0.6) -- node[label=(90+\cmdKV@general@rotate):\cmdKV@gLabels@iLA] {} (0.9,0.9);
    \draw[label distance=0,\cmdKV@gTypes@iB] (0.9,0.9) -- node[label=(90+\cmdKV@general@rotate):\cmdKV@gLabels@iLB] {} (1.5,0.9);
    \draw[label distance=-1,\cmdKV@gTypes@iC] (0.9,0.9) -- node[label=(180+\cmdKV@general@rotate):\cmdKV@gLabels@iLC] {} (1.5,0.3);
    \foreach \x in {5,...,14} \fill[opacity=0.12,white] (1.2,0.6) circle (\x/100);
    \draw[label distance=0,\cmdKV@gTypes@iD] (1.5,0.3) -- node[label=(-90+\cmdKV@general@rotate):\cmdKV@gLabels@iLD] {} (0.9,0.3);
    \draw[label distance=0,\cmdKV@gTypes@iE] (0.9,0.3) -- node[label=(-90+\cmdKV@general@rotate):\cmdKV@gLabels@iLE] {} (0.3,0.6);
    \draw[label distance=-1,\cmdKV@gTypes@iF] (1.5,0.9) -- node[label=(180+\cmdKV@general@rotate):\cmdKV@gLabels@iLF] {} (0.9,0.3);
    #2
  \end{tikzpicture}
  \gSetStdTypes{line}
}

\def\gTriPenta{\@ifnextchar[\@gTriPenta{\@gTriPenta[]}}
\def\@gTriPenta[#1]#2{%
  \setkeys*{pre}{#1}%
  \setrmkeys{general,gTypes,gLabels}%
  \begin{tikzpicture}
    [line width=1pt,
    baseline={([yshift=-0.5ex+\cmdKV@general@yshift]current bounding box.center)},
    scale=\cmdKV@general@scale,
    rotate=\cmdKV@general@rotate,
    font=\cmdKV@general@font]
    \draw[\cmdKV@gTypes@eA] (-108:0.5) -- (-135:1) node[label=(145+\cmdKV@general@rotate):\cmdKV@gLabels@eLA] {};
    \draw[\cmdKV@gTypes@eB] (180:0.5) -- (180:1) node[label=(180+\cmdKV@general@rotate):\cmdKV@gLabels@eLB] {};
    \draw[\cmdKV@gTypes@eC] (108:0.5) -- (135:1) node[label=(-145+\cmdKV@general@rotate):\cmdKV@gLabels@eLC] {};
    \draw[\cmdKV@gTypes@eD] (0:1.2) -- (0:1.5) node[label=(\cmdKV@general@rotate):\cmdKV@gLabels@eLD] {};
    \draw[label distance=0,\cmdKV@gTypes@iA] (-108:0.5) -- node[label=(-135+\cmdKV@general@rotate):\cmdKV@gLabels@iLA] {} (-180:0.5);
    \draw[label distance=0,\cmdKV@gTypes@iB] (180:0.5) -- node[label=(135+\cmdKV@general@rotate):\cmdKV@gLabels@iLB] {} (108:0.5);
    \draw[label distance=0,\cmdKV@gTypes@iC] (108:0.5) -- node[label=(72+\cmdKV@general@rotate):\cmdKV@gLabels@iLC] {} (36:0.5);
    \draw[label distance=0,\cmdKV@gTypes@iD] (36:0.5) -- node[label=(72+\cmdKV@general@rotate):\cmdKV@gLabels@iLD] {} (0:1.2);
    \draw[label distance=2,\cmdKV@gTypes@iE] (0:1.2) -- node[label=(-90+\cmdKV@general@rotate):\cmdKV@gLabels@iLE] {} (-36:0.5);
    \draw[label distance=2,\cmdKV@gTypes@iF] (-36:0.5) -- node[label=(-90+\cmdKV@general@rotate):\cmdKV@gLabels@iLF] {} (-108:0.5);
    \draw[label distance=0,\cmdKV@gTypes@iG] (36:0.5) -- node[label=(180+\cmdKV@general@rotate):\cmdKV@gLabels@iLG] {} (-36:0.5);
    #2
  \end{tikzpicture}
  \gSetStdTypes{line}
}
\def\gTriTri{\@ifnextchar[\@gTriTri{\@gTriTri[]}}
\def\@gTriTri[#1]#2{%
  \setkeys*{pre}{#1}%
  \setrmkeys{general,gTypes,gLabels}%
  \begin{tikzpicture}
    [line width=1pt,
    baseline={([yshift=-0.5ex+\cmdKV@general@yshift]current bounding box.center)},
    scale=\cmdKV@general@scale,
    rotate=\cmdKV@general@rotate,
    font=\cmdKV@general@font]
    \draw[\cmdKV@gTypes@eA] (0.3,0.3) -- (0,0) node[label=(-180+\cmdKV@general@rotate):\cmdKV@gLabels@eLA] {};
    \draw[\cmdKV@gTypes@eB] (0.3,1.1) -- (0,1.4) node[label=(180+\cmdKV@general@rotate):\cmdKV@gLabels@eLB] {};
    \draw[\cmdKV@gTypes@eC] (1.9,1.1) -- (2.2,1.4) node[label=(\cmdKV@general@rotate):\cmdKV@gLabels@eLC] {};
    \draw[\cmdKV@gTypes@eD] (1.9,0.3) -- (2.2,0) node[label=(\cmdKV@general@rotate):\cmdKV@gLabels@eLD] {};
    \draw[label distance=0,\cmdKV@gTypes@iA] (0.3,0.3) -- node[label=(180+\cmdKV@general@rotate):\cmdKV@gLabels@iLA] {} (0.3,1.1);
    \draw[label distance=0,\cmdKV@gTypes@iB] (0.3,1.1) -- node[label=(45+\cmdKV@general@rotate):\cmdKV@gLabels@iLB] {} (0.9,0.7);
    \draw[label distance=0,\cmdKV@gTypes@iC] (0.9,0.7) -- node[label=(-45+\cmdKV@general@rotate):\cmdKV@gLabels@iLC] {} (0.3,0.3);
    \draw[label distance=0,\cmdKV@gTypes@iD] (0.9,0.7) -- node[label=(90+\cmdKV@general@rotate):\cmdKV@gLabels@iLD] {} (1.3,0.7);
    \draw[label distance=0,\cmdKV@gTypes@iE] (1.3,0.7) -- node[label=(135+\cmdKV@general@rotate):\cmdKV@gLabels@iLE] {} (1.9,1.1);
    \draw[label distance=0,\cmdKV@gTypes@iF] (1.9,1.1) -- node[label=(\cmdKV@general@rotate):\cmdKV@gLabels@iLF] {} (1.9,0.3);
    \draw[label distance=0,\cmdKV@gTypes@iG] (1.9,0.3) -- node[label=(-135+\cmdKV@general@rotate):\cmdKV@gLabels@iLG] {} (1.3,0.7);
    #2
  \end{tikzpicture}
  \gSetStdTypes{line}
}

\def\gBoxBubA{\@ifnextchar[\@gBoxBubA{\@gBoxBubA[]}}
\def\@gBoxBubA[#1]#2{%
  \setkeys*{pre}{#1}%
  \setrmkeys{general,gTypes,gLabels}%
  \begin{tikzpicture}
    [line width=1pt,
    baseline={([yshift=-0.5ex+\cmdKV@general@yshift]current bounding box.center)},
    scale=\cmdKV@general@scale,
    rotate=\cmdKV@general@rotate,
    font=\cmdKV@general@font]
    \draw[\cmdKV@gTypes@eA] (0.1,0.1) -- (-0.1,-0.1) node[label=(-180+\cmdKV@general@rotate):\cmdKV@gLabels@eLA] {};
    \draw[\cmdKV@gTypes@eB] (0.1,0.9) -- (-0.1,1.1) node[label=(180+\cmdKV@general@rotate):\cmdKV@gLabels@eLB] {};
    \draw[\cmdKV@gTypes@eC] (0.9,0.9) -- (1.1,1.1) node[label=(\cmdKV@general@rotate):\cmdKV@gLabels@eLC] {};
    \draw[\cmdKV@gTypes@eD] (0.9,0.1) -- (1.1,-0.1) node[label=(\cmdKV@general@rotate):\cmdKV@gLabels@eLD] {};
    \draw[label distance=0,\cmdKV@gTypes@iA] (0.1,0.1) -- node[label=(180+\cmdKV@general@rotate):\cmdKV@gLabels@iLA] {} (0.1,0.9);
    \draw[label distance=0,\cmdKV@gTypes@iB] (0.1,0.9) -- node[label=(90+\cmdKV@general@rotate):\cmdKV@gLabels@iLB] {} (0.32,0.9);
    \draw[label distance=0,\cmdKV@gTypes@iC] (0.32,0.9) .. controls (0.32,1.15) and (0.68,1.15) .. node[label=(90+\cmdKV@general@rotate):\cmdKV@gLabels@iLC] {} (0.68,0.9);
    \draw[label distance=0,\cmdKV@gTypes@iD] (0.68,0.9) .. controls (0.68,0.65) and (0.32,0.65) .. node[label=(-90+\cmdKV@general@rotate):\cmdKV@gLabels@iLD] {} (0.32,0.9);
    \draw[label distance=0,\cmdKV@gTypes@iE] (0.68,0.9) -- node[label=(90+\cmdKV@general@rotate):\cmdKV@gLabels@iLE] {} (0.9,0.9);
    \draw[label distance=0,\cmdKV@gTypes@iF] (0.9,0.9) -- node[label=(\cmdKV@general@rotate):\cmdKV@gLabels@iLF] {} (0.9,0.1);
    \draw[label distance=0,\cmdKV@gTypes@iG] (0.9,0.1) -- node[label=(-90+\cmdKV@general@rotate):\cmdKV@gLabels@iLG] {} (0.1,0.1);
    #2
  \end{tikzpicture}
  \gSetStdTypes{line}
}

\def\gBoxBubB{\@ifnextchar[\@gBoxBubB{\@gBoxBubB[]}}
\def\@gBoxBubB[#1]#2{%
  \setkeys*{pre}{#1}%
  \setrmkeys{general,gTypes,gLabels}%
  \begin{tikzpicture}
    [line width=1pt,
    baseline={([yshift=-0.5ex+\cmdKV@general@yshift]current bounding box.center)},
    scale=\cmdKV@general@scale,
    rotate=\cmdKV@general@rotate,
    font=\cmdKV@general@font]
    \draw[label distance=-1,\cmdKV@gTypes@eA] (0,-0.4) -- (0,-0.7) node[label=(180+\cmdKV@general@rotate):\cmdKV@gLabels@eLA] {};
    \draw[\cmdKV@gTypes@eB] (-0.4,0) -- (-0.7,0) node[label=(180+\cmdKV@general@rotate):\cmdKV@gLabels@eLB] {};
    \draw[label distance=-1,\cmdKV@gTypes@eC] (0,0.4) -- (0,0.7) node[label=(180+\cmdKV@general@rotate):\cmdKV@gLabels@eLC] {};
    \draw[\cmdKV@gTypes@eD] (1.1,0) -- (1.4,0) node[label=(\cmdKV@general@rotate):\cmdKV@gLabels@eLD] {};
    \draw[label distance=0,\cmdKV@gTypes@iA] (0,-0.4) -- node[label=(-135+\cmdKV@general@rotate):\cmdKV@gLabels@iLA] {} (-0.4,0);
    \draw[label distance=0,\cmdKV@gTypes@iB] (-0.4,0) -- node[label=(135+\cmdKV@general@rotate):\cmdKV@gLabels@iLB] {} (0,0.4);
    \draw[label distance=0,\cmdKV@gTypes@iC] (0,0.4) -- node[label=(45+\cmdKV@general@rotate):\cmdKV@gLabels@iLC] {} (0.4,0);
    \draw[label distance=0,\cmdKV@gTypes@iD] (0.4,0) -- node[label=(-45+\cmdKV@general@rotate):\cmdKV@gLabels@iLD] {} (0,-0.4);
    \draw[label distance=0,\cmdKV@gTypes@iE] (0.4,0) --  node[label=(90+\cmdKV@general@rotate):\cmdKV@gLabels@iLE] {} (0.7,0);
    \draw[label distance=0,\cmdKV@gTypes@iF] (0.7,0) .. controls (0.7,0.26) and (1.1,0.26) .. node[label=(90+\cmdKV@general@rotate):\cmdKV@gLabels@iLF] {} (1.1,0);
    \draw[label distance=0,\cmdKV@gTypes@iG] (1.1,0) .. controls (1.1,-0.26) and (0.7,-0.26) ..  node[label=(-90+\cmdKV@general@rotate):\cmdKV@gLabels@iLG] {} (0.7,0);
    #2
  \end{tikzpicture}
  \gSetStdTypes{line}
}

\def\gPentaTad{\@ifnextchar[\@gPentaTad{\@gPentaTad[]}}
\def\@gPentaTad[#1]#2{%
  \setkeys*{pre}{#1}%
  \setrmkeys{general,gTypes,gLabels}%
  \begin{tikzpicture}
    [line width=1pt,
    baseline={([yshift=-0.5ex+\cmdKV@general@yshift]current bounding box.center)},
    scale=\cmdKV@general@scale,
    rotate=\cmdKV@general@rotate,
    font=\cmdKV@general@font]
    \draw[label distance=-0.5,\cmdKV@gTypes@eA] (108:0.35) -- (108:0.75) node[label=(180+\cmdKV@general@rotate):\cmdKV@gLabels@eLA] {};
    \draw[\cmdKV@gTypes@eB] (36:0.35) -- (36:0.75) node[label=(36+\cmdKV@general@rotate):\cmdKV@gLabels@eLB] {};
    \draw[\cmdKV@gTypes@eC] (-36:0.35) -- (-36:0.75) node[label=(-36+\cmdKV@general@rotate):\cmdKV@gLabels@eLC] {};
    \draw[label distance=-0.5,\cmdKV@gTypes@eD] (-108:0.35) -- (-108:0.75) node[label=(-180+\cmdKV@general@rotate):\cmdKV@gLabels@eLD] {};
    \draw[label distance=0,\cmdKV@gTypes@iA] (108:0.35) -- node[label=(72+\cmdKV@general@rotate):\cmdKV@gLabels@iLA] {} (36:0.35);
    \draw[label distance=0,\cmdKV@gTypes@iB] (36:0.35) -- node[label=(\cmdKV@general@rotate):\cmdKV@gLabels@iLB] {} (-36:0.35);
    \draw[label distance=0,\cmdKV@gTypes@iC] (-36:0.35) -- node[label=(-72+\cmdKV@general@rotate):\cmdKV@gLabels@iLC] {} (-108:0.35);
    \draw[label distance=0,\cmdKV@gTypes@iD] (-108:0.35) -- node[label=(-144+\cmdKV@general@rotate):\cmdKV@gLabels@iLD] {} (-180:0.35);
    \draw[label distance=0,\cmdKV@gTypes@iE] (180:0.35) -- node[label=(144+\cmdKV@general@rotate):\cmdKV@gLabels@iLE] {} (108:0.35);
    \draw[label distance=0,\cmdKV@gTypes@iF] (180:0.35) -- node[label=(90+\cmdKV@general@rotate):\cmdKV@gLabels@iLF] {} (180:0.75);
    \draw[label distance=-0.5,\cmdKV@gTypes@iG] (180:0.75) .. controls ($(180:0.75)+(0,-0.18)$) and ($(180:1)+(0,-0.18)$) .. (180:1) node[label=(180+\cmdKV@general@rotate):\cmdKV@gLabels@iLG] {} .. controls ($(180:1)+(0,0.18)$) and ($(180:0.75)+(0,0.18)$) .. (180:0.75);
    #2
  \end{tikzpicture}
  \gSetStdTypes{line}
}

\def\gBoxTadA{\@ifnextchar[\@gBoxTadA{\@gBoxTadA[]}}
\def\@gBoxTadA[#1]#2{%
  \setkeys*{pre}{#1}%
  \setrmkeys{general,gTypes,gLabels}%
  \begin{tikzpicture}
    [line width=1pt,
    baseline={([yshift=-0.5ex+\cmdKV@general@yshift]current bounding box.center)},
    scale=\cmdKV@general@scale,
    rotate=\cmdKV@general@rotate,
    font=\cmdKV@general@font]
    \draw[label distance=-0.5,\cmdKV@gTypes@eA] (0,0.4) -- (0,0.7) node[label=(\cmdKV@general@rotate):\cmdKV@gLabels@eLA] {};
    \draw[\cmdKV@gTypes@eB] (0.4,0) -- (0.7,0) node[label=(\cmdKV@general@rotate):\cmdKV@gLabels@eLB] {};
    \draw[label distance=-0.5,\cmdKV@gTypes@eC] (0,-0.4) -- (0,-0.7) node[label=(\cmdKV@general@rotate):\cmdKV@gLabels@eLC] {};
    \draw[\cmdKV@gTypes@eD] (-0.7,0) -- (-0.7,-0.3) node[label=(-90+\cmdKV@general@rotate):\cmdKV@gLabels@eLD] {};
    \draw[label distance=0,\cmdKV@gTypes@iA] (0,0.4) -- node[label=(45+\cmdKV@general@rotate):\cmdKV@gLabels@iLA] {} (0.4,0);
    \draw[label distance=0,\cmdKV@gTypes@iB] (0.4,0) -- node[label=(-45+\cmdKV@general@rotate):\cmdKV@gLabels@iLB] {} (0,-0.4);
    \draw[label distance=0,\cmdKV@gTypes@iC] (0,-0.4) -- node[label=(-135+\cmdKV@general@rotate):\cmdKV@gLabels@iLC] {} (-0.4,0);
    \draw[label distance=0,\cmdKV@gTypes@iD] (-0.4,0) -- node[label=(135+\cmdKV@general@rotate):\cmdKV@gLabels@iLD] {} (0,0.4);
    \draw[label distance=0,\cmdKV@gTypes@iE] (-0.4,0) --  node[label=(90+\cmdKV@general@rotate):\cmdKV@gLabels@iLE] {} (-0.7,0);
    \draw[label distance=0,\cmdKV@gTypes@iF] (-0.7,0) -- node[label=(90+\cmdKV@general@rotate):\cmdKV@gLabels@iLF] {} (-1.1,0);
    \draw[label distance=-0.5,\cmdKV@gTypes@iG] (-1.1,0) .. controls (-1.1,-0.18) and (-1.35,-0.18) .. (-1.35,0) node[label=(180+\cmdKV@general@rotate):\cmdKV@gLabels@iLG] {} .. controls (-1.35,0.18) and (-1.1,0.18) .. (-1.1,0);
    #2
  \end{tikzpicture}
  \gSetStdTypes{line}
}

\def\gSunsetA{\@ifnextchar[\@gSunsetA{\@gSunsetA[]}}
\def\@gSunsetA[#1]#2{%
  \setkeys*{pre}{#1}%
  \setrmkeys{general,gTypes,gLabels}%
  \begin{tikzpicture}
    [line width=1pt,
    baseline={([yshift=-0.5ex+\cmdKV@general@yshift]current bounding box.center)},
    scale=\cmdKV@general@scale,
    rotate=\cmdKV@general@rotate,
    font=\cmdKV@general@font]
    \draw[\cmdKV@gTypes@eA] (0.3,0.6) -- (0,0) node[label=(-180+\cmdKV@general@rotate):\cmdKV@gLabels@eLA] {};
    \draw[\cmdKV@gTypes@eB] (0.3,0.6) -- (0,1.2) node[label=(180+\cmdKV@general@rotate):\cmdKV@gLabels@eLB] {};
    \draw[\cmdKV@gTypes@eC] (1.5,0.6) -- (1.8,1.2) node[label=(\cmdKV@general@rotate):\cmdKV@gLabels@eLC] {};
    \draw[\cmdKV@gTypes@eD] (1.5,0.6) -- (1.8,0) node[label=(\cmdKV@general@rotate):\cmdKV@gLabels@eLD] {};
    \draw[label distance=0,\cmdKV@gTypes@iA] (0.3,0.6) to[bend left=90] node[label=(90+\cmdKV@general@rotate):\cmdKV@gLabels@iLA] {} (1.5,0.6);
    \draw[label distance=0,\cmdKV@gTypes@iB] (1.5,0.6) to[bend left=90] node[label=(-90+\cmdKV@general@rotate):\cmdKV@gLabels@iLB] {} (0.3,0.6);
    \draw[label distance=-2,\cmdKV@gTypes@iC] (0.3,0.6) -- node[label=(90+\cmdKV@general@rotate):\cmdKV@gLabels@iLC] {} (1.5,0.6);
    #2
  \end{tikzpicture}
  \gSetStdTypes{line}
}

\def\gTriBubA{\@ifnextchar[\@gTriBubA{\@gTriBubA[]}}
\def\@gTriBubA[#1]#2{%
  \setkeys*{pre}{#1}%
  \setrmkeys{general,gTypes,gLabels}%
  \begin{tikzpicture}
    [line width=1pt,
    baseline={([yshift=-0.5ex+\cmdKV@general@yshift]current bounding box.center)},
    scale=\cmdKV@general@scale,
    rotate=\cmdKV@general@rotate,
    font=\cmdKV@general@font]
    \draw[\cmdKV@gTypes@eA] (0.4,0.3) -- (0,0) node[label=(-180+\cmdKV@general@rotate):\cmdKV@gLabels@eLA] {};
    \draw[\cmdKV@gTypes@eB] (0.4,0.9) -- (0,1.2) node[label=(180+\cmdKV@general@rotate):\cmdKV@gLabels@eLB] {};
    \draw[\cmdKV@gTypes@eC] (1.4,0.6) -- (1.8,1.2) node[label=(\cmdKV@general@rotate):\cmdKV@gLabels@eLC] {};
    \draw[\cmdKV@gTypes@eD] (1.4,0.6) -- (1.8,0) node[label=(\cmdKV@general@rotate):\cmdKV@gLabels@eLD] {};
    \draw[label distance=0,\cmdKV@gTypes@iA] (0.4,0.3) to[bend left=40] node[label=(180+\cmdKV@general@rotate):\cmdKV@gLabels@iLA] {} (0.4,0.9);
    \draw[label distance=0,\cmdKV@gTypes@iB] (0.4,0.9) -- node[label=(90+\cmdKV@general@rotate):\cmdKV@gLabels@iLB] {} (1.4,0.6);
    \draw[label distance=0,\cmdKV@gTypes@iC] (1.4,0.6) -- node[label=(-90+\cmdKV@general@rotate):\cmdKV@gLabels@iLC] {} (0.4,0.3);
    \draw[label distance=-1,\cmdKV@gTypes@iD] (0.4,0.9) to[bend left=40] node[label=(\cmdKV@general@rotate):\cmdKV@gLabels@iLD] {} (0.4,0.3);
    #2
  \end{tikzpicture}
  \gSetStdTypes{line}
}

\def\gBubBubA{\@ifnextchar[\@gBubBubA{\@gBubBubA[]}}
\def\@gBubBubA[#1]#2{%
  \setkeys*{pre}{#1}%
  \setrmkeys{general,gTypes,gLabels}%
  \begin{tikzpicture}
    [line width=1pt,
    baseline={([yshift=-0.5ex+\cmdKV@general@yshift]current bounding box.center)},
    scale=\cmdKV@general@scale,
    rotate=\cmdKV@general@rotate,
    font=\cmdKV@general@font]
    \draw[\cmdKV@gTypes@eA] (0.3,0.6) -- (0,0) node[label=(-180+\cmdKV@general@rotate):\cmdKV@gLabels@eLA] {};
    \draw[\cmdKV@gTypes@eB] (0.3,0.6) -- (0,1.2) node[label=(180+\cmdKV@general@rotate):\cmdKV@gLabels@eLB] {};
    \draw[\cmdKV@gTypes@eC] (1.5,0.6) -- (1.8,1.2) node[label=(\cmdKV@general@rotate):\cmdKV@gLabels@eLC] {};
    \draw[\cmdKV@gTypes@eD] (1.5,0.6) -- (1.8,0) node[label=(\cmdKV@general@rotate):\cmdKV@gLabels@eLD] {};
    \draw[label distance=0,\cmdKV@gTypes@iA] (0.3,0.6) to[bend left=90] node[label=(90+\cmdKV@general@rotate):\cmdKV@gLabels@iLA] {} (0.9,0.6);
    \draw[label distance=0,\cmdKV@gTypes@iB] (0.9,0.6) to[bend left=90] node[label=(90+\cmdKV@general@rotate):\cmdKV@gLabels@iLB] {} (1.5,0.6);
    \draw[label distance=0,\cmdKV@gTypes@iC] (1.5,0.6) to[bend left=90] node[label=(-90+\cmdKV@general@rotate):\cmdKV@gLabels@iLC] {} (0.9,0.6);
    \draw[label distance=0,\cmdKV@gTypes@iD] (0.9,0.6) to[bend left=90] node[label=(-90+\cmdKV@general@rotate):\cmdKV@gLabels@iLD] {} (0.3,0.6);
    #2
  \end{tikzpicture}
  \gSetStdTypes{line}
}

\def\gBoxBubC{\@ifnextchar[\@gBoxBubC{\@gBoxBubC[]}}
\def\@gBoxBubC[#1]#2{%
  \setkeys*{pre}{#1}%
  \setrmkeys{general,gTypes,gLabels}%
  \begin{tikzpicture}
    [line width=1pt,
    baseline={([yshift=-0.5ex+\cmdKV@general@yshift]current bounding box.center)},
    scale=\cmdKV@general@scale,
    rotate=\cmdKV@general@rotate,
    font=\cmdKV@general@font]
    \draw[\cmdKV@gTypes@eA] (0.1,0.1) -- (-0.1,-0.1) node[label=(-180+\cmdKV@general@rotate):\cmdKV@gLabels@eLA] {};
    \draw[\cmdKV@gTypes@eB] (0.1,0.9) -- (-0.1,1.1) node[label=(180+\cmdKV@general@rotate):\cmdKV@gLabels@eLB] {};
    \draw[\cmdKV@gTypes@eC] (0.9,0.9) -- (1.1,1.1) node[label=(\cmdKV@general@rotate):\cmdKV@gLabels@eLC] {};
    \draw[\cmdKV@gTypes@eD] (0.9,0.1) -- (1.1,-0.1) node[label=(\cmdKV@general@rotate):\cmdKV@gLabels@eLD] {};
    \draw[label distance=0,\cmdKV@gTypes@iA] (0.1,0.1) -- node[label=(180+\cmdKV@general@rotate):\cmdKV@gLabels@iLA] {} (0.1,0.9);
    \draw[label distance=0,\cmdKV@gTypes@iB] (0.1,0.9) to[bend left] node[label=(90+\cmdKV@general@rotate):\cmdKV@gLabels@iLB] {} (0.9,0.9);
    \draw[label distance=0,\cmdKV@gTypes@iC] (0.9,0.9) to[bend left] node[label=(-90+\cmdKV@general@rotate):\cmdKV@gLabels@iLC] {} (0.1,0.9);
    \draw[label distance=0,\cmdKV@gTypes@iD] (0.9,0.9) -- node[label=(\cmdKV@general@rotate):\cmdKV@gLabels@iLD] {} (0.9,0.1);
    \draw[label distance=0,\cmdKV@gTypes@iE] (0.9,0.1) -- node[label=(-90+\cmdKV@general@rotate):\cmdKV@gLabels@iLE] {} (0.1,0.1);
    #2
  \end{tikzpicture}
  \gSetStdTypes{line}
}

\def\gBoxZ{\@ifnextchar[\@gBoxZ{\@gBoxZ[]}}
\def\@gBoxZ[#1]#2{%
  \setkeys*{pre}{#1}%
  \setrmkeys{general,gTypes,gLabels}%
  \begin{tikzpicture}
    [line width=1pt,
    baseline={([yshift=-0.5ex+\cmdKV@general@yshift]current bounding box.center)},
    scale=\cmdKV@general@scale,
    rotate=\cmdKV@general@rotate,
    font=\cmdKV@general@font]
    \draw[\cmdKV@gTypes@eA] (0.15,0.15) -- (-0.1,-0.1) node[label=(-180+\cmdKV@general@rotate):\cmdKV@gLabels@eLA] {};
    \draw[\cmdKV@gTypes@eB] (0.15,0.85) -- (-0.1,1.1) node[label=(180+\cmdKV@general@rotate):\cmdKV@gLabels@eLB] {};
    \draw[\cmdKV@gTypes@eC] (0.85,0.85) -- (1.1,1.1) node[label=(\cmdKV@general@rotate):\cmdKV@gLabels@eLC] {};
    \draw[\cmdKV@gTypes@eD] (0.85,0.15) -- (1.1,-0.1) node[label=(\cmdKV@general@rotate):\cmdKV@gLabels@eLD] {};
    \draw[label distance=0,\cmdKV@gTypes@iA] (0.15,0.15) -- node[label=(180+\cmdKV@general@rotate):\cmdKV@gLabels@iLA] {} (0.15,0.85);
    \draw[label distance=0,\cmdKV@gTypes@iB] (0.15,0.85) -- node[label=(90+\cmdKV@general@rotate):\cmdKV@gLabels@iLB] {} (0.85,0.85);
    \draw[label distance=0,\cmdKV@gTypes@iC] (0.85,0.85) -- node[label=(\cmdKV@general@rotate):\cmdKV@gLabels@iLC] {} (0.85,0.15);
    \draw[label distance=0,\cmdKV@gTypes@iD] (0.85,0.15) -- node[label=(-90+\cmdKV@general@rotate):\cmdKV@gLabels@iLD] {} (0.15,0.15);
    \draw[label distance=-4,\cmdKV@gTypes@iE] (0.15,0.15) -- node[label=(135+\cmdKV@general@rotate):\cmdKV@gLabels@iLE] {} (0.85,0.85);
    #2
  \end{tikzpicture}
  \gSetStdTypes{line}
}

\def\gBoxBoxBoxA{\@ifnextchar[\@gBoxBoxBoxA{\@gBoxBoxBoxA[]}}
\def\@gBoxBoxBoxA[#1]#2{%
  \setkeys*{pre}{#1}%
  \setrmkeys{general,gTypes,gLabels}%
  \begin{tikzpicture}
    [line width=1pt,
    baseline={([yshift=-0.5ex+\cmdKV@general@yshift]current bounding box.center)},
    scale=\cmdKV@general@scale,
    rotate=\cmdKV@general@rotate,
    font=\cmdKV@general@font]
    \draw[\cmdKV@gTypes@eA] (-0.3,0.3) -- (-0.6,0) node[label=(-180+\cmdKV@general@rotate):\cmdKV@gLabels@eLA] {};
    \draw[\cmdKV@gTypes@eB] (-0.3,0.9) -- (-0.6,1.2) node[label=(180+\cmdKV@general@rotate):\cmdKV@gLabels@eLB] {};
    \draw[\cmdKV@gTypes@eC] (1.5,0.9) -- (1.8,1.2) node[label=(\cmdKV@general@rotate):\cmdKV@gLabels@eLC] {};
    \draw[\cmdKV@gTypes@eD] (1.5,0.3) -- (1.8,0) node[label=(\cmdKV@general@rotate):\cmdKV@gLabels@eLD] {};
    \draw[label distance=0,\cmdKV@gTypes@iA] (-0.3,0.3) -- node[label=(180+\cmdKV@general@rotate):\cmdKV@gLabels@iLA] {} (-0.3,0.9);
    \draw[label distance=0,\cmdKV@gTypes@iB] (-0.3,0.9) -- node[label=(90+\cmdKV@general@rotate):\cmdKV@gLabels@iLB] {} (0.3,0.9);
    \draw[label distance=0,\cmdKV@gTypes@iC] (0.3,0.9) -- node[label=(90+\cmdKV@general@rotate):\cmdKV@gLabels@iLC] {} (0.9,0.9);
    \draw[label distance=0,\cmdKV@gTypes@iD] (0.9,0.9) -- node[label=(90+\cmdKV@general@rotate):\cmdKV@gLabels@iLD] {} (1.5,0.9);
    \draw[label distance=0,\cmdKV@gTypes@iE] (1.5,0.9) -- node[label=(\cmdKV@general@rotate):\cmdKV@gLabels@iLE] {} (1.5,0.3);
    \draw[label distance=0,\cmdKV@gTypes@iF] (1.5,0.3) -- node[label=(-90+\cmdKV@general@rotate):\cmdKV@gLabels@iLF] {} (0.9,0.3);
    \draw[label distance=0,\cmdKV@gTypes@iG] (0.9,0.3) -- node[label=(-90+\cmdKV@general@rotate):\cmdKV@gLabels@iLG] {} (0.3,0.3);
    \draw[label distance=0,\cmdKV@gTypes@iH] (0.3,0.3) -- node[label=(-90+\cmdKV@general@rotate):\cmdKV@gLabels@iLH] {} (-0.3,0.3);
    \draw[label distance=0,\cmdKV@gTypes@iI] (0.3,0.9) -- node[label=(\cmdKV@general@rotate):\cmdKV@gLabels@iLI] {} (0.3,0.3);
    \draw[label distance=0,\cmdKV@gTypes@iJ] (0.9,0.9) -- node[label=(\cmdKV@general@rotate):\cmdKV@gLabels@iLJ] {} (0.9,0.3);
    #2
  \end{tikzpicture}
  \gSetStdTypes{line}
}

\def\gBoxBoxBoxB{\@ifnextchar[\@gBoxBoxBoxB{\@gBoxBoxBoxB[]}}
\def\@gBoxBoxBoxB[#1]#2{%
  \setkeys*{pre}{#1}%
  \setrmkeys{general,gTypes,gLabels}%
  \begin{tikzpicture}
    [line width=1pt,
    baseline={([yshift=-0.5ex+\cmdKV@general@yshift]current bounding box.center)},
    scale=\cmdKV@general@scale,
    rotate=\cmdKV@general@rotate,
    font=\cmdKV@general@font]
    \draw[\cmdKV@gTypes@eA] (0.3,0.3) -- (0,0) node[label=(-180+\cmdKV@general@rotate):\cmdKV@gLabels@eLA] {};
    \draw[\cmdKV@gTypes@eB] (0.3,1.5) -- (0,1.8) node[label=(180+\cmdKV@general@rotate):\cmdKV@gLabels@eLB] {};
    \draw[\cmdKV@gTypes@eC] (1.5,1.5) -- (1.8,1.8) node[label=(\cmdKV@general@rotate):\cmdKV@gLabels@eLC] {};
    \draw[\cmdKV@gTypes@eD] (1.5,0.3) -- (1.8,0) node[label=(\cmdKV@general@rotate):\cmdKV@gLabels@eLD] {};
    \draw[label distance=0,\cmdKV@gTypes@iA] (0.3,0.3) -- node[label=(180+\cmdKV@general@rotate):\cmdKV@gLabels@iLA] {} (0.3,0.9);
    \draw[label distance=0,\cmdKV@gTypes@iB] (0.3,0.9) -- node[label=(180+\cmdKV@general@rotate):\cmdKV@gLabels@iLB] {} (0.3,1.5);
    \draw[label distance=0,\cmdKV@gTypes@iC] (0.3,1.5) -- node[label=(90+\cmdKV@general@rotate):\cmdKV@gLabels@iLC] {} (0.9,1.5);
    \draw[label distance=0,\cmdKV@gTypes@iD] (0.9,1.5) -- node[label=(90+\cmdKV@general@rotate):\cmdKV@gLabels@iLD] {} (1.5,1.5);
    \draw[label distance=0,\cmdKV@gTypes@iE] (1.5,1.5) -- node[label=(\cmdKV@general@rotate):\cmdKV@gLabels@iLE] {} (1.5,0.3);
    \draw[label distance=0,\cmdKV@gTypes@iF] (1.5,0.3) -- node[label=(-90+\cmdKV@general@rotate):\cmdKV@gLabels@iLF] {} (0.9,0.3);
    \draw[label distance=0,\cmdKV@gTypes@iG] (0.9,0.3) -- node[label=(-90+\cmdKV@general@rotate):\cmdKV@gLabels@iLG] {} (0.3,0.3);
    \draw[label distance=0,\cmdKV@gTypes@iH] (0.9,1.5) -- node[label=(\cmdKV@general@rotate):\cmdKV@gLabels@iLH] {} (0.9,0.9);
    \draw[label distance=0,\cmdKV@gTypes@iI] (0.9,0.9) -- node[label=(\cmdKV@general@rotate):\cmdKV@gLabels@iLI] {} (0.9,0.3);
    \draw[label distance=0,\cmdKV@gTypes@iJ] (0.9,0.9) -- node[label=(90+\cmdKV@general@rotate):\cmdKV@gLabels@iLJ] {} (0.3,0.9);
    #2
  \end{tikzpicture}
  \gSetStdTypes{line}
}

\def\gEFTTree{\@ifnextchar[\@gEFTTree{\@gEFTTree[]}}
\def\@gEFTTree[#1]#2{%
  \setkeys*{pre}{#1}%
  \setrmkeys{general,gTypes,gLabels}%
  \begin{tikzpicture}
    [line width=1pt,
    baseline={([yshift=-0.5ex+\cmdKV@general@yshift]current bounding box.center)},
    scale=\cmdKV@general@scale,
    rotate=\cmdKV@general@rotate,
    font=\cmdKV@general@font]
    \draw[\cmdKV@gTypes@iA] (0,0.5) node[label=(90+\cmdKV@general@rotate):\cmdKV@gLabels@eLA] {} -- node[label=(\cmdKV@general@rotate):\cmdKV@gLabels@iLA] {} (0,-0.5) node[label=(-90+\cmdKV@general@rotate):\cmdKV@gLabels@eLB] {};
    \draw[fill] (0,0.5) circle (0.02);
    \draw[fill] (0,-0.5) circle (0.02);
    #2
  \end{tikzpicture}
  \gSetStdTypes{line}
}

\def\gEFTV{\@ifnextchar[\@gEFTV{\@gEFTV[]}}
\def\@gEFTV[#1]#2{%
  \setkeys*{pre}{#1}%
  \setrmkeys{general,gTypes,gLabels}%
  \begin{tikzpicture}
    [line width=1pt,
    baseline={([yshift=-0.5ex+\cmdKV@general@yshift]current bounding box.center)},
    scale=\cmdKV@general@scale,
    rotate=\cmdKV@general@rotate,
    font=\cmdKV@general@font]
    \draw[\cmdKV@gTypes@iA] (0,0.5) node[label=(90+\cmdKV@general@rotate):\cmdKV@gLabels@eLA] {} -- node[label=(200+\cmdKV@general@rotate):\cmdKV@gLabels@iLA] {} (0.3,-0.5) node[label=(-90+\cmdKV@general@rotate):\cmdKV@gLabels@eLC] {};
    \draw[\cmdKV@gTypes@iB] (0.6,0.5) node[label=(90+\cmdKV@general@rotate):\cmdKV@gLabels@eLB] {} -- node[label=(-20+\cmdKV@general@rotate):\cmdKV@gLabels@iLB] {} (0.3,-0.5);
    \draw[fill] (0,0.5) circle (0.02);
    \draw[fill] (0.3,-0.5) circle (0.02);
    \draw[fill] (0.6,0.5) circle (0.02);
    #2
  \end{tikzpicture}
  \gSetStdTypes{line}
}

\def\gEFTY{\@ifnextchar[\@gEFTY{\@gEFTY[]}}
\def\@gEFTY[#1]#2{%
  \setkeys*{pre}{#1}%
  \setrmkeys{general,gTypes,gLabels}%
  \begin{tikzpicture}
    [line width=1pt,
    baseline={([yshift=-0.5ex+\cmdKV@general@yshift]current bounding box.center)},
    scale=\cmdKV@general@scale,
    rotate=\cmdKV@general@rotate,
    font=\cmdKV@general@font]
    \draw[\cmdKV@gTypes@iA] (0,0.5) node[label=(90+\cmdKV@general@rotate):\cmdKV@gLabels@eLA] {} -- node[label=(225+\cmdKV@general@rotate):\cmdKV@gLabels@iLA] {} (0.3,0);
    \draw[\cmdKV@gTypes@iB] (0.6,0.5) node[label=(90+\cmdKV@general@rotate):\cmdKV@gLabels@eLB] {} -- node[label=(-45+\cmdKV@general@rotate):\cmdKV@gLabels@iLB] {} (0.3,0);
    \draw[\cmdKV@gTypes@iC] (0.3,0) -- node[label=(\cmdKV@general@rotate):\cmdKV@gLabels@iLC] {} (0.3,-0.5) node[label=(-90+\cmdKV@general@rotate):\cmdKV@gLabels@eLC] {};
    \draw[fill] (0,0.5) circle (0.02);
    \draw[fill] (0.6,0.5) circle (0.02);
    \draw[fill] (0.3,-0.5) circle (0.02);
    #2
  \end{tikzpicture}
  \gSetStdTypes{line}
}

\def\gEFTPsi{\@ifnextchar[\@gEFTPsi{\@gEFTPsi[]}}
\def\@gEFTPsi[#1]#2{%
  \setkeys*{pre}{#1}%
  \setrmkeys{general,gTypes,gLabels}%
  \begin{tikzpicture}
    [line width=1pt,
    baseline={([yshift=-0.5ex+\cmdKV@general@yshift]current bounding box.center)},
    scale=\cmdKV@general@scale,
    rotate=\cmdKV@general@rotate,
    font=\cmdKV@general@font]
    \draw[\cmdKV@gTypes@iA] (0,0.5) node[label=(90+\cmdKV@general@rotate):\cmdKV@gLabels@eLA] {} -- node[label=(225+\cmdKV@general@rotate):\cmdKV@gLabels@iLA] {} (0.5,0);
    \draw[\cmdKV@gTypes@iB] (0.5,0.5) node[label=(90+\cmdKV@general@rotate):\cmdKV@gLabels@eLB] {} -- node[label=(\cmdKV@general@rotate):\cmdKV@gLabels@iLB] {} (0.5,0);
    \draw[\cmdKV@gTypes@iC] (1,0.5) node[label=(90+\cmdKV@general@rotate):\cmdKV@gLabels@eLC] {} -- node[label=(-45+\cmdKV@general@rotate):\cmdKV@gLabels@iLC] {} (0.5,0);
    \draw[\cmdKV@gTypes@iD] (0.5,0) -- node[label=(\cmdKV@general@rotate):\cmdKV@gLabels@iLD] {} (0.5,-0.5) node[label=(-90+\cmdKV@general@rotate):\cmdKV@gLabels@eLD] {};
    \draw[fill] (0,0.5) circle (0.02);
    \draw[fill] (0.5,0.5) circle (0.02);
    \draw[fill] (1,0.5) circle (0.02);
    \draw[fill] (0.5,-0.5) circle (0.02);
    #2
  \end{tikzpicture}
  \gSetStdTypes{line}
}

\def\gEFTVinY{\@ifnextchar[\@gEFTVinY{\@gEFTVinY[]}}
\def\@gEFTVinY[#1]#2{%
  \setkeys*{pre}{#1}%
  \setrmkeys{general,gTypes,gLabels}%
  \begin{tikzpicture}
    [line width=1pt,
    baseline={([yshift=-0.5ex+\cmdKV@general@yshift]current bounding box.center)},
    scale=\cmdKV@general@scale,
    rotate=\cmdKV@general@rotate,
    font=\cmdKV@general@font]
    \draw[\cmdKV@gTypes@iA] (0,0.5) node[label=(90+\cmdKV@general@rotate):\cmdKV@gLabels@eLA] {} -- node[label=(225+\cmdKV@general@rotate):\cmdKV@gLabels@iLA] {} (0.25,0.25);
    \draw[\cmdKV@gTypes@iE] (0.25,0.25) -- node[label=(225+\cmdKV@general@rotate):\cmdKV@gLabels@iLE] {} (0.5,0);
    \draw[\cmdKV@gTypes@iB] (0.5,0.5) node[label=(90+\cmdKV@general@rotate):\cmdKV@gLabels@eLB] {} -- node[label=(\cmdKV@general@rotate):\cmdKV@gLabels@iLB] {} (0.25,0.25);
    \draw[\cmdKV@gTypes@iC] (1,0.5) node[label=(90+\cmdKV@general@rotate):\cmdKV@gLabels@eLC] {} -- node[label=(-45+\cmdKV@general@rotate):\cmdKV@gLabels@iLC] {} (0.5,0);
    \draw[\cmdKV@gTypes@iD] (0.5,0) -- node[label=(\cmdKV@general@rotate):\cmdKV@gLabels@iLD] {} (0.5,-0.5) node[label=(-90+\cmdKV@general@rotate):\cmdKV@gLabels@eLD] {};
    \draw[fill] (0,0.5) circle (0.02);
    \draw[fill] (0.5,0.5) circle (0.02);
    \draw[fill] (1,0.5) circle (0.02);
    \draw[fill] (0.5,-0.5) circle (0.02);
    #2
  \end{tikzpicture}
  \gSetStdTypes{line}
}

\def\gEFTYinvY{\@ifnextchar[\@gEFTYinvY{\@gEFTYinvY[]}}
\def\@gEFTYinvY[#1]#2{%
  \setkeys*{pre}{#1}%
  \setrmkeys{general,gTypes,gLabels}%
  \begin{tikzpicture}
    [line width=1pt,
    baseline={([yshift=-0.5ex+\cmdKV@general@yshift]current bounding box.center)},
    scale=\cmdKV@general@scale,
    rotate=\cmdKV@general@rotate,
    font=\cmdKV@general@font]
    \draw[\cmdKV@gTypes@iA] (0,0.5) node[label=(90+\cmdKV@general@rotate):\cmdKV@gLabels@eLA] {} -- node[label=(225+\cmdKV@general@rotate):\cmdKV@gLabels@iLA] {} (0.3,0.166);
    \draw[\cmdKV@gTypes@iB] (0.6,0.5) node[label=(90+\cmdKV@general@rotate):\cmdKV@gLabels@eLB] {} -- node[label=(-45+\cmdKV@general@rotate):\cmdKV@gLabels@iLB] {} (0.3,0.166);
    \draw[\cmdKV@gTypes@iC] (0.3,-0.166) -- node[label=(135+\cmdKV@general@rotate):\cmdKV@gLabels@iLC] {} (0,-0.5) node[label=(-90+\cmdKV@general@rotate):\cmdKV@gLabels@eLC] {}; 
    \draw[\cmdKV@gTypes@iD] (0.3,-0.166) -- node[label=(45+\cmdKV@general@rotate):\cmdKV@gLabels@iLD] {} (0.6,-0.5) node[label=(-90+\cmdKV@general@rotate):\cmdKV@gLabels@eLD] {};
    \draw[\cmdKV@gTypes@iE] (0.3,0.166) -- node[label=(\cmdKV@general@rotate):\cmdKV@gLabels@iLE] {} (0.3,-0.166);
    \draw[fill] (0,0.5) circle (0.02);
    \draw[fill] (0.6,0.5) circle (0.02);
    \draw[fill] (0,-0.5) circle (0.02);
    \draw[fill] (0.6,-0.5) circle (0.02);
    #2
  \end{tikzpicture}
  \gSetStdTypes{line}
}

\def\gEFTX{\@ifnextchar[\@gEFTX{\@gEFTX[]}}
\def\@gEFTX[#1]#2{%
  \setkeys*{pre}{#1}%
  \setrmkeys{general,gTypes,gLabels}%
  \begin{tikzpicture}
    [line width=1pt,
    baseline={([yshift=-0.5ex+\cmdKV@general@yshift]current bounding box.center)},
    scale=\cmdKV@general@scale,
    rotate=\cmdKV@general@rotate,
    font=\cmdKV@general@font]
    \draw[\cmdKV@gTypes@iA] (0,0.5) node[label=(90+\cmdKV@general@rotate):\cmdKV@gLabels@eLA] {} -- node[label=(225+\cmdKV@general@rotate):\cmdKV@gLabels@iLA] {} (0.3,0);
    \draw[\cmdKV@gTypes@iB] (0.6,0.5) node[label=(90+\cmdKV@general@rotate):\cmdKV@gLabels@eLB] {} -- node[label=(-45+\cmdKV@general@rotate):\cmdKV@gLabels@iLB] {} (0.3,0);
    \draw[\cmdKV@gTypes@iC] (0.3,0) -- node[label=(135+\cmdKV@general@rotate):\cmdKV@gLabels@iLC] {} (0,-0.5) node[label=(-90+\cmdKV@general@rotate):\cmdKV@gLabels@eLC] {}; 
    \draw[\cmdKV@gTypes@iD] (0.3,0) -- node[label=(45+\cmdKV@general@rotate):\cmdKV@gLabels@iLD] {} (0.6,-0.5) node[label=(-90+\cmdKV@general@rotate):\cmdKV@gLabels@eLD] {};
    \draw[fill] (0,0.5) circle (0.02);
    \draw[fill] (0.6,0.5) circle (0.02);
    \draw[fill] (0,-0.5) circle (0.02);
    \draw[fill] (0.6,-0.5) circle (0.02);
    #2
  \end{tikzpicture}
  \gSetStdTypes{line}
}

\def\gEFTH{\@ifnextchar[\@gEFTH{\@gEFTH[]}}
\def\@gEFTH[#1]#2{%
  \setkeys*{pre}{#1}%
  \setrmkeys{general,gTypes,gLabels}%
  \begin{tikzpicture}
    [line width=1pt,
    baseline={([yshift=-0.5ex+\cmdKV@general@yshift]current bounding box.center)},
    scale=\cmdKV@general@scale,
    rotate=\cmdKV@general@rotate,
    font=\cmdKV@general@font]
    \draw[\cmdKV@gTypes@iA] (0,0.5) node[label=(90+\cmdKV@general@rotate):\cmdKV@gLabels@eLA] {} -- node[label=(180+\cmdKV@general@rotate):\cmdKV@gLabels@iLA] {} (0,0);
    \draw[\cmdKV@gTypes@iB] (0.6,0.5) node[label=(90+\cmdKV@general@rotate):\cmdKV@gLabels@eLB] {} -- node[label=(\cmdKV@general@rotate):\cmdKV@gLabels@iLB] {} (0.6,0);
    \draw[\cmdKV@gTypes@iC] (0.0,0) -- node[label=(180+\cmdKV@general@rotate):\cmdKV@gLabels@iLC] {} (0,-0.5) node[label=(-90+\cmdKV@general@rotate):\cmdKV@gLabels@eLC] {}; 
    \draw[\cmdKV@gTypes@iD] (0.6,0) -- node[label=(\cmdKV@general@rotate):\cmdKV@gLabels@iLD] {} (0.6,-0.5) node[label=(-90+\cmdKV@general@rotate):\cmdKV@gLabels@eLD] {};
    \draw[\cmdKV@gTypes@iE] (0,0) -- node[label=(90+\cmdKV@general@rotate):\cmdKV@gLabels@iLE] {} (0.6,0);
    \draw[fill] (0,0.5) circle (0.02);
    \draw[fill] (0.6,0.5) circle (0.02);
    \draw[fill] (0,-0.5) circle (0.02);
    \draw[fill] (0.6,-0.5) circle (0.02);
    #2
  \end{tikzpicture}
  \gSetStdTypes{line}
}

\def\cTree{\@ifnextchar[\@cTree{\@cTree[]}}
\def\@cTree[#1]#2{%
  \setkeys*{pre}{#1}%
  \setrmkeys{general,gTypes,gLabels,blobs}%
  \begin{tikzpicture}
    [line width=1pt,
    baseline={([yshift=-0.5ex+\cmdKV@general@yshift]current bounding box.center)},
    scale=\cmdKV@general@scale,
    rotate=\cmdKV@general@rotate,
    font=\cmdKV@general@font]
    \fill[\cmdKV@blobs@blobA] (0,0) ellipse (0.25 and 0.25);
    \draw[\cmdKV@gTypes@eA] ($ (0,0) + (-135:0.25) $) -- ($ (0,0) + (-135:0.6) $) node[label=(-180+\cmdKV@general@rotate):\cmdKV@gLabels@eLA] {};
    \draw[\cmdKV@gTypes@eB] ($ (0,0) + (135:0.25) $) -- ($ (0,0) + (135:0.6) $) node[label=(180+\cmdKV@general@rotate):\cmdKV@gLabels@eLB] {};
    \draw[\cmdKV@gTypes@eC] ($ (0,0) + (45:0.25) $) -- ($ (0,0) + (45:0.6) $) node[label=(\cmdKV@general@rotate):\cmdKV@gLabels@eLC] {};
    \draw[\cmdKV@gTypes@eD] ($ (0,0) + (-45:0.25) $) -- ($ (0,0) + (-45:0.6) $) node[label=(\cmdKV@general@rotate):\cmdKV@gLabels@eLD] {};
    \node[label=(90+\cmdKV@general@rotate):\cmdKV@gLabels@iLA] at (0,0.3) {};
    \node[label=(\cmdKV@general@rotate):\cmdKV@gLabels@iLB] at (0.3,0) {};
    #2
  \end{tikzpicture}
  \gSetStdTypes{line}
}

\def\cBoxA{\@ifnextchar[\@cBoxA{\@cBoxA[]}}
\def\@cBoxA[#1]#2{%
  \setkeys*{pre}{#1}%
  \setrmkeys{general,gTypes,gLabels,dots,blobs}%
  \begin{tikzpicture}
    [line width=1pt,
    baseline={([yshift=-0.5ex+\cmdKV@general@yshift]current bounding box.center)},
    scale=\cmdKV@general@scale,
    rotate=\cmdKV@general@rotate,
    font=\cmdKV@general@font]
    \fill[\cmdKV@blobs@blobA] (-0.4,0) ellipse (0.2 and 0.4);
    \fill[\cmdKV@blobs@blobB] (0.4,0) ellipse (0.2 and 0.4);
    \draw[\cmdKV@gTypes@eA] ($ (-0.4,0) + (-135:0.26) $) -- ($ (-0.4,0) + (-135:0.7) $) node[label=(-180+\cmdKV@general@rotate):\cmdKV@gLabels@eLA] {};
    \draw[\cmdKV@gTypes@eB] ($ (-0.4,0) + (135:0.26) $) -- ($ (-0.4,0) + (135:0.7) $) node[label=(180+\cmdKV@general@rotate):\cmdKV@gLabels@eLB] {};
    \draw[\cmdKV@gTypes@eC] ($ (0.4,0) + (45:0.26) $) -- ($ (0.4,0) + (45:0.7) $) node[label=(\cmdKV@general@rotate):\cmdKV@gLabels@eLC] {};
    \draw[\cmdKV@gTypes@eD] ($ (0.4,0) + (-45:0.26) $) -- ($ (0.4,0) + (-45:0.7) $) node[label=(\cmdKV@general@rotate):\cmdKV@gLabels@eLD] {};
    \draw[label distance=1,\cmdKV@gTypes@iA] ($ (-0.4,0) + (45:0.26) $) -- node[label=(90+\cmdKV@general@rotate):\cmdKV@gLabels@iLA] {}($ (0.4,0) + (135:0.26) $);
    \draw[label distance=1,\cmdKV@gTypes@iB] ($ (0.4,0) + (-135:0.26) $) -- node[label=(-90+\cmdKV@general@rotate):\cmdKV@gLabels@iLB] {}($ (-0.4,0) + (-45:0.26) $);
    \node[label=(180+\cmdKV@general@rotate):\cmdKV@gLabels@iLC] at (-0.6,0) {};
    \node[label=(\cmdKV@general@rotate):\cmdKV@gLabels@iLD] at (0.6,0) {};
    \ifKV@dots@lDots
      \draw[dotted] ($(-0.4,0) + (-150:0.7)$) to[bend left] ($(-0.4,0) + (150:0.7)$);%
    \fi
    \ifKV@dots@rDots
      \draw[dotted] ($(0.4,0) + (30:0.7)$) to[bend left] ($(0.4,0) + (-30:0.7)$);%
    \fi
    #2
  \end{tikzpicture}
  \gSetStdTypes{line}
}

\def\cBoxAA{\@ifnextchar[\@cBoxAA{\@cBoxAA[]}}
\def\@cBoxAA[#1]#2{%
  \setkeys*{pre}{#1}%
  \setrmkeys{general,gTypes,gLabels,dots,blobs}%
  \begin{tikzpicture}
    [line width=1pt,
    baseline={([yshift=-0.5ex+\cmdKV@general@yshift]current bounding box.center)},
    scale=\cmdKV@general@scale,
    rotate=\cmdKV@general@rotate,
    font=\cmdKV@general@font]
    \fill[\cmdKV@blobs@blobA] (-0.4,0) ellipse (0.2 and 0.4);
    \fill[\cmdKV@blobs@blobB] (0.4,0) ellipse (0.2 and 0.4);
    \draw[\cmdKV@gTypes@eA] ($ (-0.4,0) + (-135:0.26) $) -- ($ (-0.4,0) + (-135:0.7) $) node[label=(-180+\cmdKV@general@rotate):\cmdKV@gLabels@eLA] {};
    \draw[\cmdKV@gTypes@eB] ($ (-0.4,0) + (135:0.26) $) -- ($ (-0.4,0) + (135:0.7) $) node[label=(180+\cmdKV@general@rotate):\cmdKV@gLabels@eLB] {};
    \draw[\cmdKV@gTypes@eC] ($ (0.4,0) + (45:0.26) $) -- ($ (0.4,0) + (45:0.7) $) node[label=(\cmdKV@general@rotate):\cmdKV@gLabels@eLC] {};
    \draw[\cmdKV@gTypes@eD] ($ (0.4,0) + (-45:0.26) $) -- ($ (0.4,0) + (-45:0.7) $) node[label=(\cmdKV@general@rotate):\cmdKV@gLabels@eLD] {};
    \draw[label distance=1,\cmdKV@gTypes@iA] ($ (-0.4,0) + (60:0.32) $) to[bend left] node[label=(90+\cmdKV@general@rotate):\cmdKV@gLabels@iLA] {}($ (0.4,0) + (120:0.32) $);
    \draw[label distance=1,\cmdKV@gTypes@iB] ($ (0.4,0) + (-120:0.32) $) to[bend left]  node[label=(-90+\cmdKV@general@rotate):\cmdKV@gLabels@iLB] {}($ (-0.4,0) + (-60:0.32) $);
    \draw[dotted] (0,0.25) -- (0,-0.25);
    \node[label=(180+\cmdKV@general@rotate):\cmdKV@gLabels@iLC] at (-0.6,0) {};
    \node[label=(\cmdKV@general@rotate):\cmdKV@gLabels@iLD] at (0.6,0) {};
    \ifKV@dots@lDots
      \draw[dotted] ($(-0.4,0) + (-150:0.7)$) to[bend left] ($(-0.4,0) + (150:0.7)$);%
    \fi
    \ifKV@dots@rDots
      \draw[dotted] ($(0.4,0) + (30:0.7)$) to[bend left] ($(0.4,0) + (-30:0.7)$);%
    \fi
    #2
  \end{tikzpicture}
  \gSetStdTypes{line}
}

\def\cBoxB{\@ifnextchar[\@cBoxB{\@cBoxB[]}}
\def\@cBoxB[#1]#2{%
  \setkeys*{pre}{#1}%
  \setrmkeys{general,gTypes,gLabels,blobs}%
  \begin{tikzpicture}
    [line width=1pt,
    baseline={([yshift=-0.5ex+\cmdKV@general@yshift]current bounding box.center)},
    scale=\cmdKV@general@scale,
    rotate=\cmdKV@general@rotate,
    font=\cmdKV@general@font]
    \fill[\cmdKV@blobs@blobA] (0,-0.3) ellipse (0.4 and 0.2);
    \fill[\cmdKV@blobs@blobB] (0,0.3) ellipse (0.4 and 0.2);
    \draw[\cmdKV@gTypes@eA] ($ (0,-0.3) + (180:0.4) $) -- ($ (0,-0.3) + (-170:0.7) $) node[label=(180+\cmdKV@general@rotate):\cmdKV@gLabels@eLA] {};
    \draw[\cmdKV@gTypes@eB] ($ (0,0.3) + (180:0.4) $) -- ($ (0,0.3) + (170:0.7) $) node[label=(180+\cmdKV@general@rotate):\cmdKV@gLabels@eLB] {};
    \draw[\cmdKV@gTypes@eC] ($ (0,0.3) + (0:0.4) $) -- ($ (0,0.3) + (10:0.7) $) node[label=(\cmdKV@general@rotate):\cmdKV@gLabels@eLC] {};
    \draw[\cmdKV@gTypes@eD] ($ (0,-0.3) + (0:0.4) $) -- ($ (0,-0.3) + (-10:0.7) $) node[label=(\cmdKV@general@rotate):\cmdKV@gLabels@eLD] {};
    \draw[label distance=1,\cmdKV@gTypes@iA] ($ (0,-0.3) + (135:0.26) $) -- node[label=(-180+\cmdKV@general@rotate):\cmdKV@gLabels@iLA] {}($ (0,0.3) + (-135:0.26) $);
    \draw[label distance=1,\cmdKV@gTypes@iB] ($ (0,0.3) + (-45:0.26) $) -- node[label=(\cmdKV@general@rotate):\cmdKV@gLabels@iLB] {}($ (0,-0.3) + (45:0.26) $);
    \node[label=(90+\cmdKV@general@rotate):\cmdKV@gLabels@iLC] at (0,0.5) {};
    \node[label=(-90+\cmdKV@general@rotate):\cmdKV@gLabels@iLD] at (0,-0.5) {};
    #2
  \end{tikzpicture}
  \gSetStdTypes{line}
}

\def\cBoxBoxA{\@ifnextchar[\@cBoxBoxA{\@cBoxBoxA[]}}
\def\@cBoxBoxA[#1]#2{%
  \setkeys*{pre}{#1}%
  \setrmkeys{general,gTypes,gLabels,blobs}%
  \begin{tikzpicture}
    [line width=1pt,
    baseline={([yshift=-0.5ex+\cmdKV@general@yshift]current bounding box.center)},
    scale=\cmdKV@general@scale,
    rotate=\cmdKV@general@rotate,
    font=\cmdKV@general@font]
    \fill[\cmdKV@blobs@blobA] (-0.4,0) ellipse (0.2 and 0.4);
    \fill[\cmdKV@blobs@blobB] (0.4,0) ellipse (0.2 and 0.4);
    \fill[\cmdKV@blobs@blobC] (1.2,0) ellipse (0.2 and 0.4);
    \draw[\cmdKV@gTypes@eA] ($ (-0.4,0) + (-135:0.26) $) -- ($ (-0.4,0) + (-135:0.7) $) node[label=(-180+\cmdKV@general@rotate):\cmdKV@gLabels@eLA] {};
    \draw[\cmdKV@gTypes@eB] ($ (-0.4,0) + (135:0.26) $) -- ($ (-0.4,0) + (135:0.7) $) node[label=(180+\cmdKV@general@rotate):\cmdKV@gLabels@eLB] {};
    \draw[\cmdKV@gTypes@eC] ($ (1.2,0) + (45:0.26) $) -- ($ (1.2,0) + (45:0.7) $) node[label=(\cmdKV@general@rotate):\cmdKV@gLabels@eLC] {};
    \draw[\cmdKV@gTypes@eD] ($ (1.2,0) + (-45:0.26) $) -- ($ (1.2,0) + (-45:0.7) $) node[label=(\cmdKV@general@rotate):\cmdKV@gLabels@eLD] {};
    \draw[label distance=1,\cmdKV@gTypes@iA] ($ (-0.4,0) + (45:0.26) $) -- node[label=(90+\cmdKV@general@rotate):\cmdKV@gLabels@iLA] {}($ (0.4,0) + (135:0.26) $);
    \draw[label distance=1,\cmdKV@gTypes@iB] ($ (0.4,0) + (45:0.26) $) -- node[label=(90+\cmdKV@general@rotate):\cmdKV@gLabels@iLB] {}($ (1.2,0) + (135:0.26) $);
    \draw[label distance=1,\cmdKV@gTypes@iC] ($ (1.2,0) + (-135:0.26) $) -- node[label=(-90+\cmdKV@general@rotate):\cmdKV@gLabels@iLC] {}($ (0.4,0) + (-45:0.26) $);
    \draw[label distance=1,\cmdKV@gTypes@iD] ($ (0.4,0) + (-135:0.26) $) -- node[label=(-90+\cmdKV@general@rotate):\cmdKV@gLabels@iLD] {}($ (-0.4,0) + (-45:0.26) $);
    \node[label=(180+\cmdKV@general@rotate):\cmdKV@gLabels@iLE] at (-0.6,0) {};
    \node[label=(\cmdKV@general@rotate):\cmdKV@gLabels@iLF] at (0.6,0) {};
    \node[label=(\cmdKV@general@rotate):\cmdKV@gLabels@iLG] at (1.4,0) {};
    #2
  \end{tikzpicture}
  \gSetStdTypes{line}
}

\def\cBoxBoxB{\@ifnextchar[\@cBoxBoxB{\@cBoxBoxB[]}}
\def\@cBoxBoxB[#1]#2{%
  \setkeys*{pre}{#1}%
  \setrmkeys{general,gTypes,gLabels}%
  \begin{tikzpicture}
    [line width=1pt,
    baseline={([yshift=-0.5ex+\cmdKV@general@yshift]current bounding box.center)},
    scale=\cmdKV@general@scale,
    rotate=\cmdKV@general@rotate,
    font=\cmdKV@general@font]
    \fill[\cmdKV@blobs@blobA] (0,0.3) ellipse (0.4 and 0.2);
    \fill[\cmdKV@blobs@blobB] (0,-0.3) ellipse (0.4 and 0.2);
    \fill[\cmdKV@blobs@blobC] (1,0) ellipse (0.2 and 0.4);
    \draw[\cmdKV@gTypes@eA] ($ (0,-0.3) + (180:0.4) $) -- ($ (0,-0.3) + (-170:0.7) $) node[label=(180+\cmdKV@general@rotate):\cmdKV@gLabels@eLA] {};
    \draw[\cmdKV@gTypes@eB] ($ (0,0.3) + (180:0.4) $) -- ($ (0,0.3) + (170:0.7) $) node[label=(180+\cmdKV@general@rotate):\cmdKV@gLabels@eLB] {};
    \draw[\cmdKV@gTypes@eC] ($ (1,0) + (45:0.26) $) -- ($ (1,0) + (45:0.7) $) node[label=(\cmdKV@general@rotate):\cmdKV@gLabels@eLC] {};
    \draw[\cmdKV@gTypes@eD] ($ (1,0) + (-45:0.26) $) -- ($ (1,0) + (-45:0.7) $) node[label=(\cmdKV@general@rotate):\cmdKV@gLabels@eLD] {};
    \draw[label distance=1,\cmdKV@gTypes@iA] ($ (0,-0.3) + (135:0.26) $) -- node[label=(-180+\cmdKV@general@rotate):\cmdKV@gLabels@iLA] {}($ (0,0.3) + (-135:0.26) $);
    \draw[label distance=1,\cmdKV@gTypes@iB] ($ (0,0.3) + (0:0.4) $) -- node[label=(90+\cmdKV@general@rotate):\cmdKV@gLabels@iLB] {} ($ (1,0) + (135:0.26) $);
    \draw[label distance=1,\cmdKV@gTypes@iC] ($ (1,0) + (-135:0.26) $) -- node[label=(-90+\cmdKV@general@rotate):\cmdKV@gLabels@iLC] {} ($ (0,-0.3) + (0:0.4) $);
    \draw[label distance=1,\cmdKV@gTypes@iD] ($ (0,0.3) + (-45:0.26) $) -- node[label=(\cmdKV@general@rotate):\cmdKV@gLabels@iLD] {}($ (0,-0.3) + (45:0.26) $);
    \node[label=(90+\cmdKV@general@rotate):\cmdKV@gLabels@iLE] at (0,0.5) {};
    \node[label=(\cmdKV@general@rotate):\cmdKV@gLabels@iLF] at (1.2,0) {};
    \node[label=(-90+\cmdKV@general@rotate):\cmdKV@gLabels@iLG] at (0,-0.5) {};
    #2
  \end{tikzpicture}
  \gSetStdTypes{line}
}

\def\cNPBox{\@ifnextchar[\@cNPBox{\@cNPBox[]}}
\def\@cNPBox[#1]#2{%
  \setkeys*{pre}{#1}%
  \setrmkeys{general,gTypes,gLabels,blob}%
  \begin{tikzpicture}
    [line width=1pt,
    baseline={([yshift=-0.5ex+\cmdKV@general@yshift]current bounding box.center)},
    scale=\cmdKV@general@scale,
    rotate=\cmdKV@general@rotate,
    font=\cmdKV@general@font]
    \fill[\cmdKV@blobs@blobA] (-0.6,0) ellipse (0.2 and 0.4);
    \fill[\cmdKV@blobs@blobB] (0.4,0) ellipse (0.2 and 0.4);
    \draw[\cmdKV@gTypes@eA] ($ (-0.6,0) + (-180:0.2) $) -- ($ (-0.6,0) + (-180:0.6) $) node[label=(-180+\cmdKV@general@rotate):\cmdKV@gLabels@eLA] {};
    \draw[\cmdKV@gTypes@eB] ($ (-0.6,0) + (0:0.2) $) -- (-0.18,0) node[label=(\cmdKV@general@rotate):\cmdKV@gLabels@eLB] {};
    \draw[\cmdKV@gTypes@eC] ($ (0.4,0) + (45:0.26) $) -- ($ (0.4,0) + (45:0.7) $) node[label=(\cmdKV@general@rotate):\cmdKV@gLabels@eLC] {};
    \draw[\cmdKV@gTypes@eD] ($ (0.4,0) + (-45:0.26) $) -- ($ (0.4,0) + (-45:0.7) $) node[label=(\cmdKV@general@rotate):\cmdKV@gLabels@eLD] {};
    \draw[label distance=1,\cmdKV@gTypes@iA] ($ (-0.6,0) + (60:0.3) $) -- node[label=(90+\cmdKV@general@rotate):\cmdKV@gLabels@iLA] {}($ (0.4,0) + (135:0.26) $);
    \draw[label distance=1,\cmdKV@gTypes@iB] ($ (0.4,0) + (-135:0.26) $) -- node[label=(-90+\cmdKV@general@rotate):\cmdKV@gLabels@iLB] {}($ (-0.6,0) + (-60:0.3) $);
    \node[label=(\cmdKV@general@rotate):\cmdKV@gLabels@iLD] at (0.6,0) {};
    #2
  \end{tikzpicture}
  \gSetStdTypes{line}
}

\def\cFourPtRec{\@ifnextchar[\@cFourPtRec{\@cFourPtRec[]}}
\def\@cFourPtRec[#1]#2{%
  \setkeys*{pre}{#1}%
  \setrmkeys{general,gTypes,gLabels,blobs}%
  \begin{tikzpicture}
    [line width=1pt,
    baseline={([yshift=-0.5ex+\cmdKV@general@yshift]current bounding box.center)},
    scale=\cmdKV@general@scale,
    rotate=\cmdKV@general@rotate,
    font=\cmdKV@general@font]
    \fill[\cmdKV@blobs@blobA] (-0.5,0) ellipse (0.2 and 0.3);
    \fill[\cmdKV@blobs@blobB] (0.5,0) ellipse (0.3 and 0.4);
    \filldraw[fill=white,draw=black] (0.55,0.15) circle (0.1);
    \filldraw[fill=white,draw=black] (0.45,-0.1) circle (0.1);
    \draw[\cmdKV@gTypes@eA] ($(-0.5,0) + (135:0.23)$) -- ($(-0.5,0) + (135:0.5)$) node[label=(180+\cmdKV@general@rotate):\cmdKV@gLabels@eLA] {};
    \draw[\cmdKV@gTypes@eB] ($(-0.5,0) + (-135:0.23)$) -- ($(-0.5,0) + (-135:0.5)$) node[label=(180+\cmdKV@general@rotate):\cmdKV@gLabels@eLB] {};
    \draw[\cmdKV@gTypes@eC] ($(0.5,0) + (45:0.35)$) -- ($(0.5,0) + (45:0.7)$) node[label=(\cmdKV@general@rotate):\cmdKV@gLabels@eLC] {};
    \draw[\cmdKV@gTypes@eD] ($(0.5,0) + (-45:0.35)$) -- ($(0.5,0) + (-45:0.7)$) node[label=(\cmdKV@general@rotate):\cmdKV@gLabels@eLD] {};
    \draw[label distance=1,\cmdKV@gTypes@iA] ($(-0.5,0) + (30:0.23)$) -- node[label=(90+\cmdKV@general@rotate):\cmdKV@gLabels@iLA] {} ($(0.5,0) + (157:0.3)$);
    \draw[label distance=1,\cmdKV@gTypes@iB] ($(0.5,0) + (-157:0.3)$) -- node[label=(-90+\cmdKV@general@rotate):\cmdKV@gLabels@iLB] {} ($(-0.5,0) + (-30:0.23)$);
    #2
 \end{tikzpicture}
  \gSetStdTypes{line}
}

\makeatother

\usepackage{mathtools}
\usepackage{amssymb}
\usepackage{amsmath}
\usepackage{epsfig}
\usepackage{hyperref}
\usepackage{breakurl}
\usepackage{bm}
\usepackage{color}
\usetikzlibrary{decorations.markings}
\usepackage[english]{babel}
\def\ash{%
  {
    a%
    \kern-.045em%
    \lower.3ex\hbox{sh}%
}%
}

\newcommand\beq{\begin{equation}}
\newcommand\eeq{\end{equation}}
\def\bea{\begin{eqnarray}}
\def\eea{\end{eqnarray}}

\DeclareRobustCommand{\SkipTocEntry}[4]{}

\DeclareMathOperator{\Arcsinh}{sinh^{-1}}

\newcommand{\nn}{\nonumber}

\newcommand\beal{\begin{aligned}}
\newcommand\eeal{\end{aligned}}

\newcommand\dd{{\mathrm d}}
\usepackage{xcolor}

\newcommand{\bp}{{\boldsymbol p}}


\begin{document}

\preprint{\tt DESY 20-131} 
\preprint{\tt SLAC-PUB-17555}

\title{Conservative Tidal Effects in Compact Binary Systems \\ [0.2cm] to Next-to-Leading Post-Minkowskian Order}
\author{Gregor K\"alin}
\email{greka@slac.stanford.edu}
\affiliation{SLAC National Accelerator Laboratory, Stanford University, Stanford, CA 94309, USA}
\author{Zhengwen Liu}
\email{zhengwen.liu@desy.de}
\author{Rafael A. Porto}
\email{rafael.porto@desy.de}
\affiliation{ Deutsches Elektronen-Synchrotron DESY,
Notkestrasse 85, 22607 Hamburg, Germany}

\begin{abstract}
Using the Effective Field Theory approach together with the Boundary-to-Bound map, we compute the next-to-leading order (NLO) Post-Minkowskian (PM) tidal effects in the conservative dynamics of compact binary systems. We derive the mass \& current quadrupole and, for the first time, octupole corrections to the binding energy for circular orbits at ${\cal O}(G^3)$. Our results are consistent with the test-body limit as well as the existent Post-Newtonian literature. We also reconstruct a Hamiltonian incorporating tidal effects to NLO in the PM expansion and find complete agreement with the recent derivation of its quadrupolar part using the classical limit of scattering amplitudes. 

\end{abstract}

\maketitle

\emph{Introduction.} The demonstrated feasibility of direct detection of gravitational waves (GWs) from binary systems \cite{LIGOScientific:2018mvr,LIGO}, and in particular the observation of neutron star inspirals \cite{ns}, has revealed a new window to explore compact objects in an unprecedented fashion~\cite{buosathya,tune,music}. Not only carry GWs the imprint of the equation of state of nuclear matter through tidal effects \cite{tanja1,tanja2,Abbott:2018exr}, they have also opened new frontiers for beyond the standard model searches~\cite{axiverse,gcollider1,gcollider2} as well as the exploration of the remarkable properties of black holes in Einstein's gravity~\cite{music,tune}.  On the other hand, distinguishing the properties of compact objects from tidal disruptions is a daunting task requiring a high level of analytic control, to at least fifth Post-Newtonian (5PN) order \cite{tune,music}, while lifting several degeneracies may also require an even a higher level of precision for waveform modeling.\vskip 2pt The Effective Field Theory (EFT) formalism for PN sources introduced in \cite{nrgr}, which has already achieved a high level of analytic accuracy both for non-spinning \cite{dis1,nrgr2pn,nrgr3pn,Foffa:2012rn,tail,nrgrG5,apparent,nrgr4pn1,nrgr4pn2,5pn1,5pn2,blum, andirad,adamchad1,radnrgr} and spinning binaries \cite{nrgrs,prl,Porto:2007tt,dis2,nrgrss,nrgrs2,nrgrso,rads1,amps,maiaso,maiass,levi,Levi:2020kvb,Levi:2020uwu}, is tailor-made to incorporate finite-size effects, see e.g.\,\cite{nrgr,walterLH,foffa,iragrg,grg13,review}. For instance, it was used in \cite{luc7pn} to obtain the next-to-next-to-leading-order (NNLO) contributions to the equations of motion to 7PN order. However, partially due~to~the repurposing of powerful tools from the amplitudes program, e.g. \cite{ira1,cheung,bohr,zvi1,zvi2,cristof1,donal,Cheung:2020gyp,Bern:2020buy,Parra}, the `Boundary-to-Bound' dictionary~\cite{paper1,paper2} as well as other developments, e.g.\,\cite{damour1,damour2,damour3n,damour3,binit}, it has become apparent that the study of scattering processes in the Post-Minkowskian (PM) expansion may ultimately push even further the frontiers of analytic understanding of binary systems. With these tools at hand an EFT framework in the PM regime was developed in~\cite{pmeft} and readily implemented in \cite{3pmeft} to reach the present state-of-the-art at 3PM~\cite{zvi1,zvi2,Cheung:2020gyp}. Our purpose here is to~extend the calculation of leading tidal effects in~\cite{pmeft} (see also \cite{binit,Haddad:2020que}), and compute the~mass \& current quadrupolar and octupolar tidal effects to NLO in the PM expansion. While the latter are presented~for the first time, we find agreement for the former with the recent results in~\cite{soloncheung}. The derivation in~\cite{soloncheung} uses the classical limit of the scattering amplitude augmented with higher-derivative interactions and standard Feynman diagrams, together with the `impetus formula'~\cite{paper1}. Although Feynman's tools are also at the core of~our approach, the formalisms are rather different. In particular, unlike the derivations in~\cite{zvi1,zvi2,Cheung:2020gyp,soloncheung}, ours is reduced to (massless) integrals whose velocity-dependence is bootstrapped via differential equation from the EFT with static classical  sources~\cite{3pmeft}, which greatly simplifies the calculations.\vskip 2pt

\emph{Extended objects in the EFT approach.} Following \cite{nrgr}, tidal effects are incorporated in~\cite{pmeft} by including a series of higher-derivative terms in the worldline action,\footnote{The action in \eqref{act} is equivalent to the reparameterization-invariant one in \cite{nrgr}, up to higher orders in the curvature. In~the presence of finite-size terms~the~gauge choice  $e_a=1$ for the einbein sets $\tau_a$ as the proper time at future and past infinity. The relative signs are due to our flat-metric convention.}
 \begin{align}
&S_{\rm pp} = \sum_{a = 1,2} \int \dd\tau_a\Big(-\frac{m_a}{2}\,g_{\mu\nu} v_a^\mu v_a^\nu +  c^{(a)}_{E^2} E_{\mu\nu} E^{\mu\nu}  \label{act}\\
&+  c^{(a)}_{B^2} B_{\mu\nu} B^{\mu\nu} - c^{(a)}_{{\tilde E}^2} E_{\mu\nu\alpha}  E^{\mu\nu\alpha} -  c^{(a)}_{{\tilde B}^2} B_{\mu\nu\alpha} B^{\mu\nu\alpha}+\cdots \Big)\,,\nn
\end{align}
with $\big(c^{(a)}_{E^2}, c^{(a)}_{B^2}\big)$ and $\big(c^{(a)}_{{\tilde E}^2}, c^{(a)}_{{\tilde B}^2}\big)$ the mass \& current quadrupole and octupole tidal `Love numbers', respectively. The couplings are written in terms of the electric- and magnetic-type components of the Riemann (Weyl) tensor and its dual, \begin{align}
&E_{\alpha\beta} = R_{\mu\alpha\nu\beta} u^\mu u^\nu,\,\, B_{\alpha\beta} = R^\star_{\mu\alpha\nu\beta} u^\mu u^\nu, \\
&E_{\alpha\beta\gamma} = \nabla^\perp_{\{\alpha} R_{\beta\rho\gamma \} \nu} u^\rho u^\nu,\,\, B_{\alpha\beta\gamma} = \nabla^\perp_{\{\alpha} R^\star_{\beta\rho\gamma \} \nu} u^\rho u^\nu,\nn \end{align}  where~$\nabla^\perp_\alpha$ is the covariant derivative projected orthogonal to the velocity, and $\{\ldots\}$ stands for symmetrization. The two-body effective action is obtained by integrating out the metric field in the weak-field and saddle-point approximation via Feynman diagrams \cite{pmeft,3pmeft}. We use the convention $\eta_{\mu\nu}= {\rm diag}(+,-,-,-)$ for the Minkowski metric. Intermediate divergences are handled by dimensional~regularization in $D=4-2\epsilon$ dimensions.
\vskip 2pt 
\emph{Scattering angle.} In the EFT formalism of \cite{pmeft}, the scattering angle is computed via the impulse. The latter follows iteratively from the effective Lagrangian,
\beq
\Delta p_a^\mu = -\eta^{\mu\nu} \int_{-\infty}^{+\infty} \dd\tau_a \frac{\partial {\cal L_{\rm eff}}}{\partial x^\nu_a}(x_a(\tau_a))\,,\label{dp}
\eeq
by inputting the PM expansion of the trajectories
\beq
\label{pmexp}
\begin{aligned}
x^\mu_a(\tau_a) &= b^\mu_a + u^\mu_a \tau_a + \sum_n \delta^{(n)} x^\mu_a (\tau_a)\,,
\end{aligned}
\eeq
with $b^\mu \equiv b^\mu_1 - b_2^\mu$ the impact parameter and $u_a$ the incoming velocities. The leading ${\cal L}_{\rm eff}$ also contributes to NLO when evaluated on \eqref{pmexp}. We refer to these corrections as {\it iterations} \cite{pmeft,3pmeft}. The deflection angle is given~by 
\beq
2\sin \frac{\chi}{2}= \chi+ {\cal O}(\chi^3) = {\sqrt{-\Delta p_a^2} \over p_\infty}\,,\label{chi2}
\eeq
where $p_\infty = \mu \tfrac{\sqrt{\gamma^2-1}}{\Gamma}$, with $\Gamma \equiv \tfrac{E}{M} = \sqrt{1+ 2\nu (\gamma-1)}\,,$
 $(M, E)$ the total mass/energy, $\mu = m_1m_2/M$ the reduced mass, and  $\nu = \mu/M$ the symmetric mass ratio. Throughout this letter we use the notation \beq \gamma \equiv u_1\cdot u_2 = \frac{p_1\cdot p_2}{m_1m_2} =  1 + {\cal E} + \frac{\nu}{2}{\cal E}^2\,,
\eeq  
where ${\cal E}=(E-M)/\mu$ is the (reduced) binding energy.\vskip 2pt
  \begin{figure}[t!] 
\includegraphics[width=0.42\textwidth]{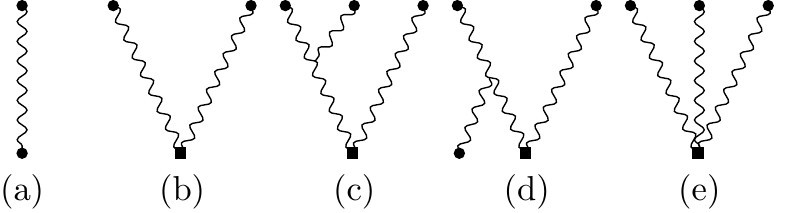} 
     \caption{Feynman diagrams needed for tidal effects to NLO. The square represents the finite-size couplings in~\eqref{act}.} 
      \label{fig1}
            \vspace{-0.4cm}
\end{figure}

\emph{Feynman master integrals.} At leading PM order only the diagram in Fig.\,\ref{fig1}\,(b) contributes. The derivation for the quadrupole coupling was carried out in \cite{pmeft} and can be easily extended to the octupole term. For the NLO effects,  the remaining diagrams in~Fig.\,\ref{fig1} are needed, including Fig.\,\ref{fig1}\,(a) which is required to compute the iterations.  As discussed in \cite{pmeft,3pmeft}, in addition to the standard massless $1/k^2$ propagators for the gravitational field, we have linear ones, $(k\cdot u_a \mp i0)^{-1}$, which arise from the expansion in \eqref{pmexp}. We restrict ourselves to the impulse in the direction of the impact parameter, which is sufficient to obtain the scattering angle \cite{pmeft}. As in \cite{3pmeft}, the computation is reduced to terms involving the (transverse) Fourier transform in the transfer momentum of a series of `two loop' (cut) integrals.\vskip 2pt 
 
 As it turns out, a subset of the family of master integrals in \cite{3pmeft} is sufficient to compute all of the diagrams in Fig.\,\ref{fig1}, including the iterations. As it was discussed in~\cite{3pmeft}, the $\gamma$-dependence is obtained either  by going to the rest frame of one of the particles, or through differential equations whose boundary conditions are extracted from the static limit. In all cases, the integrals are reduced to the same type that appear in the computation of tidal effects in the EFT with PN sources \cite{nrgr2pn}. The intermediate divergences either cancel out or yield contact-terms which do not enter in the classical limit. Hence, we do not encounter ultraviolet poles requiring a counter-term contribution in the effective action in \eqref{act} at this order. Not surprisingly, at the end of the day the resulting tidal effects also feature the (in)famous factor of $\Arcsinh\sqrt{(\gamma-1)/2}$, first observed in the monopole contributions at NNLO~\cite{zvi1,zvi2,Parra,3pmeft}.

\vskip 2pt
\emph{Scattering data.} Garnering all the ingredients for the impulse projected in the direction of the impact parameter, the scattering angle then follows from \eqref{chi2}, yielding for the quadrupolar tidal effects (with $1/j \equiv GM\mu/(p_\infty b)$)
\begin{widetext}
\begin{align}\label{dchi1}
\frac{\Delta \chi_{(E,B)}}{\Gamma}  & = \frac{45\pi}{64} \frac{(\gamma^2-1)^2}{(\Gamma j)^6} 
\Big[ \left(35 \gamma^4-30 \gamma^2-5\right) \lambda_{B^2}
+\left(35 \gamma^4-30 \gamma^2+11\right) \lambda_{E^2} \Big] \\
&+ \frac{192}{35} \frac{(\gamma^2-1)^{3/2}}{(\Gamma j)^7} 
\Big[\left(160 \gamma^6-192 \gamma^4+30 \gamma^2+2\right) \lambda_{B^2} +\left(160 \gamma^6-192 \gamma^4+72 \gamma^2-5\right) \lambda_{E^2} \Big]
\nonumber\\
&+ \frac{96 \nu }{35} \frac{\sqrt{\gamma^2-1}}{(\Gamma j)^7}\kappa_{B^2}
 \Big[224 \gamma^9-320 \gamma^8-728 \gamma^7+704 \gamma^6+5488 \gamma^5-444 \gamma^4+66262 \gamma^3 +56 \gamma^2+28084 \gamma +4\Big]
\nonumber\\
&+ \frac{96 \nu }{35} \frac{\sqrt{\gamma^2-1}}{(\Gamma j)^7}\kappa_{E^2}
 \Big[224 \gamma^9-320 \gamma^8-728 \gamma^7+704 \gamma^6+5628 \gamma^5-528 \gamma^4+65982 \gamma^3 +154 \gamma^2+28329 \gamma -10\Big]
\nonumber\\
&- \frac{576 \nu \sqrt{\gamma^2-1}}{(\Gamma j)^7}
\Big[\left(440 \gamma^4+474 \gamma^2+32\right)\, \kappa_{B^2} + \left(440 \gamma^4+474 \gamma^2+33\right)\kappa_{E^2} \Big]\ash(\gamma)\,,
\nonumber
\end{align}
where we used the shortened notation $\ash(\gamma) \equiv (\gamma^2 {-} 1)^{-1/2}\Arcsinh\sqrt{ (\gamma -1) / 2}$; whereas for the octupolar contribution, computed here for the first time, we arrive at
\begin{align}\label{dchi2}
 \frac{\Delta \chi_{({\tilde E},{\tilde B})}}{\Gamma} = \frac{525\pi}{512(\Gamma j)^8}(\gamma^2 &- 1)^3
 \Big[(21\gamma^6+385\gamma^4-305\gamma^2+91)\lambda_{{\tilde E}^2}
 +(21\gamma^6+385\gamma^4-385\gamma^2-21)\lambda_{{\tilde B}^2}\Big]
 \\ 
+\frac{512 (\gamma^2-1)^{5/2}}{3003(\Gamma j)^9}
&\Big[\left(4800 \gamma^8+77520 \gamma^6-74888 \gamma^4+17707 \gamma^2+1888\right) \lambda_{{\tilde E}^2} 
\nonumber\\
&+\left(4800 \gamma^8+77520 \gamma^6-87472 \gamma^4+5552 \gamma^2-400\right) \lambda_{{\tilde B}^2}\Big]\nonumber
\\ 
+\frac{128\nu \sqrt{\gamma^2 -1}\kappa_{{\tilde B}^2}}{3003(\Gamma j)^9}
&\Big[27456 \gamma^{13}-19200 \gamma^{12}+205920 \gamma^{11}-271680 \gamma^{10}-1589016 \gamma^9
+950848 \gamma^8+22048884 \gamma^7
\nonumber\\ 
&-1032064\gamma^6+579540390 \gamma^5+395904 \gamma^4 +826613931 \gamma^3-25408 \gamma^2+148331040 \gamma +1600\Big]
\nonumber\\ 
+\frac{128 \nu\sqrt{\gamma^2 -1} \kappa_{{\tilde E}^2} }{3003(\Gamma j)^9}
&\Big[27456 \gamma^{13}-19200 \gamma^{12}+205920 \gamma^{11}-271680 \gamma^{10}-1468896 \gamma^9 +900512 \gamma^8+21724560 \gamma^7
\nonumber\\ 
&-980012\gamma^6+580453302 \gamma^5 +433656 \gamma^4+837773079 \gamma^3-55724 \gamma^2+155291994 \gamma -7552\Big]
\nonumber\\ 
-\frac{3840\nu \sqrt{\gamma^2-1}}{(\Gamma j)^9}
&\Big[(7292 \gamma^6+19484 \gamma^4+7905 \gamma^2+288)\kappa_{{\tilde B}^2}  
+ (7292 \gamma^6+19644 \gamma^4+8141 \gamma^2+310)\kappa_{{\tilde E}^2} 
\Big] \ash(\gamma)\,.
\nonumber
\end{align}
\end{widetext}
In these expressions we introduced the parameters
\beq
\begin{aligned}
 \lambda_{E^2} &\equiv \frac{1}{G^4 M^5} \left(m_2\frac{ c_{E^2}^{(1)}}{m_1}+m_1\frac{ c_{E^2}^{(2)}}{m_2} \right)\,, \\ 
  \kappa_{E^2} &\equiv \lambda_{E^2}+  \frac{c_{E^2}^{(1)}+c_{E^2}^{(2)}}{G^4 M^5} =  \frac{1}{G^4 M^4} \left(\frac{c_{E^2}^{(1)}}{m_1} +\frac{c_{E^2}^{(2)}}{m_2} \right)\,, \,\,
\end{aligned}
\eeq
and similarly for all the other couplings, normalized with $1/G^6M^7$ for the octupole Love numbers. We find the expression in \eqref{dchi1} to be fully equivalent to Eq.\,(13)~in~\cite{soloncheung}. Notice that, as it happens also at leading PM order \cite{pmeft, binit}, the electric- and magnetic-type tidal effects have a strikingly similar behavior in the high-energy limit. (Likewise, this is encoded in the rather simple factor in the $\lambda_1$ coupling to the Kretschmann scalar in~\cite{soloncheung}.)

\vskip 2pt
{\it Probe limit.} A non-trivial test for our results is the consistency with the test-particle limit. We computed the scattering angle for a tidally-deformed object by means of the on-shell condition in a Schwarzschild background, \bea
 g_{\rm Sch}^{\mu\nu}\, p^{(a)}_\mu p^{(a)}_\nu &=& m_a^2- 2m_a\left( c^{(a)}_{E^2} \left(E^{\rm Sch}_{\mu\nu}\right)^2 +  c^{(a)}_{B^2} \left(B^{\rm Sch}_{\mu\nu}\right)^2  \right.   \nn \\  &-& \left. c^{(a)}_{{\tilde E}^2}  \left(E^{\rm Sch}_{\alpha\mu\nu}\right)^2 - c^{(a)}_{{\tilde B}^2} \left(B^{\rm Sch}_{\alpha\mu\nu}\right)^2 \right)\,.\label{on-shell}
\eea
We constructed the radial action for hyperbolic motion from which we derived the scattering angle in the PM expansion via differentiation. Identifying the incoming (reduced) energy with the boost factor (${\cal E}_0 \to \gamma$), we found that the expressions in \eqref{dchi1} \& \eqref{dchi2} are consistent with the deflection of a tidally-disrupted test-body in a black hole background. Not surprisingly, the probe limit also fixes the leading tidal effects~\cite{paper1}. See the supplemental material for more details.\vskip 2pt

\vskip 2pt
\emph{B2B dictionary.} The (reduced) radial action \cite{paper1,paper2}
\beq
i_r \equiv  \frac{p_\infty}{\sqrt{-p_\infty^2}}
\chi^{(1)}_j - j \left(1 + \frac{2}{\pi} \sum_{n=1}^{\infty}  \frac{\chi^{(2n)}_j}{(1-2n)j^{2n}}\right)  \,,\label{eq:ir}
\eeq
is built from the PM expansion of the deflection angle, 
\beq
\frac{\chi}{2}= \sum_n \chi^{(n)}_b\left(\frac{GM}{b}\right)^n = \sum_n \frac{\chi^{(n)}_j}{j^n}\,,\label{pmchi}
\eeq 
via analytic continuation in the binding energy. Similarly to what occurs at 3PM with the monopole term \cite{paper1,paper2,3pmeft}, we can incorporate the information in the NLO tidal effects by performing a PN-truncation. To do so we use the map in \cite{paper1} to write the $P_n$'s in the  expansion of the square of the center-of-mass momentum for each particle,
\beq
\bp^2 = p_\infty^2 + \sum_{n=1}^\infty P_n(E) \left(\frac{G}{r}\right)^n\label{pinfpm}\,,
\eeq
as a function of the PM coefficients in \eqref{pmchi}. This allows us to read off the finite-size contributions to $\{P_6,P_7,P_8,P_9\}$ from \eqref{dchi1}~and~\eqref{dchi2}. We then use the inverse map~\cite{paper1},~e.g. 
\begin{align}
\chi^{(8)}_j &=\frac{105\pi }{64}\Big( \frac{\bar P_2^4}{12} + \frac{\bar P_1^2\bar P_3^2}{2}+ \bar P_1 \bar P_2^2\bar P_3+ \hat p_\infty^2 (\bar P_1^2 \bar P_6 \label{chin2}  \\ &+ P_2P_3^2)+\hat p_\infty^4\left(\bar P_1\bar P_7+ \bar P_2\bar P_6\right)+ \frac{\hat p_\infty^6}{3} \bar P_8 + \cdots \Big)\,,\nn\\
 \chi^{(10)}_j &=  \frac{315\pi }{512 }\Big(\frac{\bar P_2^5}{5} + \bar P_1^4 \bar P_6+ \cdots + 4 \hat p_\infty^2\big(3\bar P_1^2\bar P_2\bar P_6 \label{chin3} \\ &+ \bar P_1^3\bar P_7+\cdots  \big) +  6\hat p_\infty^4\big( \bar P_1^2\bar P_8 + 2\bar P_1\bar P_2\bar P_7+\cdots \big)\nn \\ &+ 4 \hat p_\infty^6 (\bar P_2\bar P_8 +\bar P_1\bar P_9+\bar P_3\bar P_7) + \hat p_\infty^8 \bar P_{10} + \cdots\Big)\,, \nn
\end{align}
with $\bar P_n \equiv P_n/(\mu^2M^n)$ and $\hat p_\infty = p_\infty/\mu$, to input the known information into the $\chi^{(2n)}_j$'s in \eqref{eq:ir}. We~have displayed only a subset of the relevant coefficients and their respective dependence on the $\{P_1,P_2, P_3\}$ at 3PM, as well as the $\{P_6,P_7,P_8,P_9\}$ whose tidal contributions we have computed. Notice we are still missing the quadrupole corrections to $\{P_8,P_9\}$ as well as $P_{n \geq 10}$. However, the reader will immediately notice the factors of $\hat p_\infty^2$ attached to each term in \eqref{chin2}-\eqref{chin3} (depending on the number of $P_n$'s involved) \cite{paper1,paper2}. After analytic continuation, the $\hat p_\infty^2$ scales with the (reduced) binding energy of the binary. Hence, since the static limit of \eqref{pinfpm} is well defined, we can consistently truncate~\eqref{eq:ir} by ignoring terms which enter at higher PN orders. There is still one subtlety~left. While the analytic continuation formally maps the $1/j$ expansion of the observables between hyperbolic and elliptic motion, for the latter case we have the additional PN scaling $j^{-1}\simeq |\hat p_\infty|$ which mixes the power-counting. Therefore, we have to retain also higher orders in the $1/j$ expansion of \eqref{eq:ir}. For instance, by keeping the $\{P_1,P_6\}$ contributions to the deflection angle in \eqref{chin2}-\eqref{chin3} we recover the exact value of the periastron advance~in the Newtonian limit in \cite{binit}. Different powers in $1/j$ are also necessary to match the PN results for the monopole terms at higher PM/PN orders \cite{paper1,paper2}.\vskip 2pt The procedure is now straightforward, allowing us to derive gauge-invariant observables directly from the analytically continued radial action \cite{paper1,paper2,pmeft,3pmeft}. For instance, we readily obtain the azimuthal orbital frequency, $\Omega_\phi(j,\gamma)$, by taking derivatives of \eqref{eq:ir} with respect to the binding energy and angular momentum. For the case of circular orbits we proceed as follows \cite{paper1,paper2,pmeft}. First,~we solve for $j_{\rm circ}(\gamma)$ with the condition $i_r=0$ in \eqref{eq:ir}, including also the 3PM monopole corrections \cite{paper1,paper2,pmeft,3pmeft}. We  plug it back into $\Omega_\phi (j_{\rm circ}(\gamma),\gamma) \equiv \Omega_{\rm circ}(\gamma)$, which can then be inverted to extract the binding energy as a function of the orbital frequency. (Alternatively, we have checked that the first law of binary dynamics~\cite{letiec} holds for tidal effects.) 
Bundling~the terms together, and keeping up to 2PN corrections in each sector, we find \begin{widetext}
\begin{align}
&\Delta {\cal E}_{\rm T}= x \Bigg[ 18\, \lambda_{E^2}x^5 + 11\Big( 3(1-\nu)\lambda_{E^2} +6\,\lambda_{B^2}  +5\nu \,\kappa_{E^2}\Big)x^6 +  \left(390\lambda_{{\tilde E}^2}-\frac{13}{28}(161\nu^2-161\nu-132)\lambda_{E^2}- \frac{1326\nu}{7}\kappa_{B^2}\right.\nn\\
& +\left.\frac{13}{28}(616\nu+699)\lambda_{B^2}+\frac{13\nu}{84}(490\nu-729)\kappa_{E^2}  +\frac{13}{6} \Delta \bar P_{8,\rm stc}^{(E,B)}\right)x^7 +75  \big(  45 \nu \kappa_{{\tilde E}^2} - (13 \nu+3)\lambda_{{\tilde E}^2} + 16 \lambda_{{\tilde B}^2}\big)x^8\label{binding} \\
 &-\Big(
  \frac{85}{36} \left(1083 \nu^2+1539 \nu +163\right)\lambda_{{\tilde E}^2} +\frac{27200 \nu }{3}\kappa_{{\tilde B}^2} - \frac{85}{4} (270 \nu +383) \nu\kappa_{{\tilde E}^2}-\frac{680}{9} (90 \nu+173)\lambda_{{\tilde B}^2}  - \frac{17}{6}  \Delta \bar P_{10,\rm stc}^{(\tilde E,\tilde B)} \Big)x^9 \Bigg]  \nn
\end{align}
\end{widetext}
with $x \equiv (GM\Omega_{\rm circ})^{2/3}$ the standard PN parameter. The $\Delta \bar P_{8,\rm stc}^{(E,B)}$ and $\Delta \bar P_{10,\rm stc}^{(\tilde E,\tilde B)}$ are the contributions in the static limit $(\gamma \to 1)$ from the quadrupole and octupole couplings at NNLO in $G$, respectively.
The expression in \eqref{binding} agrees with the results in Eq. 6.5b of \cite{luc7pn} for the quadrupole couplings to 6PN, as well as the leading octupole~at~7PN. Moreover, the difference at 7PN in \cite{luc7pn} is only due to the static limit~of~$P_8$. We can extract its value using \eqref{chin2} with the~$\chi^{(8)}_j$ which follows from the Lagrangian to 7PN obtained in \cite{luc7pn}, yielding 
\begin{align}\label{p8st}
\Delta \bar P_{8,\rm stc}^{(E,B)}& =  \frac{1326}{7}\nu\kappa_{B^2}+\big(243-90\nu\big)\nu \kappa_{E^2} \\
+&\Big( 45\nu^2  - \frac{885\nu}{7}+\frac{675}{14}\Big)\lambda_{E^2} -\Big(234\nu +\frac{837}{14}\Big)\lambda_{B^2}\,.\nonumber
\end{align}
We have checked that its ${\cal O}(\nu^0)$ part is consistent with the probe limit (see the supplemental material). The correction in \eqref{p8st} gives us the last ingredient for the binding energy at~7PN, while at the same time proves the equivalence of the derivations in \cite{luc7pn} with a truncation of the PM results in the quadrupole sector to ${\cal O}(G^3v^2)$. \vskip 2pt As~advertised, the octupolar contributions at 8PN are presented for the first time. We also included the partial results at~9PN order, missing only the static corrections at NNLO in $G$ from $\Delta P_{10}$, whose ${\cal O}(\nu^0)$ part can be extracted from the probe limit (see supplemental material) 
\beq
\label{p10st}
 \Delta \bar P_{10,\rm stc}^{(\tilde E,\tilde B)} =  \frac{1}{3}\big( 2050 \lambda_{{\tilde E}^2} -13120 \lambda_{{\tilde B}^2}\big)+ {\cal O}(\nu) \,.
\eeq
\vskip 2pt \emph{Hamiltonian.} The B2B map allows us to directly produce observables without a Hamiltonian. However, it~is still instructive to reconstruct it using our dictionary \cite{paper1}. We~do so in the center-of-mass (isotropic) frame, where
\beq
H(r,\bp^2) = \sum_{n=0}^{\infty} \frac{c_n(\bp^2)}{n!} \left(\frac{G}{r}\right)^n\,,\label{Ham}
\eeq
with $c_0 \equiv \sum_a \sqrt{\bp^2+m_a^2}$. The $c_n$ coefficients~in~\eqref{Ham} can be then obtained iteratively from the~$P_n$'s in~\eqref{pinfpm}. The tidal contributions to the latter are collected in the supplemental material (see \eqref{p6}-\eqref{p9}), from which we derive the (lengthier) finite-size contributions to the former, \begin{widetext}
\begin{align}\label{c6}
&  \Delta c_6 = -\frac{270M^7 \nu^2}{\Gamma^2 \xi } \Big[(35 \gamma^4-30 \gamma^2+11) \lambda_{E^2}
+ 5 (7 \gamma^4-6 \gamma^2-1) \lambda_{B^2}\Big],
\nonumber\\[0.35 em]
&\begin{aligned}
    \Delta c_7 =
    \frac{270 M^8 \nu^2}{(\gamma^2-1) \Gamma^2 \xi }
    \bigg\{ & {\nu \over \gamma^2-1}
    \Big[ 
    \big(D_{7,1}  {\kappa_{E^2}} + D_{7,2} {\kappa_{B^2}}\big)
    + \big(D_{7,3} {\kappa_{E^2}} + D_{7,4} {\kappa_{B^2}} \big) \ash(\gamma)
    \Big]
    \\
    &+\Big[D_{7,5}
      + \frac{\nu^2}{\Gamma^7\xi^2}
      \big(D_{7,7} +(\gamma -1)D_{7,9} \nu +(\gamma -1)^3 D_{7,11} \nu^{2}\big) \Big]\lambda_{E^2}\\
    &+\Big[D_{7,6}
      + \frac{(\gamma^2-1)\nu^2}{\Gamma^7\xi^2}
      \big(D_{7,8} + (\gamma -1) D_{7,10}\nu + (\gamma -1)^3 D_{7,12} \nu^2\big)\Big]\lambda_{B^2}
\bigg\},
\end{aligned}\nonumber\\[0.35 em]
&\Delta c_8 = - \frac{18900 M^9 \nu^2}{\Gamma^2 \xi }
\Big[ (21 \gamma^6+385 \gamma^4-305 \gamma^2+91) \lambda_{{\tilde E}^2}
+ 7(3 \gamma^6+55 \gamma^4-55 \gamma^2-3) \lambda_{{\tilde B}^2}\Big],
\\[0.35 em]
&\begin{aligned}
   \Delta c_9 =
    \frac{302400 M^{10} \nu^{2}}{143 (\gamma^2-1) \Gamma^2 \xi}
    \bigg\{& {\nu \over (\gamma^2-1)^2}
    \Big[ \big(D_{9,1}\kappa_{{\tilde E}^2} + D_{9,2}\kappa_{{\tilde B}^2}\big)
    + \big( D_{9,3} \kappa_{{\tilde E}^2} + D_{9,4} \kappa_{{\tilde B}^2}\big)\ash(\gamma) \Big]
    \\
    &+\Big[D_{9,5}
      + \frac{1287\nu^2}{16\Gamma^7\xi^2}\big(D_{9,7} +(\gamma -1)D_{9,9}\nu +(\gamma -1)^3 D_{9,11} \nu^2\big)
      \Big] \lambda_{{\tilde E}^2}\\
    &+\Big[D_{9,6} + \frac{1287\nu^2(\gamma^2-1)}{16\Gamma^7\xi^2}
      \big(D_{9,8} +(\gamma -1) D_{9,10}\nu + (\gamma -1)^3 D_{9,12} \nu^2\big)
      \Big]\lambda_{{\tilde B}^2}
      \bigg\}\,.
\end{aligned}\nonumber
\end{align}
\end{widetext}
The $D_{i,j}$ are polynomials in $\gamma$ which we display in the supplemental material (see \eqref{dij} and \eqref{dij2}). We find agreement with the results for the quadrupolar contributions computed~in~\cite{soloncheung}, while the octupolar corrections beyond leading PN/PM order are derived here for the first time.

\vskip 2pt
{\it Conclusions}. Motivated by probing compact objects via GW observations \cite{buosathya,tune,music,tanja1,tanja2,Abbott:2018exr,axiverse,gcollider1,gcollider2}, we computed tidal effects in the conservative dynamics to NLO in the PM expansion. We used the EFT approach and B2B map developed in \cite{pmeft,3pmeft,paper1,paper2} to calculate the mass \& current quadrupolar and, for the first time, octupolar corrections to the scattering angle to NLO, from which we derived the binding energy for circular orbits. Ultimately, it is through the accurate reconstruction of finite-size effects that we will constrain the nature of compact objects, notably the one(s) recently observed in the so-called `mass-gap'~\cite{gap1,gap2}. Measuring tidal responses is especially relevant for (non-rotating) black holes, due to their vanishing Love numbers \cite{Binnington:2009bb,Damour:2009vw,Kol:2011vg} (see also \cite{altspin}), which offers a unique opportunity to search for new physics \cite{tune,gcollider1,gcollider2}.  Our results thus provide new ingredients for accurate waveform modeling including tidal corrections.\vskip 2pt Our derivation is also interesting with regards to the high-energy limit ($\gamma \to \infty$). Remarkably, there is a pattern between the electric and magnetic quadrupolar as well as octupolar corrections, notably for the impulse at fixed impact parameter. For instance, the difference at leading PM order is ${\cal O}(1/\gamma)$ in the quadrupole, as noted in \cite{binit}, and ${\cal O}(\gamma)$ for the octupole. This feature extends~to~all orders in the probe limit, whereas at ${\cal O}(\nu)$ the  electric/magnetic split picks an extra factor of $\gamma$ for each multipole; except for the~$\ash(\gamma)$. The mismatch in the impulse for the latter goes as $G^3\gamma^{-6(4)} \log \gamma$ for the quadrupole (octupole) coupling. We also find a softer behavior for the individual terms in comparison with the monopole, which instead scales as $G^3\gamma^2 \log\gamma$ in the high-energy limit \cite{3pmeft}. It~would be interesting to understand these features and whether they persist at higher~orders.\vskip 2pt

\vskip 2pt
{\it Acknowledgments}. R.A.P. is supported by the ERC-CoG ``Precision Gravity: From the LHC~to LISA"  provided by the European Research Council (ERC) under the European Union's H2020 research and innovation programme (grant No.\,817791). Z.L.~and R.A.P.~are supported by the DFG under Germany's Excellence Strategy `Quantum Universe' (No.\,390833306). G.K. is supported by the Knut and Alice Wallenberg Foundation (grant KAW 2018.0441), and in part by the US DoE under contract DE-AC02-76SF00515. We also acknowledge extensive use of the \texttt{xAct} computer algebra packages~\cite{martin2019xact}. \vskip 8pt
\begin{center} {\bf Supplemental Material}\end{center}
{\em Schwarzschild Background.} Consistency with the test-particle limit can be shown directly in terms of the (gauge-invariant) scattering angle. Using the on-shell condition in \eqref{on-shell} we solve for $p^{(1)}_r$ as a function of the distance,~the (reduced) energy, ${\cal E}^{(1)}_0$, angular momentum,~$J^{(1)}_0$,~and~tidal Love numbers of the test body (which we take as particle~1). We~construct the radial action, $\int p^{(1)}_r \dd r$, such that the scattering angle follows via differentiation w.r.t.~the angular momentum. Expanding in powers of $1/j_0 \equiv  (Gm_2m_1)/J^{(1)}_0$, replacing the energy by the boost factor, ${\cal E}^{(1)}_0 \to \gamma$, and following the integration procedure described in e.g.\,\cite{binit}, we arrive at
\begin{widetext}
\begin{align}
\label{chisch}
\Delta\chi^\text{Sch}_{(1)} &=  {45\pi  \over 64 j_0^6}\,(\gamma^2 {-} 1)^2
\Big( (35 \gamma^4-30 \gamma^2+11)\lambda^{(1)}_{E^2} + (35 \gamma^4-30 \gamma^2-5)\lambda^{(1)}_{B^2}\Big)  \\
&+ \frac{192}{35 j_0^7} (\gamma^2 {-} 1)^{3/2}
\Big( (160 \gamma^6-192 \gamma^4+72 \gamma^2-5)\lambda^{(1)}_{E^2} 
+ (160 \gamma^6-192 \gamma^4+30 \gamma^2+2)\lambda^{(1)}_{B^2}\Big)\nn\\
&+  {525\pi  \over  512 j_0^8} (\gamma^2 {-} 1)^3
\Big( (21 \gamma^6+385 \gamma^4-305 \gamma^2+91)\lambda^{(1)}_{\tilde{E}^2} 
+ (21 \gamma^6+385 \gamma^4-385 \gamma^2-21)\lambda^{(1)}_{\tilde{B}^2}\Big)\nn\\
&+ {63\pi \over  256 j_0^8}\,(\gamma^2 {-} 1)\, \Big( 
(9009 \gamma^8-15246 \gamma^6+8484 \gamma^4-1666 \gamma^2+59)\lambda^{(1)}_{E^2} 
+ 21 (\gamma^2 {-} 1) (429 \gamma^6-297 \gamma^4+27 \gamma^2+1)\lambda^{(1)}_{B^2}\Big)\nn\\
&+{256  \over  77 j_0^9} \sqrt{\gamma^2 {-} 1} \, \Big( 
(14336 \gamma^{10}-32256 \gamma^8+25792 \gamma^6-8720 \gamma^4+1104 \gamma^2-25)\lambda^{(1)}_{E^2}\nn \\
&\qquad\qquad\qquad\quad+ (14336 \gamma^{10}-32256 \gamma^8+23680 \gamma^6-6080 \gamma^4+312 \gamma^2+8)\lambda^{(1)}_{B^2}\Big)\nn \\
&+{512  \over  3003 j_0^9}\,(\gamma^2 {-} 1)^{5/2}
\Big( (4800 \gamma^8+77520 \gamma^6-74888 \gamma^4+17707 \gamma^2+1888)\lambda^{(1)}_{\tilde{E}^2} \nn
\\
&\qquad\qquad\qquad\qquad\quad + (4800 \gamma^8+77520 \gamma^6-87472 \gamma^4+5552 \gamma^2-400)\lambda^{(1)}_{\tilde{B}^2}\Big)+{\cal O}(1/j_0^{10})\nn\,,
\end{align}
\end{widetext}
for the corrections due to tidal effects in the deflection angle in a Schwarzschild background. The tidal parameters are given by $\lambda^{(1)}_{E^2(B^2)} \equiv G^{-4} m_2^{-5} (m_2/m_1) c_{E^2(B^2)}^{(1)}$ and $\lambda^{(1)}_{\tilde E^2(\tilde B^2)} \equiv G^{-6} m_2^{-7} (m_2/m_1) c_{\tilde E^2(\tilde B^2)}^{(1)}$. The~expression in \eqref{chisch}, which to our knowledge is presented here for the first time, must be symmetrized to obtain the mirror image. The result neatly agrees with the test-body limit of \eqref{dchi1} \& \eqref{dchi2}. Moreover, using \eqref{chisch} and inverting \eqref{chin2}-\eqref{chin3}, we can then solve for the tidal correction to the momentum coefficients $\bar P_n^{\rm Sch}$ in a Schwarzschild background, e.g.
\begin{widetext}
\beq
\begin{aligned}
\Delta \bar{P}_6^\text{Sch} \,=\,  &\frac{3}{4} \left(35 \gamma ^4-30 \gamma^2+11\right) \lambda^{(1)}_{E^2} +\frac{15}{4} \left(7 \gamma ^4-6 \gamma^2-1\right) \lambda^{(1)}_{B^2}\,,
\\[0.2 em]
\Delta \bar{P}_7^\text{Sch} \,=\, &\frac{3}{28} \left(110 \gamma^4+363 \gamma ^2-305\right) \lambda^{(1)}_{E^2} +\frac{33}{28} \left(10 \gamma ^4+33\gamma ^2+13\right) \lambda^{(1)}_{B^2}\,,
\\[0.2 em]
\Delta \bar{P}_8^\text{Sch} \,=\,& {9 \over 140} (544 \gamma ^4-933\gamma ^2+1139) \lambda^{(1)}_{E^2}
+ {9 \over 140} (544 \gamma ^4-933 \gamma^2-541) \lambda^{(1)}_{B^2}
\\
&+ {15 \over 16} (21\gamma ^6+385 \gamma ^4-305 \gamma ^2+91) \lambda^{(1)}_{\tilde{E}^2}
+ {105 \over 16} (3 \gamma ^6+55\gamma ^4-55 \gamma ^2-3) \lambda^{(1)}_{\tilde{B}^2}\,,
 \\[0.2 em]
\Delta \bar{P}_9^\text{Sch} \,=\, 
& {1 \over 880} \left(14302 \gamma ^4+59187 \gamma^2-107149\right) \lambda^{(1)}_{E^2}
+ {1 \over 880}\left(14302 \gamma ^4+59187 \gamma ^2+51251\right) \lambda^{(1)}_{B^2}
\\
&- {5 \over 6864} (71178 \gamma ^6+1857639 \gamma^4-2313484 \gamma ^2+940651)\,\lambda^{(1)}_{\tilde{E}^2}
\\
&- {5 \over 6864} (71178 \gamma ^6+1857639\gamma ^4-2949548 \gamma ^2-214789)\, \lambda^{(1)}_{\tilde{B}^2}\,,
\\[0.2 em]
\Delta \bar{P}_{10}^\text{Sch} \,=\,   
& {(42008 \gamma^4-108497 \gamma ^2+257471)  \over  1540}  \lambda^{(1)}_{E^2}
+ {(42008 \gamma ^4-108497 \gamma ^2-123679)  \over 1540} \lambda^{(1)}_{B^2}
\\
&
+ {(872265 \gamma ^6+37951761 \gamma^4-65175713 \gamma ^2+34559887)  \over 12012} \lambda^{(1)}_{\tilde{E}^2}
\\
&
+ {(872265 \gamma^6+37951761 \gamma ^4-83522041 \gamma ^2-7834465) \over 12012} \lambda^{(1)}_{\tilde{B^2}}\,.
\end{aligned}
\eeq
\end{widetext}
After adding the mirror images, the quadrupole contribution to $\bar P_8^{\rm Sch}(\gamma \to 1)$ exactly matches the static limit of the full $\Delta P_8$ in \eqref{p8st}  at ${\cal O}(\nu^0)$, as expected. Similarly, the $\Delta P_{10}^{\rm Sch}(\gamma \to 1)$ yields the ${\cal O}(\nu^0)$ part shown in \eqref{p10st}. Finally, notice that the probe limit also fixes the leading PM deflection for comparable masses. This is clear in impact parameter space, where the impulse remains the same and we only have to add the mirror image. Hence, replacing $p^{(1)}_\infty \to \mu\sqrt{\gamma^2-1}/{\Gamma}$ we obtain the two-body deflection angle at leading PM order. \vskip 8pt

 {\em Momentum \& Hamiltonian PM-coefficients.} The corrections to the center-of-mass momentum in \eqref{pinfpm} are obtained from the map in \cite{paper1}. Using the value of the NLO scattering angle due to tidal effects in \eqref{dchi1}~\&~\eqref{dchi2}, we find 
 \begin{widetext}
\begin{align}
 \label{p6}
\Delta P_6 = \frac{3 M^8 \nu^2}{4\Gamma} \Big(5 (7 &\gamma^4-6 \gamma^2-1) \lambda_{B^2}
+(35 \gamma^4-30 \gamma^2+11) \lambda_{E^2}\Big),
\\
    \Delta P_7 =
      \frac{3 M^9 \nu^2}{28 (\gamma^2 {-} 1) \Gamma }\bigg\{
        &16 \Big( (160 \gamma^6-192 \gamma^4+30 \gamma^2+2) \lambda_{B^2} + (160 \gamma^6-192 \gamma^4+72 \gamma^2-5) \lambda_{E^2}\Big) \\
        &-35 \Gamma\, (2 \gamma^2-1) \Big( (35 \gamma^4-30 \gamma^2-5) \lambda_{B^2} + (35 \gamma^4-30 \gamma^2+11) \lambda_{E^2}\Big)
        \nn\\
        +\frac{8\nu}{\gamma^2 {-} 1} \Big[
          &(224 \gamma^9-320 \gamma^8-728 \gamma^7+704 \gamma^6+5488 \gamma^5-444 \gamma^4+66262 \gamma^3+56 \gamma^2+28084 \gamma +4) \kappa_{B^2}\nn\\
          + &(224 \gamma^9-320 \gamma^8-728 \gamma^7+704 \gamma^6+5628 \gamma^5-528 \gamma^4+65982 \gamma^3+154 \gamma^2+28329 \gamma -10) \kappa_{E^2}\nonumber\\
          -&210 \Big( (440 \gamma^4+474 \gamma^2+32)\kappa_{B^2} + (440 \gamma^4+474 \gamma^2+33) \kappa_{E^2}\Big)\ash(\gamma)
          \Big]
        \bigg\}\nn\,,
\end{align}
for the quadrupole contributions, whereas for the octupolar corrections we have
\begin{align}\label{p8}
\Delta P_8 = \frac{15 M^{10} \nu^2}{16\Gamma} \Big( (21 \gamma&^6+ 385\gamma^4 - 305\gamma^2+91) \lambda_{\tilde E}^2
+ 7(3 \gamma^6+55 \gamma^4-55 \gamma^2-3) \lambda_{{\tilde B}^2}\Big),
\\
\label{p9}
\Delta P_9 =
\frac{5 M^{11} \nu^2}{6864 (\gamma^2 {-} 1)\, \Gamma}
\bigg\{
      &64 \Big[(4800 \gamma^8+77520 \gamma^6-74888 \gamma^4+17707 \gamma^2+1888)\lambda_{{\tilde E}^2} 
      \\[-0.35 em]
      &\quad +16 (300 \gamma^8+4845 \gamma^6-5467 \gamma^4+347 \gamma^2-25)\lambda_{{\tilde B}^2}\Big]\nonumber\\
      &-9009(2\gamma^2 - 1)\, \Gamma\, \Big[(21 \gamma^6+385 \gamma^4-305 \gamma^2+91) \lambda_{{\tilde E}^2}
        +7(3 \gamma^6+55 \gamma^4-55 \gamma^2-3) \lambda_{{\tilde B}^2}\Big] 
       \nonumber\\
+ {16\nu \over (\gamma^2 {-} 1)^2}\Big[
        \big(27&456 \gamma^{13}-19200 \gamma^{12}+205920 \gamma^{11}-271680 \gamma^{10}-1468896 \gamma^9+900512 \gamma^8+21724560 \gamma^7  \nonumber\\[-0.35 em]
     -&980012 \gamma^6+580453302 \gamma^5+433656 \gamma^4+837773079 \gamma^3-55724 \gamma^2+155291994 \gamma -7552\big) \kappa_{{\tilde E}^2} \nonumber\\
      +\big(27&456 \gamma^{13}-19200 \gamma^{12}+205920 \gamma^{11}-271680 \gamma^{10}-1589016 \gamma^9+950848 \gamma^8 +22048884 \gamma^7  \nonumber\\
      -&1032064 \gamma^6+579540390 \gamma^5+395904 \gamma^4+826613931 \gamma^3-25408 \gamma^2+148331040 \gamma +1600\big) \kappa_{{\tilde B}^2 } \nonumber\\
      +90090 &\big( (7292 \gamma^6+19644 \gamma^4+8141 \gamma^2+310)\kappa_{{\tilde E}^2}
      + (7292 \gamma^6+19484 \gamma^4+7905 \gamma^2+288) \kappa_{{\tilde B}^2}\big)\ash(\gamma)
        \Big]
        \bigg\}\nn\,.
\end{align}
From here we then read off the PM coefficient of the Hamiltonian \cite{paper1} shown in \eqref{c6}, with
\begin{align}\label{dij}
    D_{7,1} &= -8 \left(224 \gamma^9-320 \gamma^8-728 \gamma^7+704 \gamma^6+5628 \gamma^5-528 \gamma^4+65982 \gamma^3+154 \gamma^2+28329 \gamma -10\right)\,, \\
    D_{7,2} &= -8 \left(224 \gamma^9-320 \gamma^8-728 \gamma^7+704 \gamma^6+5488 \gamma^5-444 \gamma^4+66262 \gamma^3+56 \gamma^2+28084 \gamma +4\right)\,, \nonumber\\
    D_{7,3} &= 1680 \left(440 \gamma^4+474 \gamma^2+33\right),\quad
    D_{7,4} = 1680 \left(440 \gamma^4+474 \gamma^2+32\right)\,, \nonumber\\
    D_{7,5} &= -16 \left(160 \gamma^6-192 \gamma^4+72 \gamma^2-5\right),\quad
    D_{7,6} = -16 \left(160 \gamma^6-192 \gamma^4+30 \gamma^2+2\right)\,, \nonumber\\
    D_{7,7} &= 7\left(700 \gamma^8-1110 \gamma^6+597 \gamma^4-96 \gamma^2-11\right),\quad
    D_{7,8} = 35\left(140 \gamma^6-82 \gamma^4-\gamma^2-1\right)\,, \nonumber\\
    D_{7,9} &= 7\left(1750 \gamma^8-1050 \gamma^7-2785 \gamma^6+1655 \gamma^5+1436 \gamma^4-952 \gamma^3-181 \gamma^2+203 \gamma -44\right)\,, \nonumber\\
    D_{7,10} &= 35\left(350 \gamma^6-210 \gamma^5-207 \gamma^4+121 \gamma^3-3 \gamma^2+\gamma -4\right)\,, \nonumber\\
    D_{7,11} &= 14 \left(490 \gamma^7-280 \gamma^6-895 \gamma^5+380 \gamma^4+536 \gamma^3-208 \gamma^2-115 \gamma +44\right)\,, \nonumber\\
    D_{7,12} &= 70\left(98 \gamma^5-56 \gamma^4-81 \gamma^3+20 \gamma^2+7 \gamma +4\right)\,, \nonumber
\end{align}
and
\begin{align}\label{dij2}
    D_{9,1} &= -27456 \gamma^{13}+19200 \gamma^{12}-205920 \gamma^{11}+271680 \gamma^{10}+1468896 \gamma^9-900512 \gamma^8-21724560 \gamma^7\\
    &\quad+980012 \gamma^6-580453302 \gamma^5-433656 \gamma^4-837773079 \gamma^3+55724 \gamma^2-155291994 \gamma +7552\,, \nonumber\\
    D_{9,2} &= -27456 \gamma^{13}+19200 \gamma^{12}-205920 \gamma^{11}+271680 \gamma^{10}+1589016 \gamma^9-950848 \gamma^8-22048884 \gamma^7 \nonumber\\
    &\quad+1032064 \gamma^6-579540390 \gamma^5-395904 \gamma^4-826613931 \gamma^3+25408 \gamma^2-148331040 \gamma -1600\,, \nonumber\\
    D_{9,3} &= 90090 \left(7292 \gamma^6+19644 \gamma^4+8141 \gamma^2+310\right), \quad
    D_{9,4}= 90090 \left(7292 \gamma^6+19484 \gamma^4+7905 \gamma^2+288\right),  \nonumber\\
    D_{9,5} &= -4 \left(4800 \gamma^8+77520 \gamma^6-74888 \gamma^4+17707 \gamma^2+1888\right)\,,   \nonumber\\
    D_{9,6} &= -4 \left(4800 \gamma^8+77520 \gamma^6-87472 \gamma^4+5552 \gamma^2-400\right)\,,  \nonumber\\
    D_{9,7} &= 588 \gamma^{10}+8694 \gamma^8-13695 \gamma^6+6881 \gamma^4-1033 \gamma^2-91,\,\,
    D_{9,8} = 588 \gamma^8+9282 \gamma^6-6013 \gamma^4+196 \gamma^2-21\,,  \nonumber\\
    D_{9,9} &= 1470 \gamma^{10}-882 \gamma^9+20881 \gamma^8-13895 \gamma^7-33371 \gamma^6+21409 \gamma^5+16239 \gamma^4-11285 \gamma^3-1975 \gamma^2+2157 \gamma -364\,,  \nonumber\\
    D_{9,10} &= 7 \left(210 \gamma^8-126 \gamma^7+3193 \gamma^6-2111 \gamma^5-2100 \gamma^4+1336 \gamma^3+53 \gamma^2-59 \gamma -12\right)\,,  \nonumber\\
    D_{9,11} &= 2 \left(420 \gamma^9-210 \gamma^8+5656 \gamma^7-3745 \gamma^6-10464 \gamma^5+4975 \gamma^4+5928 \gamma^3-2435 \gamma^2-1156 \gamma +455\right)\,, \nonumber\\
    D_{9,12} &= 14 \left(60 \gamma^7-30 \gamma^6+868 \gamma^5-565 \gamma^4-764 \gamma^3+260 \gamma^2+92 \gamma +15\right). \nonumber
\end{align}

\end{widetext}

\bibliography{ref3PM}

\begin{thebibliography}{77}%
\makeatletter
\providecommand \@ifxundefined [1]{%
 \@ifx{#1\undefined}
}%
\providecommand \@ifnum [1]{%
 \ifnum #1\expandafter \@firstoftwo
 \else \expandafter \@secondoftwo
 \fi
}%
\providecommand \@ifx [1]{%
 \ifx #1\expandafter \@firstoftwo
 \else \expandafter \@secondoftwo
 \fi
}%
\providecommand \natexlab [1]{#1}%
\providecommand \enquote  [1]{``#1''}%
\providecommand \bibnamefont  [1]{#1}%
\providecommand \bibfnamefont [1]{#1}%
\providecommand \citenamefont [1]{#1}%
\providecommand \href@noop [0]{\@secondoftwo}%
\providecommand \href [0]{\begingroup \@sanitize@url \@href}%
\providecommand \@href[1]{\@@startlink{#1}\@@href}%
\providecommand \@@href[1]{\endgroup#1\@@endlink}%
\providecommand \@sanitize@url [0]{\catcode `\\12\catcode `\$12\catcode
  `\&12\catcode `\#12\catcode `\^12\catcode `\_12\catcode `\%12\relax}%
\providecommand \@@startlink[1]{}%
\providecommand \@@endlink[0]{}%
\providecommand \url  [0]{\begingroup\@sanitize@url \@url }%
\providecommand \@url [1]{\endgroup\@href {#1}{\urlprefix }}%
\providecommand \urlprefix  [0]{URL }%
\providecommand \Eprint [0]{\href }%
\providecommand \doibase [0]{http://dx.doi.org/}%
\providecommand \selectlanguage [0]{\@gobble}%
\providecommand \bibinfo  [0]{\@secondoftwo}%
\providecommand \bibfield  [0]{\@secondoftwo}%
\providecommand \translation [1]{[#1]}%
\providecommand \BibitemOpen [0]{}%
\providecommand \bibitemStop [0]{}%
\providecommand \bibitemNoStop [0]{.\EOS\space}%
\providecommand \EOS [0]{\spacefactor3000\relax}%
\providecommand \BibitemShut  [1]{\csname bibitem#1\endcsname}%
\let\auto@bib@innerbib\@empty
\bibitem [{\citenamefont {Abbott}\ \emph
  {et~al.}(2019{\natexlab{a}})\citenamefont {Abbott} \emph
  {et~al.}}]{LIGOScientific:2018mvr}%
  \BibitemOpen
  \bibfield  {author} {\bibinfo {author} {\bibfnamefont {B.~P.}\ \bibnamefont
  {Abbott}} \emph {et~al.} (\bibinfo {collaboration} {LIGO Scientific,
  Virgo}),\ }\href {\doibase 10.1103/PhysRevX.9.031040} {\bibfield  {journal}
  {\bibinfo  {journal} {Phys. Rev.}\ }\textbf {\bibinfo {volume} {X9}},\
  \bibinfo {pages} {031040} (\bibinfo {year} {2019}{\natexlab{a}})},\ \Eprint
  {http://arxiv.org/abs/1811.12907} {arXiv:1811.12907} \BibitemShut {NoStop}%
\bibitem [{\citenamefont {Abbott}\ \emph
  {et~al.}(2019{\natexlab{b}})\citenamefont {Abbott} \emph {et~al.}}]{LIGO}%
  \BibitemOpen
  \bibfield  {author} {\bibinfo {author} {\bibfnamefont {R.}~\bibnamefont
  {Abbott}} \emph {et~al.} (\bibinfo {collaboration} {LIGO Scientific,
  Virgo}),\ }\href@noop {} {\  (\bibinfo {year} {2019}{\natexlab{b}})},\
  \Eprint {http://arxiv.org/abs/1912.11716} {arXiv:1912.11716} \BibitemShut
  {NoStop}%
\bibitem [{\citenamefont {Abbott}\ \emph {et~al.}(2017)\citenamefont {Abbott}
  \emph {et~al.}}]{ns}%
  \BibitemOpen
  \bibfield  {author} {\bibinfo {author} {\bibfnamefont {B.~P.}\ \bibnamefont
  {Abbott}} \emph {et~al.} (\bibinfo {collaboration} {LIGO Scientific,
  Virgo}),\ }\href {\doibase 10.1103/PhysRevLett.119.161101} {\bibfield
  {journal} {\bibinfo  {journal} {Phys. Rev. Lett.}\ }\textbf {\bibinfo
  {volume} {119}},\ \bibinfo {pages} {161101} (\bibinfo {year} {2017})},\
  \Eprint {http://arxiv.org/abs/1710.05832} {arXiv:1710.05832} \BibitemShut
  {NoStop}%
\bibitem [{\citenamefont {Buonanno}\ and\ \citenamefont
  {Sathyaprakash}(2014)}]{buosathya}%
  \BibitemOpen
  \bibfield  {author} {\bibinfo {author} {\bibfnamefont {A.}~\bibnamefont
  {Buonanno}}\ and\ \bibinfo {author} {\bibfnamefont {B.}~\bibnamefont
  {Sathyaprakash}},\ }\href@noop {} {\  (\bibinfo {year} {2014})},\ \Eprint
  {http://arxiv.org/abs/1410.7832} {arXiv:1410.7832} \BibitemShut {NoStop}%
\bibitem [{\citenamefont {Porto}(2016{\natexlab{a}})}]{tune}%
  \BibitemOpen
  \bibfield  {author} {\bibinfo {author} {\bibfnamefont {R.~A.}\ \bibnamefont
  {Porto}},\ }\href {\doibase 10.1002/prop.201600064} {\bibfield  {journal}
  {\bibinfo  {journal} {Fortsch. Phys.}\ }\textbf {\bibinfo {volume} {64}},\
  \bibinfo {pages} {723} (\bibinfo {year} {2016}{\natexlab{a}})},\ \Eprint
  {http://arxiv.org/abs/1606.08895} {arXiv:1606.08895} \BibitemShut {NoStop}%
\bibitem [{\citenamefont {Porto}(2017)}]{music}%
  \BibitemOpen
  \bibfield  {author} {\bibinfo {author} {\bibfnamefont {R.~A.}\ \bibnamefont
  {Porto}},\ }\href@noop {} {\  (\bibinfo {year} {2017})},\ \Eprint
  {http://arxiv.org/abs/1703.06440} {arXiv:1703.06440 [physics.pop-ph]}
  \BibitemShut {NoStop}%
\bibitem [{\citenamefont {Flanagan}\ and\ \citenamefont
  {Hinderer}(2008)}]{tanja1}%
  \BibitemOpen
  \bibfield  {author} {\bibinfo {author} {\bibfnamefont {E.~E.}\ \bibnamefont
  {Flanagan}}\ and\ \bibinfo {author} {\bibfnamefont {T.}~\bibnamefont
  {Hinderer}},\ }\href {\doibase 10.1103/PhysRevD.77.021502} {\bibfield
  {journal} {\bibinfo  {journal} {Phys. Rev. D}\ }\textbf {\bibinfo {volume}
  {77}},\ \bibinfo {pages} {021502} (\bibinfo {year} {2008})},\ \Eprint
  {http://arxiv.org/abs/0709.1915} {arXiv:0709.1915} \BibitemShut {NoStop}%
\bibitem [{\citenamefont {Dietrich}\ \emph {et~al.}(2020)\citenamefont
  {Dietrich}, \citenamefont {Hinderer},\ and\ \citenamefont
  {Samajdar}}]{tanja2}%
  \BibitemOpen
  \bibfield  {author} {\bibinfo {author} {\bibfnamefont {T.}~\bibnamefont
  {Dietrich}}, \bibinfo {author} {\bibfnamefont {T.}~\bibnamefont {Hinderer}},
  \ and\ \bibinfo {author} {\bibfnamefont {A.}~\bibnamefont {Samajdar}},\
  }\href@noop {} {\  (\bibinfo {year} {2020})},\ \Eprint
  {http://arxiv.org/abs/2004.02527} {arXiv:2004.02527} \BibitemShut {NoStop}%
\bibitem [{\citenamefont {Abbott}\ \emph {et~al.}(2018)\citenamefont {Abbott}
  \emph {et~al.}}]{Abbott:2018exr}%
  \BibitemOpen
  \bibfield  {author} {\bibinfo {author} {\bibfnamefont {B.}~\bibnamefont
  {Abbott}} \emph {et~al.} (\bibinfo {collaboration} {LIGO Scientific,
  Virgo}),\ }\href {\doibase 10.1103/PhysRevLett.121.161101} {\bibfield
  {journal} {\bibinfo  {journal} {Phys. Rev. Lett.}\ }\textbf {\bibinfo
  {volume} {121}},\ \bibinfo {pages} {161101} (\bibinfo {year} {2018})},\
  \Eprint {http://arxiv.org/abs/1805.11581} {arXiv:1805.11581} \BibitemShut
  {NoStop}%
\bibitem [{\citenamefont {Arvanitaki}\ \emph {et~al.}(2017)\citenamefont
  {Arvanitaki} \emph {et~al.}}]{axiverse}%
  \BibitemOpen
  \bibfield  {author} {\bibinfo {author} {\bibfnamefont {A.}~\bibnamefont
  {Arvanitaki}} \emph {et~al.},\ }\href {\doibase 10.1103/PhysRevD.95.043001}
  {\bibfield  {journal} {\bibinfo  {journal} {Phys. Rev. D}\ }\textbf {\bibinfo
  {volume} {95}},\ \bibinfo {pages} {043001} (\bibinfo {year} {2017})},\
  \Eprint {http://arxiv.org/abs/1604.03958} {arXiv:1604.03958} \BibitemShut
  {NoStop}%
\bibitem [{\citenamefont {Baumann}\ \emph {et~al.}(2019)\citenamefont
  {Baumann}, \citenamefont {Chia},\ and\ \citenamefont {Porto}}]{gcollider1}%
  \BibitemOpen
  \bibfield  {author} {\bibinfo {author} {\bibfnamefont {D.}~\bibnamefont
  {Baumann}}, \bibinfo {author} {\bibfnamefont {H.~S.}\ \bibnamefont {Chia}}, \
  and\ \bibinfo {author} {\bibfnamefont {R.~A.}\ \bibnamefont {Porto}},\ }\href
  {\doibase 10.1103/PhysRevD.99.044001} {\bibfield  {journal} {\bibinfo
  {journal} {Phys. Rev. D}\ }\textbf {\bibinfo {volume} {99}},\ \bibinfo
  {pages} {044001} (\bibinfo {year} {2019})},\ \Eprint
  {http://arxiv.org/abs/1804.03208} {arXiv:1804.03208} \BibitemShut {NoStop}%
\bibitem [{\citenamefont {Baumann}\ \emph {et~al.}(2020)\citenamefont
  {Baumann}, \citenamefont {Chia}, \citenamefont {Porto},\ and\ \citenamefont
  {Stout}}]{gcollider2}%
  \BibitemOpen
  \bibfield  {author} {\bibinfo {author} {\bibfnamefont {D.}~\bibnamefont
  {Baumann}}, \bibinfo {author} {\bibfnamefont {H.~S.}\ \bibnamefont {Chia}},
  \bibinfo {author} {\bibfnamefont {R.~A.}\ \bibnamefont {Porto}}, \ and\
  \bibinfo {author} {\bibfnamefont {J.}~\bibnamefont {Stout}},\ }\href
  {\doibase 10.1103/PhysRevD.101.083019} {\bibfield  {journal} {\bibinfo
  {journal} {Phys. Rev. D}\ }\textbf {\bibinfo {volume} {101}},\ \bibinfo
  {pages} {083019} (\bibinfo {year} {2020})},\ \Eprint
  {http://arxiv.org/abs/1912.04932} {arXiv:1912.04932} \BibitemShut {NoStop}%
\bibitem [{\citenamefont {Goldberger}\ and\ \citenamefont
  {Rothstein}(2006{\natexlab{a}})}]{nrgr}%
  \BibitemOpen
  \bibfield  {author} {\bibinfo {author} {\bibfnamefont {W.~D.}\ \bibnamefont
  {Goldberger}}\ and\ \bibinfo {author} {\bibfnamefont {I.~Z.}\ \bibnamefont
  {Rothstein}},\ }\href {\doibase 10.1103/PhysRevD.73.104029} {\bibfield
  {journal} {\bibinfo  {journal} {Phys. Rev.}\ }\textbf {\bibinfo {volume}
  {D73}},\ \bibinfo {pages} {104029} (\bibinfo {year} {2006}{\natexlab{a}})},\
  \Eprint {http://arxiv.org/abs/hep-th/0409156} {arXiv:hep-th/0409156}
  \BibitemShut {NoStop}%
\bibitem [{\citenamefont {Goldberger}\ and\ \citenamefont
  {Rothstein}(2006{\natexlab{b}})}]{dis1}%
  \BibitemOpen
  \bibfield  {author} {\bibinfo {author} {\bibfnamefont {W.}~\bibnamefont
  {Goldberger}}\ and\ \bibinfo {author} {\bibfnamefont {I.}~\bibnamefont
  {Rothstein}},\ }\href {\doibase 10.1103/PhysRevD.73.104030} {\bibfield
  {journal} {\bibinfo  {journal} {Phys. Rev. D}\ }\textbf {\bibinfo {volume}
  {73}},\ \bibinfo {pages} {104030} (\bibinfo {year} {2006}{\natexlab{b}})},\
  \Eprint {http://arxiv.org/abs/hep-th/0511133} {arXiv:hep-th/0511133}
  \BibitemShut {NoStop}%
\bibitem [{\citenamefont {Gilmore}\ and\ \citenamefont {Ross}(2008)}]{nrgr2pn}%
  \BibitemOpen
  \bibfield  {author} {\bibinfo {author} {\bibfnamefont {J.~B.}\ \bibnamefont
  {Gilmore}}\ and\ \bibinfo {author} {\bibfnamefont {A.}~\bibnamefont {Ross}},\
  }\href {\doibase 10.1103/PhysRevD.78.124021} {\bibfield  {journal} {\bibinfo
  {journal} {Phys. Rev. D}\ }\textbf {\bibinfo {volume} {78}},\ \bibinfo
  {pages} {124021} (\bibinfo {year} {2008})},\ \Eprint
  {http://arxiv.org/abs/0810.1328} {arXiv:0810.1328} \BibitemShut {NoStop}%
\bibitem [{\citenamefont {Foffa}\ and\ \citenamefont
  {Sturani}(2011)}]{nrgr3pn}%
  \BibitemOpen
  \bibfield  {author} {\bibinfo {author} {\bibfnamefont {S.}~\bibnamefont
  {Foffa}}\ and\ \bibinfo {author} {\bibfnamefont {R.}~\bibnamefont
  {Sturani}},\ }\href {\doibase 10.1103/PhysRevD.84.044031} {\bibfield
  {journal} {\bibinfo  {journal} {Phys. Rev. D}\ }\textbf {\bibinfo {volume}
  {84}},\ \bibinfo {pages} {044031} (\bibinfo {year} {2011})},\ \Eprint
  {http://arxiv.org/abs/1104.1122} {arXiv:1104.1122} \BibitemShut {NoStop}%
\bibitem [{\citenamefont {Foffa}\ and\ \citenamefont
  {Sturani}(2013)}]{Foffa:2012rn}%
  \BibitemOpen
  \bibfield  {author} {\bibinfo {author} {\bibfnamefont {S.}~\bibnamefont
  {Foffa}}\ and\ \bibinfo {author} {\bibfnamefont {R.}~\bibnamefont
  {Sturani}},\ }\href {\doibase 10.1103/PhysRevD.87.064011} {\bibfield
  {journal} {\bibinfo  {journal} {Phys. Rev. D}\ }\textbf {\bibinfo {volume}
  {87}},\ \bibinfo {pages} {064011} (\bibinfo {year} {2013})},\ \Eprint
  {http://arxiv.org/abs/1206.7087} {arXiv:1206.7087} \BibitemShut {NoStop}%
\bibitem [{\citenamefont {Galley}\ \emph {et~al.}(2016)\citenamefont {Galley},
  \citenamefont {Leibovich}, \citenamefont {Porto},\ and\ \citenamefont
  {Ross}}]{tail}%
  \BibitemOpen
  \bibfield  {author} {\bibinfo {author} {\bibfnamefont {C.}~\bibnamefont
  {Galley}}, \bibinfo {author} {\bibfnamefont {A.}~\bibnamefont {Leibovich}},
  \bibinfo {author} {\bibfnamefont {R.~A.}\ \bibnamefont {Porto}}, \ and\
  \bibinfo {author} {\bibfnamefont {A.}~\bibnamefont {Ross}},\ }\href {\doibase
  10.1103/PhysRevD.93.124010} {\bibfield  {journal} {\bibinfo  {journal} {Phys.
  Rev. D}\ }\textbf {\bibinfo {volume} {93}},\ \bibinfo {pages} {124010}
  (\bibinfo {year} {2016})},\ \Eprint {http://arxiv.org/abs/1511.07379}
  {arXiv:1511.07379} \BibitemShut {NoStop}%
\bibitem [{\citenamefont {Foffa}\ \emph {et~al.}(2017)\citenamefont {Foffa},
  \citenamefont {Mastrolia}, \citenamefont {Sturani},\ and\ \citenamefont
  {Sturm}}]{nrgrG5}%
  \BibitemOpen
  \bibfield  {author} {\bibinfo {author} {\bibfnamefont {S.}~\bibnamefont
  {Foffa}}, \bibinfo {author} {\bibfnamefont {P.}~\bibnamefont {Mastrolia}},
  \bibinfo {author} {\bibfnamefont {R.}~\bibnamefont {Sturani}}, \ and\
  \bibinfo {author} {\bibfnamefont {C.}~\bibnamefont {Sturm}},\ }\href
  {\doibase 10.1103/PhysRevD.95.104009} {\bibfield  {journal} {\bibinfo
  {journal} {Phys. Rev. D}\ }\textbf {\bibinfo {volume} {95}},\ \bibinfo
  {pages} {104009} (\bibinfo {year} {2017})},\ \Eprint
  {http://arxiv.org/abs/1612.00482} {arXiv:1612.00482} \BibitemShut {NoStop}%
\bibitem [{\citenamefont {Porto}\ and\ \citenamefont
  {Rothstein}(2017)}]{apparent}%
  \BibitemOpen
  \bibfield  {author} {\bibinfo {author} {\bibfnamefont {R.~A.}\ \bibnamefont
  {Porto}}\ and\ \bibinfo {author} {\bibfnamefont {I.}~\bibnamefont
  {Rothstein}},\ }\href {\doibase 10.1103/PhysRevD.96.024062} {\bibfield
  {journal} {\bibinfo  {journal} {Phys. Rev. D}\ }\textbf {\bibinfo {volume}
  {96}},\ \bibinfo {pages} {024062} (\bibinfo {year} {2017})},\ \Eprint
  {http://arxiv.org/abs/1703.06433} {arXiv:1703.06433} \BibitemShut {NoStop}%
\bibitem [{\citenamefont {Foffa}\ and\ \citenamefont
  {Sturani}(2019)}]{nrgr4pn1}%
  \BibitemOpen
  \bibfield  {author} {\bibinfo {author} {\bibfnamefont {S.}~\bibnamefont
  {Foffa}}\ and\ \bibinfo {author} {\bibfnamefont {R.}~\bibnamefont
  {Sturani}},\ }\href {\doibase 10.1103/PhysRevD.100.024047} {\bibfield
  {journal} {\bibinfo  {journal} {Phys. Rev. D}\ }\textbf {\bibinfo {volume}
  {100}},\ \bibinfo {pages} {024047} (\bibinfo {year} {2019})},\ \Eprint
  {http://arxiv.org/abs/1903.05113} {arXiv:1903.05113} \BibitemShut {NoStop}%
\bibitem [{\citenamefont {Foffa}\ \emph
  {et~al.}(2019{\natexlab{a}})\citenamefont {Foffa}, \citenamefont {Porto},
  \citenamefont {Rothstein},\ and\ \citenamefont {Sturani}}]{nrgr4pn2}%
  \BibitemOpen
  \bibfield  {author} {\bibinfo {author} {\bibfnamefont {S.}~\bibnamefont
  {Foffa}}, \bibinfo {author} {\bibfnamefont {R.~A.}\ \bibnamefont {Porto}},
  \bibinfo {author} {\bibfnamefont {I.}~\bibnamefont {Rothstein}}, \ and\
  \bibinfo {author} {\bibfnamefont {R.}~\bibnamefont {Sturani}},\ }\href
  {\doibase 10.1103/PhysRevD.100.024048} {\bibfield  {journal} {\bibinfo
  {journal} {Phys. Rev.}\ }\textbf {\bibinfo {volume} {D100}},\ \bibinfo
  {pages} {024048} (\bibinfo {year} {2019}{\natexlab{a}})},\ \Eprint
  {http://arxiv.org/abs/1903.05118} {arXiv:1903.05118} \BibitemShut {NoStop}%
\bibitem [{\citenamefont {Foffa}\ \emph
  {et~al.}(2019{\natexlab{b}})\citenamefont {Foffa}, \citenamefont {Mastrolia},
  \citenamefont {Sturani}, \citenamefont {Sturm},\ and\ \citenamefont
  {Torres~Bobadilla}}]{5pn1}%
  \BibitemOpen
  \bibfield  {author} {\bibinfo {author} {\bibfnamefont {S.}~\bibnamefont
  {Foffa}}, \bibinfo {author} {\bibfnamefont {P.}~\bibnamefont {Mastrolia}},
  \bibinfo {author} {\bibfnamefont {R.}~\bibnamefont {Sturani}}, \bibinfo
  {author} {\bibfnamefont {C.}~\bibnamefont {Sturm}}, \ and\ \bibinfo {author}
  {\bibfnamefont {W.~J.}\ \bibnamefont {Torres~Bobadilla}},\ }\href {\doibase
  10.1103/PhysRevLett.122.241605} {\bibfield  {journal} {\bibinfo  {journal}
  {Phys. Rev. Lett.}\ }\textbf {\bibinfo {volume} {122}},\ \bibinfo {pages}
  {241605} (\bibinfo {year} {2019}{\natexlab{b}})},\ \Eprint
  {http://arxiv.org/abs/1902.10571} {arXiv:1902.10571} \BibitemShut {NoStop}%
\bibitem [{\citenamefont {Bl{\"u}mlein}\ \emph
  {et~al.}(2020{\natexlab{a}})\citenamefont {Bl{\"u}mlein}, \citenamefont
  {Maier},\ and\ \citenamefont {Marquard}}]{5pn2}%
  \BibitemOpen
  \bibfield  {author} {\bibinfo {author} {\bibfnamefont {J.}~\bibnamefont
  {Bl{\"u}mlein}}, \bibinfo {author} {\bibfnamefont {A.}~\bibnamefont {Maier}},
  \ and\ \bibinfo {author} {\bibfnamefont {P.}~\bibnamefont {Marquard}},\
  }\href@noop {} {\bibfield  {journal} {\bibinfo  {journal} {Phys. Lett. B}\
  }\textbf {\bibinfo {volume} {800}},\ \bibinfo {pages} {135100} (\bibinfo
  {year} {2020}{\natexlab{a}})},\ \Eprint {http://arxiv.org/abs/1902.11180}
  {arXiv:1902.11180} \BibitemShut {NoStop}%
\bibitem [{\citenamefont {Bl{\"u}mlein}\ \emph
  {et~al.}(2020{\natexlab{b}})\citenamefont {Bl{\"u}mlein}, \citenamefont
  {Maier}, \citenamefont {Marquard},\ and\ \citenamefont {Sch{\"a}fer}}]{blum}%
  \BibitemOpen
  \bibfield  {author} {\bibinfo {author} {\bibfnamefont {J.}~\bibnamefont
  {Bl{\"u}mlein}}, \bibinfo {author} {\bibfnamefont {A.}~\bibnamefont {Maier}},
  \bibinfo {author} {\bibfnamefont {P.}~\bibnamefont {Marquard}}, \ and\
  \bibinfo {author} {\bibfnamefont {G.}~\bibnamefont {Sch{\"a}fer}},\ }\href
  {\doibase 10.1016/j.physletb.2020.135496} {\bibfield  {journal} {\bibinfo
  {journal} {Phys. Lett. B}\ }\textbf {\bibinfo {volume} {807}},\ \bibinfo
  {pages} {135496} (\bibinfo {year} {2020}{\natexlab{b}})},\ \Eprint
  {http://arxiv.org/abs/2003.07145} {arXiv:2003.07145} \BibitemShut {NoStop}%
\bibitem [{\citenamefont {Goldberger}\ and\ \citenamefont
  {Ross}(2010)}]{andirad}%
  \BibitemOpen
  \bibfield  {author} {\bibinfo {author} {\bibfnamefont {W.~D.}\ \bibnamefont
  {Goldberger}}\ and\ \bibinfo {author} {\bibfnamefont {A.}~\bibnamefont
  {Ross}},\ }\href {\doibase 10.1103/PhysRevD.81.124015} {\bibfield  {journal}
  {\bibinfo  {journal} {Phys. Rev.}\ }\textbf {\bibinfo {volume} {D81}},\
  \bibinfo {pages} {124015} (\bibinfo {year} {2010})},\ \Eprint
  {http://arxiv.org/abs/0912.4254} {arXiv:0912.4254} \BibitemShut {NoStop}%
\bibitem [{\citenamefont {Galley}\ and\ \citenamefont
  {Leibovich}(2012)}]{adamchad1}%
  \BibitemOpen
  \bibfield  {author} {\bibinfo {author} {\bibfnamefont {C.~R.}\ \bibnamefont
  {Galley}}\ and\ \bibinfo {author} {\bibfnamefont {A.~K.}\ \bibnamefont
  {Leibovich}},\ }\href {\doibase 10.1103/PhysRevD.86.044029} {\bibfield
  {journal} {\bibinfo  {journal} {Phys. Rev. D}\ }\textbf {\bibinfo {volume}
  {86}},\ \bibinfo {pages} {044029} (\bibinfo {year} {2012})},\ \Eprint
  {http://arxiv.org/abs/1205.3842} {arXiv:1205.3842} \BibitemShut {NoStop}%
\bibitem [{\citenamefont {Leibovich}\ \emph {et~al.}(2020)\citenamefont
  {Leibovich}, \citenamefont {Maia}, \citenamefont {Rothstein},\ and\
  \citenamefont {Yang}}]{radnrgr}%
  \BibitemOpen
  \bibfield  {author} {\bibinfo {author} {\bibfnamefont {A.~K.}\ \bibnamefont
  {Leibovich}}, \bibinfo {author} {\bibfnamefont {N.~T.}\ \bibnamefont {Maia}},
  \bibinfo {author} {\bibfnamefont {I.~Z.}\ \bibnamefont {Rothstein}}, \ and\
  \bibinfo {author} {\bibfnamefont {Z.}~\bibnamefont {Yang}},\ }\href {\doibase
  10.1103/PhysRevD.101.084058} {\bibfield  {journal} {\bibinfo  {journal}
  {Phys. Rev. D}\ }\textbf {\bibinfo {volume} {101}},\ \bibinfo {pages}
  {084058} (\bibinfo {year} {2020})},\ \Eprint
  {http://arxiv.org/abs/1912.12546} {arXiv:1912.12546} \BibitemShut {NoStop}%
\bibitem [{\citenamefont {Porto}(2006)}]{nrgrs}%
  \BibitemOpen
  \bibfield  {author} {\bibinfo {author} {\bibfnamefont {R.~A.}\ \bibnamefont
  {Porto}},\ }\href {\doibase 10.1103/PhysRevD.73.104031} {\bibfield  {journal}
  {\bibinfo  {journal} {Phys. Rev. D}\ }\textbf {\bibinfo {volume} {73}},\
  \bibinfo {pages} {104031} (\bibinfo {year} {2006})},\ \Eprint
  {http://arxiv.org/abs/gr-qc/0511061} {arXiv:gr-qc/0511061} \BibitemShut
  {NoStop}%
\bibitem [{\citenamefont {Porto}\ and\ \citenamefont {Rothstein}(2006)}]{prl}%
  \BibitemOpen
  \bibfield  {author} {\bibinfo {author} {\bibfnamefont {R.~A.}\ \bibnamefont
  {Porto}}\ and\ \bibinfo {author} {\bibfnamefont {I.}~\bibnamefont
  {Rothstein}},\ }\href {\doibase 10.1103/PhysRevLett.97.021101} {\bibfield
  {journal} {\bibinfo  {journal} {Phys. Rev. Lett.}\ }\textbf {\bibinfo
  {volume} {97}},\ \bibinfo {pages} {021101} (\bibinfo {year} {2006})},\
  \Eprint {http://arxiv.org/abs/gr-qc/0604099} {arXiv:gr-qc/0604099}
  \BibitemShut {NoStop}%
\bibitem [{\citenamefont {Porto}\ and\ \citenamefont
  {Rothstein}(2007)}]{Porto:2007tt}%
  \BibitemOpen
  \bibfield  {author} {\bibinfo {author} {\bibfnamefont {R.~A.}\ \bibnamefont
  {Porto}}\ and\ \bibinfo {author} {\bibfnamefont {I.~Z.}\ \bibnamefont
  {Rothstein}},\ }\href@noop {} {\  (\bibinfo {year} {2007})},\ \Eprint
  {http://arxiv.org/abs/0712.2032} {arXiv:0712.2032} \BibitemShut {NoStop}%
\bibitem [{\citenamefont {Porto}(2008)}]{dis2}%
  \BibitemOpen
  \bibfield  {author} {\bibinfo {author} {\bibfnamefont {R.~A.}\ \bibnamefont
  {Porto}},\ }\href {\doibase 10.1103/PhysRevD.77.064026} {\bibfield  {journal}
  {\bibinfo  {journal} {Phys. Rev. D}\ }\textbf {\bibinfo {volume} {77}},\
  \bibinfo {pages} {064026} (\bibinfo {year} {2008})},\ \Eprint
  {http://arxiv.org/abs/0710.5150} {arXiv:0710.5150} \BibitemShut {NoStop}%
\bibitem [{\citenamefont {Porto}\ and\ \citenamefont
  {Rothstein}(2008{\natexlab{a}})}]{nrgrss}%
  \BibitemOpen
  \bibfield  {author} {\bibinfo {author} {\bibfnamefont {R.~A.}\ \bibnamefont
  {Porto}}\ and\ \bibinfo {author} {\bibfnamefont {I.~Z.}\ \bibnamefont
  {Rothstein}},\ }\href {\doibase 10.1103/PhysRevD.78.044012} {\bibfield
  {journal} {\bibinfo  {journal} {Phys.Rev.}\ }\textbf {\bibinfo {volume}
  {D78}},\ \bibinfo {pages} {044012} (\bibinfo {year} {2008}{\natexlab{a}})},\
  \Eprint {http://arxiv.org/abs/0802.0720} {arXiv:0802.0720} \BibitemShut
  {NoStop}%
\bibitem [{\citenamefont {Porto}\ and\ \citenamefont
  {Rothstein}(2008{\natexlab{b}})}]{nrgrs2}%
  \BibitemOpen
  \bibfield  {author} {\bibinfo {author} {\bibfnamefont {R.~A.}\ \bibnamefont
  {Porto}}\ and\ \bibinfo {author} {\bibfnamefont {I.~Z.}\ \bibnamefont
  {Rothstein}},\ }\href {\doibase 10.1103/PhysRevD.78.044013} {\bibfield
  {journal} {\bibinfo  {journal} {Phys.Rev.}\ }\textbf {\bibinfo {volume}
  {D78}},\ \bibinfo {pages} {044013} (\bibinfo {year} {2008}{\natexlab{b}})},\
  \Eprint {http://arxiv.org/abs/0804.0260} {arXiv:0804.0260} \BibitemShut
  {NoStop}%
\bibitem [{\citenamefont {Porto}(2010)}]{nrgrso}%
  \BibitemOpen
  \bibfield  {author} {\bibinfo {author} {\bibfnamefont {R.~A.}\ \bibnamefont
  {Porto}},\ }\href {\doibase 10.1088/0264-9381/27/20/205001} {\bibfield
  {journal} {\bibinfo  {journal} {Class. Quant. Grav.}\ }\textbf {\bibinfo
  {volume} {27}},\ \bibinfo {pages} {205001} (\bibinfo {year} {2010})},\
  \Eprint {http://arxiv.org/abs/1005.5730} {arXiv:1005.5730} \BibitemShut
  {NoStop}%
\bibitem [{\citenamefont {Porto}\ \emph {et~al.}(2011)\citenamefont {Porto},
  \citenamefont {Ross},\ and\ \citenamefont {Rothstein}}]{rads1}%
  \BibitemOpen
  \bibfield  {author} {\bibinfo {author} {\bibfnamefont {R.~A.}\ \bibnamefont
  {Porto}}, \bibinfo {author} {\bibfnamefont {A.}~\bibnamefont {Ross}}, \ and\
  \bibinfo {author} {\bibfnamefont {I.~Z.}\ \bibnamefont {Rothstein}},\ }\href
  {\doibase 10.1088/1475-7516/2011/03/009} {\bibfield  {journal} {\bibinfo
  {journal} {JCAP}\ }\textbf {\bibinfo {volume} {1103}},\ \bibinfo {pages}
  {009} (\bibinfo {year} {2011})},\ \Eprint {http://arxiv.org/abs/1007.1312}
  {arXiv:1007.1312} \BibitemShut {NoStop}%
\bibitem [{\citenamefont {Porto}\ \emph {et~al.}(2012)\citenamefont {Porto},
  \citenamefont {Ross},\ and\ \citenamefont {Rothstein}}]{amps}%
  \BibitemOpen
  \bibfield  {author} {\bibinfo {author} {\bibfnamefont {R.~A.}\ \bibnamefont
  {Porto}}, \bibinfo {author} {\bibfnamefont {A.}~\bibnamefont {Ross}}, \ and\
  \bibinfo {author} {\bibfnamefont {I.~Z.}\ \bibnamefont {Rothstein}},\ }\href
  {\doibase 10.1088/1475-7516/2012/09/028} {\bibfield  {journal} {\bibinfo
  {journal} {JCAP}\ }\textbf {\bibinfo {volume} {1209}},\ \bibinfo {pages}
  {028} (\bibinfo {year} {2012})},\ \Eprint {http://arxiv.org/abs/1203.2962}
  {arXiv:1203.2962} \BibitemShut {NoStop}%
\bibitem [{\citenamefont {Maia}\ \emph
  {et~al.}(2017{\natexlab{a}})\citenamefont {Maia}, \citenamefont {Galley},
  \citenamefont {Leibovich},\ and\ \citenamefont {Porto}}]{maiaso}%
  \BibitemOpen
  \bibfield  {author} {\bibinfo {author} {\bibfnamefont {N.~T.}\ \bibnamefont
  {Maia}}, \bibinfo {author} {\bibfnamefont {C.~R.}\ \bibnamefont {Galley}},
  \bibinfo {author} {\bibfnamefont {A.~K.}\ \bibnamefont {Leibovich}}, \ and\
  \bibinfo {author} {\bibfnamefont {R.~A.}\ \bibnamefont {Porto}},\ }\href
  {\doibase 10.1103/PhysRevD.96.084064} {\bibfield  {journal} {\bibinfo
  {journal} {Phys. Rev. D}\ }\textbf {\bibinfo {volume} {96}},\ \bibinfo
  {pages} {084064} (\bibinfo {year} {2017}{\natexlab{a}})},\ \Eprint
  {http://arxiv.org/abs/1705.07934} {arXiv:1705.07934} \BibitemShut {NoStop}%
\bibitem [{\citenamefont {Maia}\ \emph
  {et~al.}(2017{\natexlab{b}})\citenamefont {Maia}, \citenamefont {Galley},
  \citenamefont {Leibovich},\ and\ \citenamefont {Porto}}]{maiass}%
  \BibitemOpen
  \bibfield  {author} {\bibinfo {author} {\bibfnamefont {N.~T.}\ \bibnamefont
  {Maia}}, \bibinfo {author} {\bibfnamefont {C.~R.}\ \bibnamefont {Galley}},
  \bibinfo {author} {\bibfnamefont {A.~K.}\ \bibnamefont {Leibovich}}, \ and\
  \bibinfo {author} {\bibfnamefont {R.~A.}\ \bibnamefont {Porto}},\ }\href
  {\doibase 10.1103/PhysRevD.96.084065} {\bibfield  {journal} {\bibinfo
  {journal} {Phys. Rev. D}\ }\textbf {\bibinfo {volume} {96}},\ \bibinfo
  {pages} {084065} (\bibinfo {year} {2017}{\natexlab{b}})},\ \Eprint
  {http://arxiv.org/abs/1705.07938} {arXiv:1705.07938} \BibitemShut {NoStop}%
\bibitem [{\citenamefont {Levi}\ and\ \citenamefont {Steinhoff}(2016)}]{levi}%
  \BibitemOpen
  \bibfield  {author} {\bibinfo {author} {\bibfnamefont {M.}~\bibnamefont
  {Levi}}\ and\ \bibinfo {author} {\bibfnamefont {J.}~\bibnamefont
  {Steinhoff}},\ }\href@noop {} {\  (\bibinfo {year} {2016})},\ \Eprint
  {http://arxiv.org/abs/1607.04252} {arXiv:1607.04252} \BibitemShut {NoStop}%
\bibitem [{\citenamefont {Levi}\ \emph
  {et~al.}(2020{\natexlab{a}})\citenamefont {Levi}, \citenamefont {Mcleod},\
  and\ \citenamefont {Von~Hippel}}]{Levi:2020kvb}%
  \BibitemOpen
  \bibfield  {author} {\bibinfo {author} {\bibfnamefont {M.}~\bibnamefont
  {Levi}}, \bibinfo {author} {\bibfnamefont {A.~J.}\ \bibnamefont {Mcleod}}, \
  and\ \bibinfo {author} {\bibfnamefont {M.}~\bibnamefont {Von~Hippel}},\
  }\href@noop {} {\  (\bibinfo {year} {2020}{\natexlab{a}})},\ \Eprint
  {http://arxiv.org/abs/2003.02827} {arXiv:2003.02827} \BibitemShut {NoStop}%
\bibitem [{\citenamefont {Levi}\ \emph
  {et~al.}(2020{\natexlab{b}})\citenamefont {Levi}, \citenamefont {Mcleod},\
  and\ \citenamefont {Von~Hippel}}]{Levi:2020uwu}%
  \BibitemOpen
  \bibfield  {author} {\bibinfo {author} {\bibfnamefont {M.}~\bibnamefont
  {Levi}}, \bibinfo {author} {\bibfnamefont {A.~J.}\ \bibnamefont {Mcleod}}, \
  and\ \bibinfo {author} {\bibfnamefont {M.}~\bibnamefont {Von~Hippel}},\
  }\href@noop {} {\  (\bibinfo {year} {2020}{\natexlab{b}})},\ \Eprint
  {http://arxiv.org/abs/2003.07890} {arXiv:2003.07890} \BibitemShut {NoStop}%
\bibitem [{\citenamefont {Goldberger}(2007)}]{walterLH}%
  \BibitemOpen
  \bibfield  {author} {\bibinfo {author} {\bibfnamefont {W.~D.}\ \bibnamefont
  {Goldberger}},\ }in\ \href@noop {} {\emph {\bibinfo {booktitle} {Les Houches
  Summer School - Session 86}}}\ (\bibinfo {year} {2007})\ \Eprint
  {http://arxiv.org/abs/hep-ph/0701129} {arXiv:hep-ph/0701129} \BibitemShut
  {NoStop}%
\bibitem [{\citenamefont {Foffa}\ and\ \citenamefont {Sturani}(2014)}]{foffa}%
  \BibitemOpen
  \bibfield  {author} {\bibinfo {author} {\bibfnamefont {S.}~\bibnamefont
  {Foffa}}\ and\ \bibinfo {author} {\bibfnamefont {R.}~\bibnamefont
  {Sturani}},\ }\href {\doibase 10.1088/0264-9381/31/4/043001} {\bibfield
  {journal} {\bibinfo  {journal} {Class. Quant. Grav.}\ }\textbf {\bibinfo
  {volume} {31}},\ \bibinfo {pages} {043001} (\bibinfo {year} {2014})},\
  \Eprint {http://arxiv.org/abs/1309.3474} {arXiv:1309.3474} \BibitemShut
  {NoStop}%
\bibitem [{\citenamefont {Rothstein}(2014)}]{iragrg}%
  \BibitemOpen
  \bibfield  {author} {\bibinfo {author} {\bibfnamefont {I.}~\bibnamefont
  {Rothstein}},\ }\href {\doibase 10.1007/s10714-014-1726-y} {\bibfield
  {journal} {\bibinfo  {journal} {Gen. Rel. Grav.}\ }\textbf {\bibinfo {volume}
  {46}},\ \bibinfo {pages} {1726} (\bibinfo {year} {2014})}\BibitemShut
  {NoStop}%
\bibitem [{\citenamefont {Cardoso}\ and\ \citenamefont {Porto}(2014)}]{grg13}%
  \BibitemOpen
  \bibfield  {author} {\bibinfo {author} {\bibfnamefont {V.}~\bibnamefont
  {Cardoso}}\ and\ \bibinfo {author} {\bibfnamefont {R.~A.}\ \bibnamefont
  {Porto}},\ }\href {\doibase 10.1007/s10714-014-1682-6} {\bibfield  {journal}
  {\bibinfo  {journal} {Gen. Rel. Grav.}\ }\textbf {\bibinfo {volume} {46}},\
  \bibinfo {pages} {1682} (\bibinfo {year} {2014})},\ \Eprint
  {http://arxiv.org/abs/1401.2193} {arXiv:1401.2193} \BibitemShut {NoStop}%
\bibitem [{\citenamefont {Porto}(2016{\natexlab{b}})}]{review}%
  \BibitemOpen
  \bibfield  {author} {\bibinfo {author} {\bibfnamefont {R.~A.}\ \bibnamefont
  {Porto}},\ }\href {\doibase 10.1016/j.physrep.2016.04.003} {\bibfield
  {journal} {\bibinfo  {journal} {Phys. Rept.}\ }\textbf {\bibinfo {volume}
  {633}},\ \bibinfo {pages} {1} (\bibinfo {year} {2016}{\natexlab{b}})},\
  \Eprint {http://arxiv.org/abs/1601.04914} {arXiv:1601.04914} \BibitemShut
  {NoStop}%
\bibitem [{\citenamefont {Henry}\ \emph {et~al.}(2020)\citenamefont {Henry},
  \citenamefont {Faye},\ and\ \citenamefont {Blanchet}}]{luc7pn}%
  \BibitemOpen
  \bibfield  {author} {\bibinfo {author} {\bibfnamefont {Q.}~\bibnamefont
  {Henry}}, \bibinfo {author} {\bibfnamefont {G.}~\bibnamefont {Faye}}, \ and\
  \bibinfo {author} {\bibfnamefont {L.}~\bibnamefont {Blanchet}},\ }\href
  {\doibase 10.1103/PhysRevD.101.064047} {\bibfield  {journal} {\bibinfo
  {journal} {Phys. Rev. D}\ }\textbf {\bibinfo {volume} {101}},\ \bibinfo
  {pages} {064047} (\bibinfo {year} {2020})},\ \Eprint
  {http://arxiv.org/abs/1912.01920} {arXiv:1912.01920} \BibitemShut {NoStop}%
\bibitem [{\citenamefont {Neill}\ and\ \citenamefont {Rothstein}(2013)}]{ira1}%
  \BibitemOpen
  \bibfield  {author} {\bibinfo {author} {\bibfnamefont {D.}~\bibnamefont
  {Neill}}\ and\ \bibinfo {author} {\bibfnamefont {I.~Z.}\ \bibnamefont
  {Rothstein}},\ }\href {\doibase 10.1016/j.nuclphysb.2013.09.007} {\bibfield
  {journal} {\bibinfo  {journal} {Nucl. Phys.}\ }\textbf {\bibinfo {volume}
  {B877}},\ \bibinfo {pages} {177} (\bibinfo {year} {2013})},\ \Eprint
  {http://arxiv.org/abs/1304.7263} {arXiv:1304.7263} \BibitemShut {NoStop}%
\bibitem [{\citenamefont {Cheung}\ \emph {et~al.}(2018)\citenamefont {Cheung},
  \citenamefont {Rothstein},\ and\ \citenamefont {Solon}}]{cheung}%
  \BibitemOpen
  \bibfield  {author} {\bibinfo {author} {\bibfnamefont {C.}~\bibnamefont
  {Cheung}}, \bibinfo {author} {\bibfnamefont {I.~Z.}\ \bibnamefont
  {Rothstein}}, \ and\ \bibinfo {author} {\bibfnamefont {M.~P.}\ \bibnamefont
  {Solon}},\ }\href {\doibase 10.1103/PhysRevLett.121.251101} {\bibfield
  {journal} {\bibinfo  {journal} {Phys. Rev. Lett.}\ }\textbf {\bibinfo
  {volume} {121}},\ \bibinfo {pages} {251101} (\bibinfo {year} {2018})},\
  \Eprint {http://arxiv.org/abs/1808.02489} {arXiv:1808.02489} \BibitemShut
  {NoStop}%
\bibitem [{\citenamefont {Bjerrum-Bohr}\ \emph {et~al.}(2018)\citenamefont
  {Bjerrum-Bohr} \emph {et~al.}}]{bohr}%
  \BibitemOpen
  \bibfield  {author} {\bibinfo {author} {\bibfnamefont {N.~E.~J.}\
  \bibnamefont {Bjerrum-Bohr}} \emph {et~al.},\ }\href {\doibase
  10.1103/PhysRevLett.121.171601} {\bibfield  {journal} {\bibinfo  {journal}
  {Phys. Rev. Lett.}\ }\textbf {\bibinfo {volume} {121}},\ \bibinfo {pages}
  {171601} (\bibinfo {year} {2018})},\ \Eprint
  {http://arxiv.org/abs/1806.04920} {arXiv:1806.04920} \BibitemShut {NoStop}%
\bibitem [{\citenamefont {Bern}\ \emph
  {et~al.}(2019{\natexlab{a}})\citenamefont {Bern}, \citenamefont {Cheung},
  \citenamefont {Roiban}, \citenamefont {Shen}, \citenamefont {Solon},\ and\
  \citenamefont {Zeng}}]{zvi1}%
  \BibitemOpen
  \bibfield  {author} {\bibinfo {author} {\bibfnamefont {Z.}~\bibnamefont
  {Bern}}, \bibinfo {author} {\bibfnamefont {C.}~\bibnamefont {Cheung}},
  \bibinfo {author} {\bibfnamefont {R.}~\bibnamefont {Roiban}}, \bibinfo
  {author} {\bibfnamefont {C.-H.}\ \bibnamefont {Shen}}, \bibinfo {author}
  {\bibfnamefont {M.~P.}\ \bibnamefont {Solon}}, \ and\ \bibinfo {author}
  {\bibfnamefont {M.}~\bibnamefont {Zeng}},\ }\href {\doibase
  10.1103/PhysRevLett.122.201603} {\bibfield  {journal} {\bibinfo  {journal}
  {Phys. Rev. Lett.}\ }\textbf {\bibinfo {volume} {122}},\ \bibinfo {pages}
  {201603} (\bibinfo {year} {2019}{\natexlab{a}})},\ \Eprint
  {http://arxiv.org/abs/1901.04424} {arXiv:1901.04424} \BibitemShut {NoStop}%
\bibitem [{\citenamefont {Bern}\ \emph
  {et~al.}(2019{\natexlab{b}})\citenamefont {Bern}, \citenamefont {Cheung},
  \citenamefont {Roiban}, \citenamefont {Shen}, \citenamefont {Solon},\ and\
  \citenamefont {Zeng}}]{zvi2}%
  \BibitemOpen
  \bibfield  {author} {\bibinfo {author} {\bibfnamefont {Z.}~\bibnamefont
  {Bern}}, \bibinfo {author} {\bibfnamefont {C.}~\bibnamefont {Cheung}},
  \bibinfo {author} {\bibfnamefont {R.}~\bibnamefont {Roiban}}, \bibinfo
  {author} {\bibfnamefont {C.-H.}\ \bibnamefont {Shen}}, \bibinfo {author}
  {\bibfnamefont {M.~P.}\ \bibnamefont {Solon}}, \ and\ \bibinfo {author}
  {\bibfnamefont {M.}~\bibnamefont {Zeng}},\ }\href {\doibase
  10.1007/JHEP10(2019)206} {\bibfield  {journal} {\bibinfo  {journal} {JHEP}\
  }\textbf {\bibinfo {volume} {10}},\ \bibinfo {pages} {206} (\bibinfo {year}
  {2019}{\natexlab{b}})},\ \Eprint {http://arxiv.org/abs/1908.01493}
  {arXiv:1908.01493} \BibitemShut {NoStop}%
\bibitem [{\citenamefont {Cristofoli}\ \emph {et~al.}(2019)\citenamefont
  {Cristofoli}, \citenamefont {Bjerrum-Bohr}, \citenamefont {Damgaard},\ and\
  \citenamefont {Vanhove}}]{cristof1}%
  \BibitemOpen
  \bibfield  {author} {\bibinfo {author} {\bibfnamefont {A.}~\bibnamefont
  {Cristofoli}}, \bibinfo {author} {\bibfnamefont {N.~E.~J.}\ \bibnamefont
  {Bjerrum-Bohr}}, \bibinfo {author} {\bibfnamefont {P.~H.}\ \bibnamefont
  {Damgaard}}, \ and\ \bibinfo {author} {\bibfnamefont {P.}~\bibnamefont
  {Vanhove}},\ }\href@noop {} {\  (\bibinfo {year} {2019})},\ \Eprint
  {http://arxiv.org/abs/1906.01579} {arXiv:1906.01579} \BibitemShut {NoStop}%
\bibitem [{\citenamefont {Kosower}\ \emph {et~al.}(2019)\citenamefont
  {Kosower}, \citenamefont {Maybee},\ and\ \citenamefont {O'Connell}}]{donal}%
  \BibitemOpen
  \bibfield  {author} {\bibinfo {author} {\bibfnamefont {D.~A.}\ \bibnamefont
  {Kosower}}, \bibinfo {author} {\bibfnamefont {B.}~\bibnamefont {Maybee}}, \
  and\ \bibinfo {author} {\bibfnamefont {D.}~\bibnamefont {O'Connell}},\ }\href
  {\doibase 10.1007/JHEP02(2019)137} {\bibfield  {journal} {\bibinfo  {journal}
  {JHEP}\ }\textbf {\bibinfo {volume} {02}},\ \bibinfo {pages} {137} (\bibinfo
  {year} {2019})},\ \Eprint {http://arxiv.org/abs/1811.10950}
  {arXiv:1811.10950} \BibitemShut {NoStop}%
\bibitem [{\citenamefont {Cheung}\ and\ \citenamefont
  {Solon}(2020{\natexlab{a}})}]{Cheung:2020gyp}%
  \BibitemOpen
  \bibfield  {author} {\bibinfo {author} {\bibfnamefont {C.}~\bibnamefont
  {Cheung}}\ and\ \bibinfo {author} {\bibfnamefont {M.~P.}\ \bibnamefont
  {Solon}},\ }\href {\doibase 10.1007/JHEP06(2020)144} {\bibfield  {journal}
  {\bibinfo  {journal} {JHEP}\ }\textbf {\bibinfo {volume} {06}},\ \bibinfo
  {pages} {144} (\bibinfo {year} {2020}{\natexlab{a}})},\ \Eprint
  {http://arxiv.org/abs/2003.08351} {arXiv:2003.08351} \BibitemShut {NoStop}%
\bibitem [{\citenamefont {Bern}\ \emph {et~al.}(2020)\citenamefont {Bern},
  \citenamefont {Luna}, \citenamefont {Roiban}, \citenamefont {Shen},\ and\
  \citenamefont {Zeng}}]{Bern:2020buy}%
  \BibitemOpen
  \bibfield  {author} {\bibinfo {author} {\bibfnamefont {Z.}~\bibnamefont
  {Bern}}, \bibinfo {author} {\bibfnamefont {A.}~\bibnamefont {Luna}}, \bibinfo
  {author} {\bibfnamefont {R.}~\bibnamefont {Roiban}}, \bibinfo {author}
  {\bibfnamefont {C.-H.}\ \bibnamefont {Shen}}, \ and\ \bibinfo {author}
  {\bibfnamefont {M.}~\bibnamefont {Zeng}},\ }\href@noop {} {\  (\bibinfo
  {year} {2020})},\ \Eprint {http://arxiv.org/abs/2005.03071}
  {arXiv:2005.03071} \BibitemShut {NoStop}%
\bibitem [{\citenamefont {Parra-Martinez}\ \emph {et~al.}(2020)\citenamefont
  {Parra-Martinez}, \citenamefont {Ruf},\ and\ \citenamefont {Zeng}}]{Parra}%
  \BibitemOpen
  \bibfield  {author} {\bibinfo {author} {\bibfnamefont {J.}~\bibnamefont
  {Parra-Martinez}}, \bibinfo {author} {\bibfnamefont {M.~S.}\ \bibnamefont
  {Ruf}}, \ and\ \bibinfo {author} {\bibfnamefont {M.}~\bibnamefont {Zeng}},\
  }\href@noop {} {\  (\bibinfo {year} {2020})},\ \Eprint
  {http://arxiv.org/abs/2005.04236} {arXiv:2005.04236} \BibitemShut {NoStop}%
\bibitem [{\citenamefont {K{\"a}lin}\ and\ \citenamefont
  {Porto}(2020{\natexlab{a}})}]{paper1}%
  \BibitemOpen
  \bibfield  {author} {\bibinfo {author} {\bibfnamefont {G.}~\bibnamefont
  {K{\"a}lin}}\ and\ \bibinfo {author} {\bibfnamefont {R.~A.}\ \bibnamefont
  {Porto}},\ }\href {\doibase 10.1007/JHEP01(2020)072} {\bibfield  {journal}
  {\bibinfo  {journal} {JHEP}\ }\textbf {\bibinfo {volume} {01}},\ \bibinfo
  {pages} {072} (\bibinfo {year} {2020}{\natexlab{a}})},\ \Eprint
  {http://arxiv.org/abs/1910.03008} {arXiv:1910.03008} \BibitemShut {NoStop}%
\bibitem [{\citenamefont {K{\"a}lin}\ and\ \citenamefont
  {Porto}(2020{\natexlab{b}})}]{paper2}%
  \BibitemOpen
  \bibfield  {author} {\bibinfo {author} {\bibfnamefont {G.}~\bibnamefont
  {K{\"a}lin}}\ and\ \bibinfo {author} {\bibfnamefont {R.~A.}\ \bibnamefont
  {Porto}},\ }\href {\doibase 10.1007/JHEP02(2020)120} {\bibfield  {journal}
  {\bibinfo  {journal} {JHEP}\ }\textbf {\bibinfo {volume} {02}},\ \bibinfo
  {pages} {120} (\bibinfo {year} {2020}{\natexlab{b}})},\ \Eprint
  {http://arxiv.org/abs/1911.09130} {arXiv:1911.09130} \BibitemShut {NoStop}%
\bibitem [{\citenamefont {Damour}(2016)}]{damour1}%
  \BibitemOpen
  \bibfield  {author} {\bibinfo {author} {\bibfnamefont {T.}~\bibnamefont
  {Damour}},\ }\href {\doibase 10.1103/PhysRevD.94.104015} {\bibfield
  {journal} {\bibinfo  {journal} {Phys. Rev.}\ }\textbf {\bibinfo {volume}
  {D94}},\ \bibinfo {pages} {104015} (\bibinfo {year} {2016})},\ \Eprint
  {http://arxiv.org/abs/1609.00354} {arXiv:1609.00354} \BibitemShut {NoStop}%
\bibitem [{\citenamefont {Damour}(2018)}]{damour2}%
  \BibitemOpen
  \bibfield  {author} {\bibinfo {author} {\bibfnamefont {T.}~\bibnamefont
  {Damour}},\ }\href {\doibase 10.1103/PhysRevD.97.044038} {\bibfield
  {journal} {\bibinfo  {journal} {Phys. Rev.}\ }\textbf {\bibinfo {volume}
  {D97}},\ \bibinfo {pages} {044038} (\bibinfo {year} {2018})},\ \Eprint
  {http://arxiv.org/abs/1710.10599} {arXiv:1710.10599} \BibitemShut {NoStop}%
\bibitem [{\citenamefont {Bini}\ \emph {et~al.}(2019)\citenamefont {Bini},
  \citenamefont {Damour},\ and\ \citenamefont {Geralico}}]{damour3n}%
  \BibitemOpen
  \bibfield  {author} {\bibinfo {author} {\bibfnamefont {D.}~\bibnamefont
  {Bini}}, \bibinfo {author} {\bibfnamefont {T.}~\bibnamefont {Damour}}, \ and\
  \bibinfo {author} {\bibfnamefont {A.}~\bibnamefont {Geralico}},\ }\href@noop
  {} {\bibfield  {journal} {\bibinfo  {journal} {Phys. Rev. Lett.}\ }\textbf
  {\bibinfo {volume} {123}},\ \bibinfo {pages} {231104} (\bibinfo {year}
  {2019})},\ \Eprint {http://arxiv.org/abs/1909.02375} {arXiv:1909.02375
  [gr-qc]} \BibitemShut {NoStop}%
\bibitem [{\citenamefont {Damour}(2019)}]{damour3}%
  \BibitemOpen
  \bibfield  {author} {\bibinfo {author} {\bibfnamefont {T.}~\bibnamefont
  {Damour}},\ }\href@noop {} {\  (\bibinfo {year} {2019})},\ \Eprint
  {http://arxiv.org/abs/1912.02139} {arXiv:1912.02139} \BibitemShut {NoStop}%
\bibitem [{\citenamefont {Bini}\ \emph {et~al.}(2020)\citenamefont {Bini},
  \citenamefont {Damour},\ and\ \citenamefont {Geralico}}]{binit}%
  \BibitemOpen
  \bibfield  {author} {\bibinfo {author} {\bibfnamefont {D.}~\bibnamefont
  {Bini}}, \bibinfo {author} {\bibfnamefont {T.}~\bibnamefont {Damour}}, \ and\
  \bibinfo {author} {\bibfnamefont {A.}~\bibnamefont {Geralico}},\ }\href
  {\doibase 10.1103/PhysRevD.101.044039} {\bibfield  {journal} {\bibinfo
  {journal} {Phys. Rev. D}\ }\textbf {\bibinfo {volume} {101}},\ \bibinfo
  {pages} {044039} (\bibinfo {year} {2020})},\ \Eprint
  {http://arxiv.org/abs/2001.00352} {arXiv:2001.00352} \BibitemShut {NoStop}%
\bibitem [{\citenamefont {K{\"a}lin}\ and\ \citenamefont
  {Porto}(2020{\natexlab{c}})}]{pmeft}%
  \BibitemOpen
  \bibfield  {author} {\bibinfo {author} {\bibfnamefont {G.}~\bibnamefont
  {K{\"a}lin}}\ and\ \bibinfo {author} {\bibfnamefont {R.~A.}\ \bibnamefont
  {Porto}},\ }\href@noop {} {\  (\bibinfo {year} {2020}{\natexlab{c}})},\
  \Eprint {http://arxiv.org/abs/2006.01184} {arXiv:2006.01184} \BibitemShut
  {NoStop}%
\bibitem [{\citenamefont {K{\"a}lin}\ \emph {et~al.}(2020)\citenamefont
  {K{\"a}lin}, \citenamefont {Liu},\ and\ \citenamefont {Porto}}]{3pmeft}%
  \BibitemOpen
  \bibfield  {author} {\bibinfo {author} {\bibfnamefont {G.}~\bibnamefont
  {K{\"a}lin}}, \bibinfo {author} {\bibfnamefont {Z.}~\bibnamefont {Liu}}, \
  and\ \bibinfo {author} {\bibfnamefont {R.~A.}\ \bibnamefont {Porto}},\
  }\href@noop {} {\  (\bibinfo {year} {2020})},\ \Eprint
  {http://arxiv.org/abs/2007.04977} {arXiv:2007.04977} \BibitemShut {NoStop}%
\bibitem [{\citenamefont {Haddad}\ and\ \citenamefont
  {Helset}(2020)}]{Haddad:2020que}%
  \BibitemOpen
  \bibfield  {author} {\bibinfo {author} {\bibfnamefont {K.}~\bibnamefont
  {Haddad}}\ and\ \bibinfo {author} {\bibfnamefont {A.}~\bibnamefont
  {Helset}},\ }\href@noop {} {\  (\bibinfo {year} {2020})},\ \Eprint
  {http://arxiv.org/abs/2008.04920} {arXiv:2008.04920} \BibitemShut {NoStop}%
\bibitem [{\citenamefont {Cheung}\ and\ \citenamefont
  {Solon}(2020{\natexlab{b}})}]{soloncheung}%
  \BibitemOpen
  \bibfield  {author} {\bibinfo {author} {\bibfnamefont {C.}~\bibnamefont
  {Cheung}}\ and\ \bibinfo {author} {\bibfnamefont {M.~P.}\ \bibnamefont
  {Solon}},\ }\href@noop {} {\  (\bibinfo {year} {2020}{\natexlab{b}})},\
  \Eprint {http://arxiv.org/abs/2006.06665} {arXiv:2006.06665} \BibitemShut
  {NoStop}%
\bibitem [{\citenamefont {Le~Tiec}\ \emph {et~al.}(2012)\citenamefont
  {Le~Tiec}, \citenamefont {Blanchet},\ and\ \citenamefont {Whiting}}]{letiec}%
  \BibitemOpen
  \bibfield  {author} {\bibinfo {author} {\bibfnamefont {A.}~\bibnamefont
  {Le~Tiec}}, \bibinfo {author} {\bibfnamefont {L.}~\bibnamefont {Blanchet}}, \
  and\ \bibinfo {author} {\bibfnamefont {B.~F.}\ \bibnamefont {Whiting}},\
  }\href {\doibase 10.1103/PhysRevD.85.064039} {\bibfield  {journal} {\bibinfo
  {journal} {Phys. Rev.}\ }\textbf {\bibinfo {volume} {D85}},\ \bibinfo {pages}
  {064039} (\bibinfo {year} {2012})},\ \Eprint {http://arxiv.org/abs/1111.5378}
  {arXiv:1111.5378} \BibitemShut {NoStop}%
\bibitem [{\citenamefont {Abbott}\ \emph
  {et~al.}(2020{\natexlab{a}})\citenamefont {Abbott} \emph {et~al.}}]{gap1}%
  \BibitemOpen
  \bibfield  {author} {\bibinfo {author} {\bibfnamefont {B.}~\bibnamefont
  {Abbott}} \emph {et~al.} (\bibinfo {collaboration} {LIGO Scientific,
  Virgo}),\ }\href@noop {} {\bibfield  {journal} {\bibinfo  {journal}
  {Astrophys. J. Lett.}\ }\textbf {\bibinfo {volume} {892}},\ \bibinfo {pages}
  {L3} (\bibinfo {year} {2020}{\natexlab{a}})},\ \Eprint
  {http://arxiv.org/abs/2001.01761} {arXiv:2001.01761} \BibitemShut {NoStop}%
\bibitem [{\citenamefont {Abbott}\ \emph
  {et~al.}(2020{\natexlab{b}})\citenamefont {Abbott} \emph {et~al.}}]{gap2}%
  \BibitemOpen
  \bibfield  {author} {\bibinfo {author} {\bibfnamefont {R.}~\bibnamefont
  {Abbott}} \emph {et~al.} (\bibinfo {collaboration} {LIGO Scientific,
  Virgo}),\ }\href@noop {} {\bibfield  {journal} {\bibinfo  {journal}
  {Astrophys. J.}\ }\textbf {\bibinfo {volume} {896}},\ \bibinfo {pages} {L44}
  (\bibinfo {year} {2020}{\natexlab{b}})},\ \Eprint
  {http://arxiv.org/abs/2006.12611} {arXiv:2006.12611} \BibitemShut {NoStop}%
\bibitem [{\citenamefont {Binnington}\ and\ \citenamefont
  {Poisson}(2009)}]{Binnington:2009bb}%
  \BibitemOpen
  \bibfield  {author} {\bibinfo {author} {\bibfnamefont {T.}~\bibnamefont
  {Binnington}}\ and\ \bibinfo {author} {\bibfnamefont {E.}~\bibnamefont
  {Poisson}},\ }\href {\doibase 10.1103/PhysRevD.80.084018} {\bibfield
  {journal} {\bibinfo  {journal} {Phys. Rev. D}\ }\textbf {\bibinfo {volume}
  {80}},\ \bibinfo {pages} {084018} (\bibinfo {year} {2009})},\ \Eprint
  {http://arxiv.org/abs/0906.1366} {arXiv:0906.1366} \BibitemShut {NoStop}%
\bibitem [{\citenamefont {Damour}\ and\ \citenamefont
  {Nagar}(2009)}]{Damour:2009vw}%
  \BibitemOpen
  \bibfield  {author} {\bibinfo {author} {\bibfnamefont {T.}~\bibnamefont
  {Damour}}\ and\ \bibinfo {author} {\bibfnamefont {A.}~\bibnamefont {Nagar}},\
  }\href {\doibase 10.1103/PhysRevD.80.084035} {\bibfield  {journal} {\bibinfo
  {journal} {Phys. Rev. D}\ }\textbf {\bibinfo {volume} {80}},\ \bibinfo
  {pages} {084035} (\bibinfo {year} {2009})},\ \Eprint
  {http://arxiv.org/abs/0906.0096} {arXiv:0906.0096} \BibitemShut {NoStop}%
\bibitem [{\citenamefont {Kol}\ and\ \citenamefont
  {Smolkin}(2012)}]{Kol:2011vg}%
  \BibitemOpen
  \bibfield  {author} {\bibinfo {author} {\bibfnamefont {B.}~\bibnamefont
  {Kol}}\ and\ \bibinfo {author} {\bibfnamefont {M.}~\bibnamefont {Smolkin}},\
  }\href {\doibase 10.1007/JHEP02(2012)010} {\bibfield  {journal} {\bibinfo
  {journal} {JHEP}\ }\textbf {\bibinfo {volume} {02}},\ \bibinfo {pages} {010}
  (\bibinfo {year} {2012})},\ \Eprint {http://arxiv.org/abs/1110.3764}
  {arXiv:1110.3764} \BibitemShut {NoStop}%
\bibitem [{\citenamefont {Le~Tiec}\ and\ \citenamefont
  {Casals}(2020)}]{altspin}%
  \BibitemOpen
  \bibfield  {author} {\bibinfo {author} {\bibfnamefont {A.}~\bibnamefont
  {Le~Tiec}}\ and\ \bibinfo {author} {\bibfnamefont {M.}~\bibnamefont
  {Casals}},\ }\href@noop {} {\  (\bibinfo {year} {2020})},\ \Eprint
  {http://arxiv.org/abs/2007.00214} {arXiv:2007.00214} \BibitemShut {NoStop}%
\bibitem [{\citenamefont {Mart\'{i}n-Garc\'{i}a}\ \emph
  {et~al.}(2019)\citenamefont {Mart\'{i}n-Garc\'{i}a} \emph
  {et~al.}}]{martin2019xact}%
  \BibitemOpen
  \bibfield  {author} {\bibinfo {author} {\bibfnamefont {J.~M.}\ \bibnamefont
  {Mart\'{i}n-Garc\'{i}a}} \emph {et~al.},\ }\href@noop {} {\bibfield
  {journal} {\bibinfo  {journal} {{\href{http://www.xact.es/}{\tt
  www.xact.es}}}\ } (\bibinfo {year} {2019})}\BibitemShut {NoStop}%
\end{thebibliography}%
\end{document}